\documentclass[11pt]{article} 
\usepackage[top=20mm,bottom=20mm, left=20mm, right=20mm]{geometry}
\usepackage{graphicx,amsmath,amssymb,url,enumerate,mathrsfs,subfig,epsfig,color,transparent,cite}
\usepackage{hyperref, makecell}
 \hypersetup{
     colorlinks=true,       
     linkcolor=red,          
     citecolor=blue,        
     filecolor=magenta,      
     urlcolor=cyan           
 }

\newtheorem{rem}{\it Remark}

\newcommand{\op}[1]{\skew{5}\hat{#1}}

\title{On structured surfaces with defects: geometry, strain incompatibility, internal stress, and natural shapes}
\author{Ayan Roychowdhury and Anurag Gupta\thanks{ag@iitk.ac.in}
\\Department of Mechanical Engineering, Indian Institute of Technology Kanpur, 208016, India.
}
\date{\today}
\begin{document}
\maketitle
\begin{abstract}
Given a distribution of defects on a structured surface, such as those represented by 2-dimensional crystalline materials, liquid crystalline surfaces, and thin sandwiched shells, what is the resulting stress field and the deformed shape? Motivated by this concern, we first classify, and quantify, the translational, rotational, and metrical defects allowable over a broad class of structured surfaces. With an appropriate notion of strain, the defect densities are then shown to appear as sources of strain incompatibility. The strain incompatibility relations, with appropriate kinematical assumptions on the decomposition of strain into elastic and plastic parts, and the stress equilibrium relations, with a suitable choice of material response, provide the necessary equations for determining both the internal stress field and the deformed shape. We demonstrate this by applying our theory to Kirchhoff-Love shells with a kinematics which allows for small in-surface strains but moderately large rotations. 
\end{abstract}

{\small


}

\section{Introduction}
The aim of this article is to study geometry and mechanics of defects in structured surfaces.
The term {\it structured surface} is used to represent a variety of 2-dimensional material surfaces such as 2-dimensional crystals, with intrinsic translational, rotational, and metrical order (colloidosomes, carbon nanotubes, graphene etc.); thin sandwiched structures; liquid crystalline membranes and shells, with intrinsic crystalline order (single-layer viral capsids) or without (nematic membranes, single layers in smectics and cholesterics); and Cosserat surfaces, which are used to model a hierarchy of plate and shell theories for thin elastic structures abundant in structural engineering applications.
 The defects are anomalies within the local arrangement of entities in an ordered structure where the order is usually defined in terms of translational, rotational, and metrical symmetries of the underlying material. These anomalies are omnipresent in nature, e.g., 2-dimensional materials such as graphene are known to contain edge dislocations (translational anomalies), wedge disclinations (rotational anomalies), and point imperfections (metric anomalies) such as vacancies and self-interstitials; on the other hand, twist disclinations are commonly observed in lipid membranes. More examples are given in the following section as well as in an extensive review of the subject in \cite{bowickgiomi09}. The phenomena of thermal deformation and biological growth can also be categorized as those leading to metric anomalies. Many of the superior physicochemical properties of the 2-dimensional defective structures can be attributed to the internal stress fields resulting from the distribution of defects \cite{ZouYakobson15}, and also, unlike 3-dimensional bodies, due to their lower dimensionality, to their ability to relax by acquiring a variety of natural (stress-free) shapes, for instance, the wavy edges of growing leaves \cite{liang-mahadevan09}, the  topological corrugations present on human brain \cite{tallinenetal16}, helical strands of DNA \cite{efrati15}, among other examples. The present work is concerned with the central problem of formulating a general theory that takes under its ambit the geometric characterization of these multifarious 2-dimensional defective structures and also the determination of their internal stress fields and deformed shapes. 

Non-Euclidean differential geometry has been established to provide the necessary mathematical infrastructure to describe the geometric nature of defects in 3-dimensional solids, as well as to provide a rightful setting to discuss the related issues of strain incompatibility and residual stress distribution \cite{kondo52, bilby55, anth70, anth71, noll, kroner81a, dewit81, clayton-book}. Despite this success in 3-dimensions, the problem in lower dimensional structures is relatively less developed, primarily due to the complex interplay between the embedding geometries in the physical space, and the unavoidable non-linearities involved in the deformation as well as the constitutive response of 2-dimensional matter. We note the initial attempts made by Eshelby \cite{eshelbystroh51, eshelby79} where analytical solutions for internal stress were derived for linearly elastic plates containing isolated screw and edge dislocations. This work was extended in several directions by Chernykh \cite{chernykh} and Nabarro \cite{nabarro70, nabarrokostlan78}, among others \cite{saitoetal72, mosshoover78, SeungNelson88}. A theory of continuous distribution of defects over thin structures was first developed by Povstenko \cite{povs85, povs91} by drawing analogies from the non-Euclidean description of continuous distribution of defects in 3-dimensional elastic bodies. Povstenko introduced the notions of in-surface dislocations (characterized by the in-surface torsion tensor), disclinations (characterized by the in-surface curvature tensor), and metric anomalies (characterized by the in-surface non-metricity tensor). He also provided the non-linear conservation laws for all the in-surface defect density fields as direct consequences of the Bianchi-Padova relations in two dimensions. 

The compatibility conditions for strain fields in geometrically linear and non-linear shells, and Cosserat surfaces, are extensively discussed in existing works \cite{koiter1966, malcolm, reissner, eps1}. The presence of defects, however, introduces incompatibility in strain fields. The strain incompatibility equations for both linear elastic plates and von-K{\'a}rm{\'a}n plates, with in-surface anomalies, have been derived by Zubov and Derezin \cite{zubov1, zubov2, derezin, derezin-zubov,zhbanova2016}; these also include solutions of certain special boundary-value-problems for determining stress and natural shape of the defective plate. The strain incompatibility equations have also appeared in the recent works on non-Euclidean elastic plates \cite{efrati09,klein-efrati}, and growth and morpho-elasticity of thin biological materials \cite{mcmahonetal1, mcmahonetal2, liang-mahadevan09, lewicka-mahadevan, dervauxetal}. Without explicitly incorporating defect densities, these works consider a non-Euclidean metric, representing the distribution of plastic/growth strain field, and use the Riemannian curvature of this metric, which is the measure of strain incompatibility, along with the strain decomposition, to pose boundary-value-problems primarily for determining natural shapes. 

The concepts of material uniformity, material symmetry, and inhomogeneity in elastic Cosserat surfaces, following the pioneering works of Noll \cite{noll} and Wang \cite{wang}, are also firmly established \cite{ericksen70,wangshell,wangshell1,eps2,edeleon, EremeyevPietraszkiewicz2006}, although 
these works have neither attempted to describe the inhomogeneity distribution in terms of the curvature and non-metricity (the notion of torsion does appear in some of these works), nor have they discussed the relevant issue of strain incompatibility. A theory of materially uniform, inhomogeneous (dislocated) thin elastic films, derived from a 3-dimensional uniform, inhomogeneous (dislocated) elastic body, has been recently proposed by Steigmann \cite{steigmann14}, and applied to determining the natural shapes of plastically deformed thin sheets \cite{steigmannetal14}. Finally, we mention, only in passing, the extensive work on mechanics of topologically defective (`geometrically frustrated') liquid crystalline surfaces \cite{rossoetal12, bowickgiomi09, bowick-travesset2001, cesana2015}, which, in contrast to the local theories mentioned above, have taken a distinguished local-global (geometrical-topological) standpoint in describing the nature of defects. 

There is a clear lack of a complete non-Euclidean geometric characterization of continuously distributed material defects in 2-dimensional structured continua. While these certainly have analogous descriptions in the 3-dimensional theory, there is a considerable richness in the description of the allowable defects as well as their geometrical properties for the 2-dimensional structure. Additionally, there are no derivations of strain incompatibility relations for sufficiently general kinematic and constitutive response as afforded by most of the known 2-dimensional materials. With this in mind, we present a theory, within the natural setting of non-Euclidean differential geometry, that on one hand unifies the several seemingly different streams of research discussed above, and also provides a rigorously constructed, sufficiently general, framework for studying a large range of problems associated with geometry and mechanics of defective structured surfaces. In particular, we give a complete non-Euclidean characterization of all the translational, rotational, and metric anomalies in structured surfaces, derive the imposed restrictions from Bianchi-Padova relations, and establish general strain incompatibility relations. To illustrate our theory, we consider the specific case of Kirchhoff-Love shells and provide a framework, involving kinematics, additive decomposition, incompatibility relations, and balance laws, for posing complete boundary-value-problems for determining internal stress and deformed shapes for a class of 2-dimensional continuously defective structures undergoing small stretch but moderately large deformation.

A brief overview of the paper is as follows. In Section \ref{natdef}, we provide several illustrative examples to demonstrate the non-Euclidean character of local material defects in structured surfaces. Motivated from Section \ref{natdef}, we begin Section \ref{geom} by introducing the notion of material space, which includes a 2-dimensional body manifold, a non-Riemannian material connection, and a material metric, as our prototype to characterize continuously defective structured surfaces. Both in-surface and out-of-surface material anomalies are taken into account, and are identified with the components of the tensors of non-metricity, torsion, and Riemann-Christoffel curvature of the material connection (see Table \ref{tbl1}). Subsequently, by exploiting Bianchi-Padova relations, we obtain several restrictions, both as algebraic relations and differential equations, on these components while emphasizing their interdependence (see Table \ref{tbl2}). We conclude the section by introducing a Riemannian structure on the material space induced by the material metric. We obtain geometric relations which connect the curvatures of the Riemannian and non-Riemannian spaces. These relations are central to our formulation of strain incompatibility equations in Section \ref{strain}. A generalized notion of strain is introduced which defines the kinematical nature of our structured surface. The strain incompatibility relations lead us to pose complete boundary-value-problems for the determination of internal stress fields and deformed shapes of defective structured surfaces. This is illustrated in Section \ref{bvp-stress-shape} by restricting our attention to Kirchhoff-Love shells. We also postulate an additive decomposition for the in-surface and bending strain fields into elastic and plastic parts, while arguing that this separation of order is sufficiently general to accommodate small in-surface strains with moderately large rotations. The plastic strain fields are to be solved using the incompatibility relations with a given distribution of defects. We show that several existing formulations follow as special cases, in particular the F{\"o}ppl-von-K{\'a}rm{\'a}n equations for continuously defective thin elastic isotropic plates and the shape equations for continuously defective thin isotropic fluid films. We conclude our study in Section \ref{conclusion}.

\section{Nature of surface defects}\label{natdef}

In this section, we provide several illustrative examples of defects in structured surfaces. The defects are understood as anomalies within the \textit{local} arrangement of entities in an ordered structure where the order is usually defined in terms of rotational, translational, and metrical symmetries of the underlying material. Defects can also appear as \textit{global} anomalies which affect the topology of the surface, such as those present in multiply connected and non-orientable surfaces \cite{harris70, harris74, bowickgiomi09}; these are however not discussed in the present work. The following examples are presented with an intent to emphasize the non-Euclidean geometric nature of the defects as is incorporated in the subsequent sections.  In particular, the central idea of our work of embedding the structured surface within a 3-dimensional non-Riemannian geometric space emerges naturally as we proceed through these rudimentary illustrations.

\begin{figure}[t!]
\centering
 \subfloat[]{\includegraphics[scale=0.45]{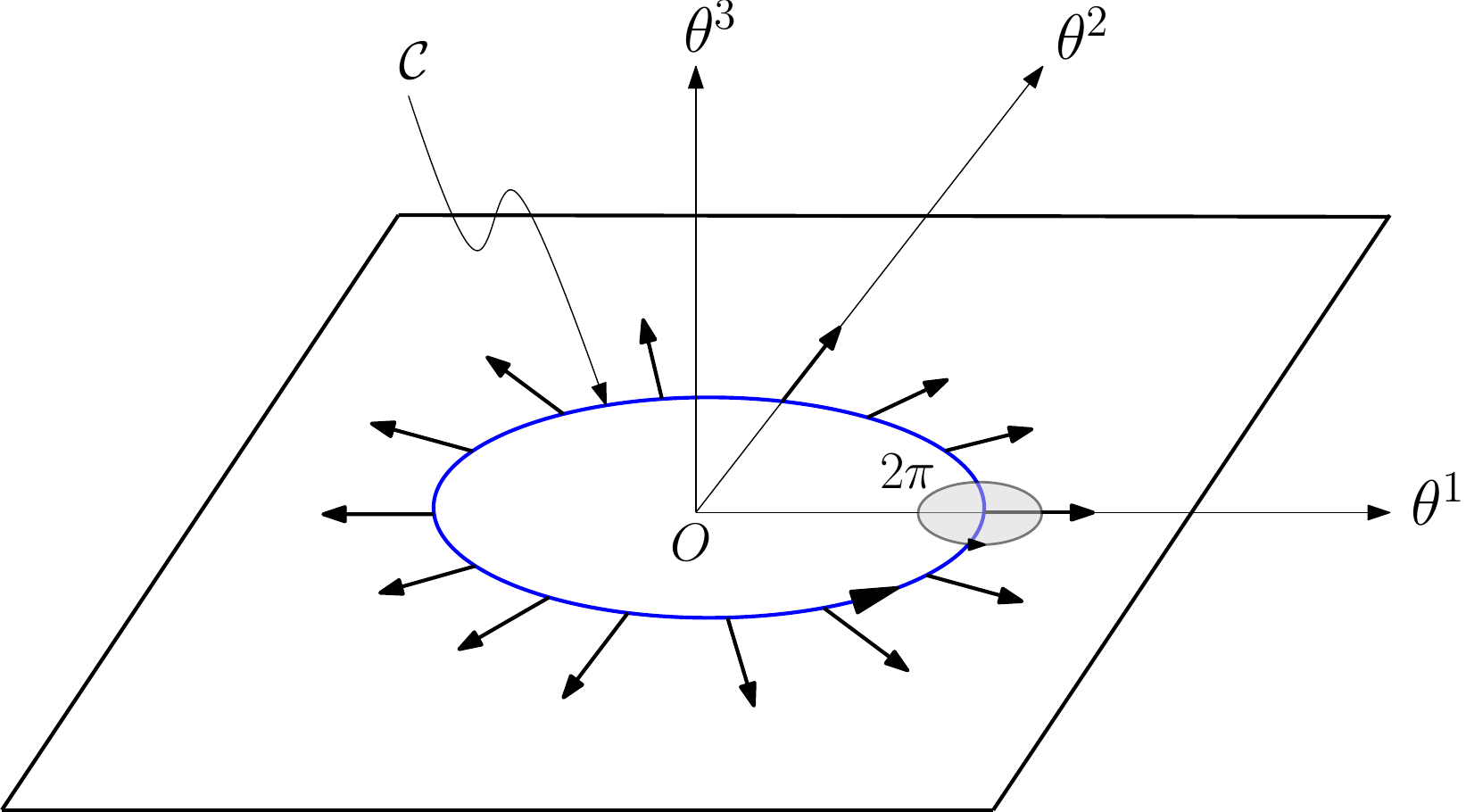}}
 \hspace{10mm}
 \subfloat[]{\includegraphics[scale=0.45]{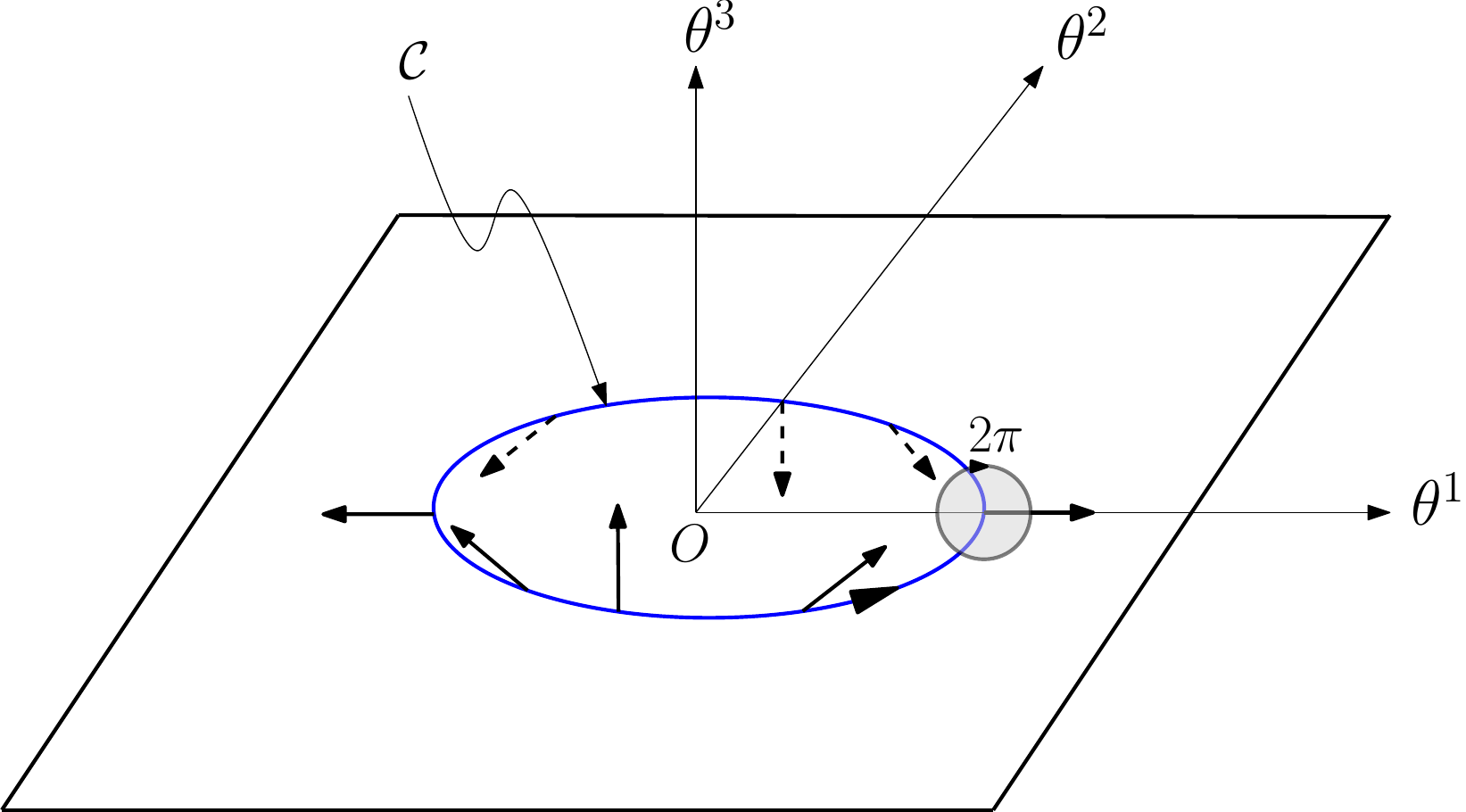}}
 \vspace{5mm}
 \subfloat[]{\includegraphics[scale=0.56]{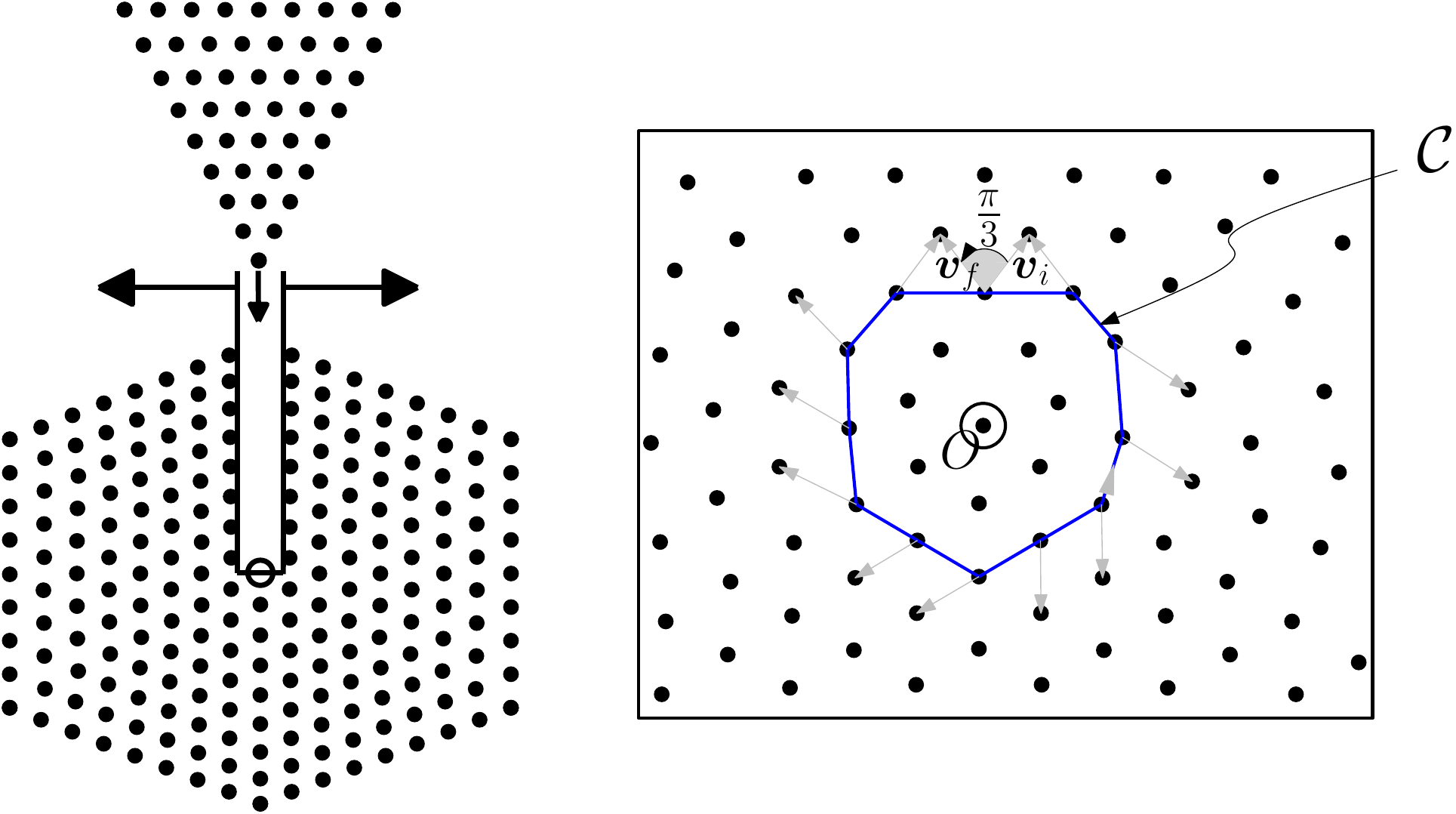}}
\caption{(a) A single wedge disclination of Frank angle $2\pi$ in a nematic membrane, located at $O$, such that $\boldsymbol{d}(\theta^1,\theta^2) = \cos\theta  \mathbf{e}_1 + \sin\theta \mathbf{e}_2$, where $\theta$ is the polar angle $\theta:=\tan^{-1}({\theta^2}/{\theta^1})$. (b) A single twist disclination  of Frank angle $2\pi$ in a nematic shell, such that $\boldsymbol{d}(\theta^1,\theta^2) = \cos\theta  \mathbf{e}_1 - \sin\theta \mathbf{e}_3$. (c) Creation of a wedge disclination of Frank angle ${\pi}/{3}$ in a 2-dimensional hexagonal lattice by cutting the surface along a line and introducing a lattice wedge of angle ${\pi}/{3}$; after \cite{anth02}. The marks on the surface represent lattice points which may carry identical atoms (in case of 2-dimensional crystals) or directors (in case of nematic shells) pointing inward/outward at the respective positions on the surface. The lattice vector, initially at $\boldsymbol{v}_i$, rotates through an angle ${\pi}/{3}$ when circumnavigated along a loop surrounding the disclination.}
\label{discli}
 \end{figure}

\begin{figure}[t!]
\centering\includegraphics[scale=0.56]{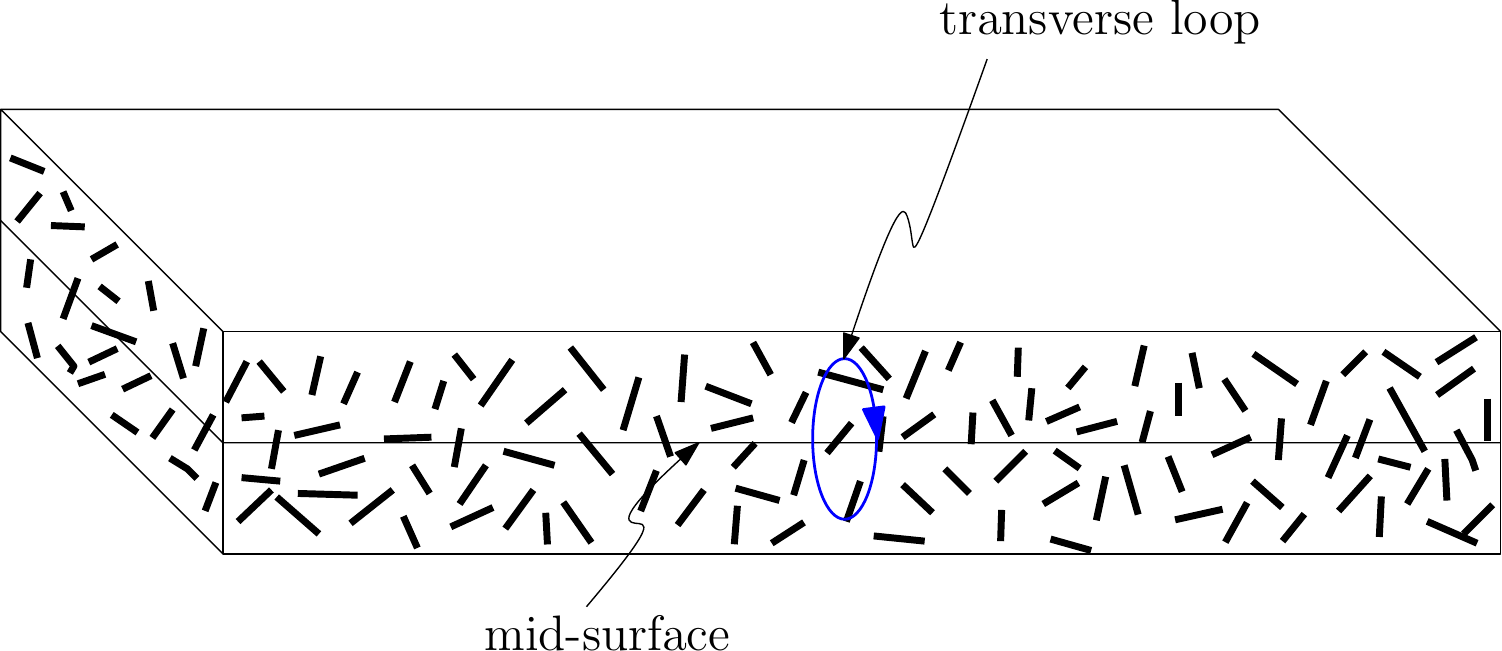}
 \caption{A transverse loop characterizing an effectively 2-dimensional representation of the 3-dimensional distribution of disclinations within a thin layered structure made up of directed media.}
\label{discli-layered}
 \end{figure}
 
The \textit{rotational anomalies} in a structured surface appear in the form of disclinations. Depending on the material nature of the surface, rotational order can be present due to intrinsic crystallinity of the surface (such as in colloidosomes, single-layer viral capsids, carbon nanotubes, and graphene) or due to an extrinsic orientation field (such as in nematic membranes, single layers in smectics, and cholesterics) \cite{bowickgiomi09}. As a result, we distinguish between rotational order, or lack thereof, appearing intrinsically and extrinsically in a surface. We also note that unlike disclinations in 3-dimensional crystalline solids, which have large formation energy and hence are rarely observed \cite{anth02},  disclinations in 2-dimensional crystals are omnipresent since the surface can now relax the energy by escaping into the third dimension. 
Isolated disclinations in structured surfaces without intrinsic crystalline order are shown in 
Figures \ref{discli}(a,b). The rotational order is here present due to a director field distribution, denoted by $\boldsymbol{d}(\theta^1,\theta^2)$, over a planar domain parametrized by Cartesian coordinates $(\theta^1,\theta^2)$. The director field in Figure \ref{discli}(a) is restricted to lie strictly in the $\theta^1\theta^2$-plane; it may represent a deformed configuration of a nematic membrane or a single layer in the cholesteric phase of some liquid crystalline material. In contrast, the directors in Figure \ref{discli}(b) are allowed to orient themselves transversely to the plane; this can model either a lipid monolayer where the director orientation represents the orientation of individual lipid molecules, or a single layer of molecules in the smectic A or C phase \cite{kleman73}. In nematics, smectics, and cholesterics, $\boldsymbol{d}$ is identifiable with $-\boldsymbol{d}$ due to the mirror symmetry about the mid-orthogonal plane of the director axis. The lack of intrinsic crystalline order (translational and rotational), within the plane, in these examples can be primarily attributed to viscous relaxation \cite{kleman-book}. Disclinations in such structured surfaces can be characterized by the signed angle through which the director rotates upon circumnavigating along a loop over the surface. The Frank vector $\boldsymbol{\omega}$ of the disclination is a precise measure of this signed angle. A disclination is of {\it wedge} or {\it twist} type depending on whether $\boldsymbol{\omega}$ is transverse or tangential, respectively, to the surface. The disclination in Figure \ref{discli}(a) is of wedge type with Frank vector $2\pi\mathbf{e}_3$ and the one in Figure \ref{discli}(b) is of twist type with Frank vector $2\pi\mathbf{e}_2$. Here, the triple $\{\mathbf{e}_1,\mathbf{e}_2,\mathbf{e}_3\}$ denote the standard basis of the Cartesian coordinate system ($\theta^1,\theta^2,\theta^3)$. Note that the wedge disclination line in Figure \ref{discli}(a) and the twist disclination line in Figure \ref{discli}(b) are both along the $\theta^3$-axis. Disclinations can also appear in surfaces with intrinsic crystalline order, e.g., an ordered arrangement of lattice sites where the directors are attached in viral capsids or hexagonal lattice structure of the carbon atoms in graphene sheets. As illustrated in Figure \ref{discli}(c), circumnavigating along a loop encircling the disclination, a lattice vector rotates through an angle which is an integral multiple of one of the rotational symmetry angles of the lattice. The wedge disclination located at $O$, in the 2-dimensional hexagonal lattice in Figure \ref{discli}(c), is characterized by its Frank vector $\boldsymbol{\omega}=({\pi}/{3}) \mathbf{e}_3$. Material surfaces can also possess twist disclinations in the form of {\it local intrinsic orientational anomalies}, which correspond to breaking of the reflectional symmetries of the 2-dimensional material with the local tangent plane of the surface as the mirror plane, e.g., hemitropic plates \cite{steigmann99a, EremeyevPietraszkiewicz2006}. They are represented mathematically as ill-defined (multi-valued) local orientation field over the surface.\footnote{An example of a {\it global} intrinsic orientational anomaly would be the global orientational anomaly present in a surface M{\"o}bius crystal due to its non-orientability.} Note that, in order to quantify the disclinations discussed so far, the loop of circumnavigation is restricted always within the surface. All the disclinations shown in Figures \ref{discli}, as well as the intrinsic orientational anomalies discussed above, are quantified using an in-surface loop $\mathcal{C}$. The case otherwise can appear in 2-dimensional homogenized models of thin 3-dimensional multi-layered structures, e.g., a stack of few monolayers of smectics or cholesterics, thin multi-walled nanotubes, or a thin slice of some 3-dimensional oriented media. In these structures, disclinations may appear over the representative base surface (often the `mid-surface' of the layered structure) as the homogenized or effective rotational anomaly of all the distributed disclinations across the thickness of the thin structure. In describing these disclinations, the 
loop of circumnavigation must be taken transversely to the base surface, see Figure \ref{discli-layered}. Depending on the direction of the resulting vector of angular mismatch, these disclinations may either be of wedge or twist type.

\begin{figure}[t!]
 \centering
 \subfloat[]{\includegraphics[scale=0.5]{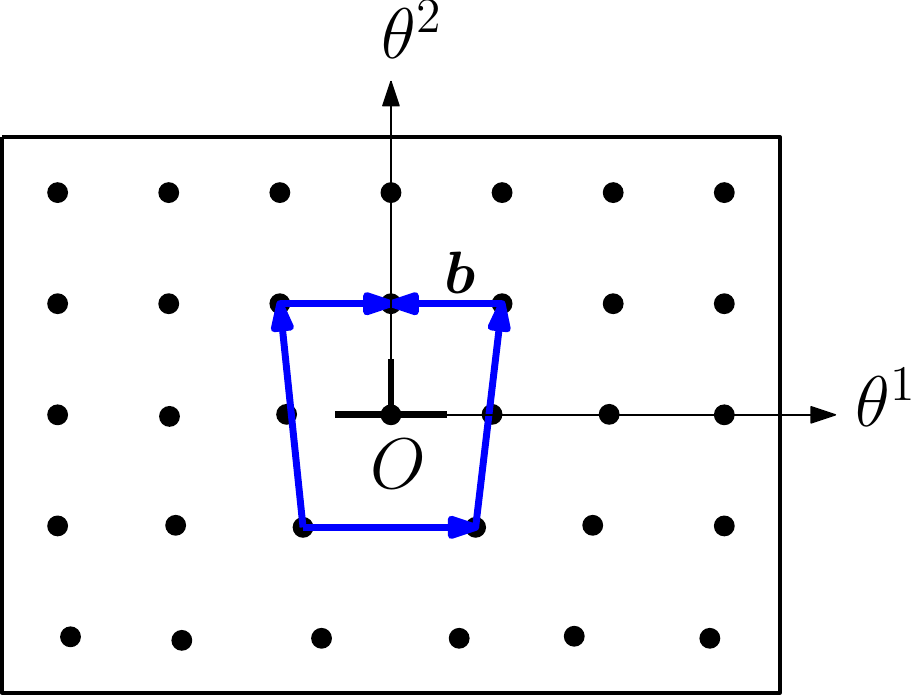}}
 \hspace{5mm}
 \subfloat[]{\includegraphics[scale=0.55]{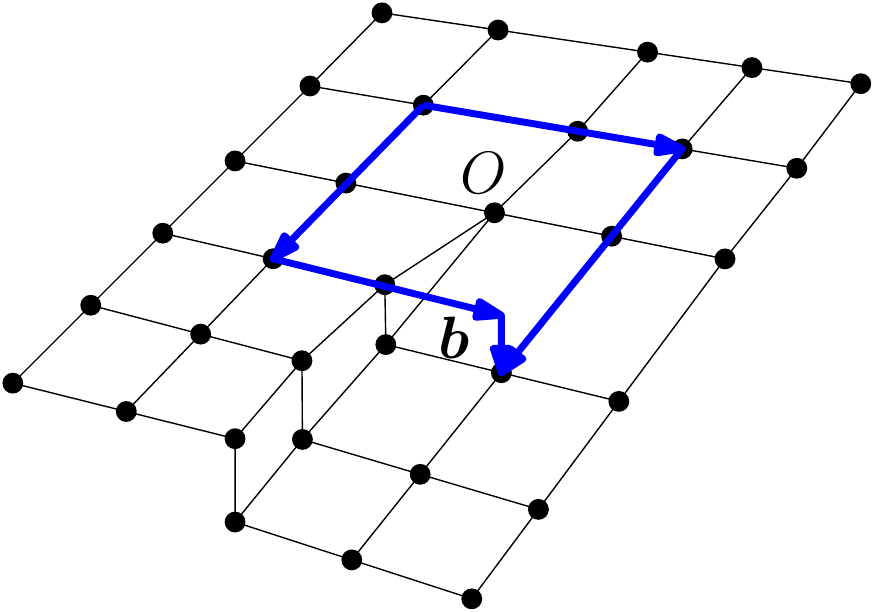}}
 \hspace{6mm}
 \subfloat[]{\includegraphics[scale=0.47]{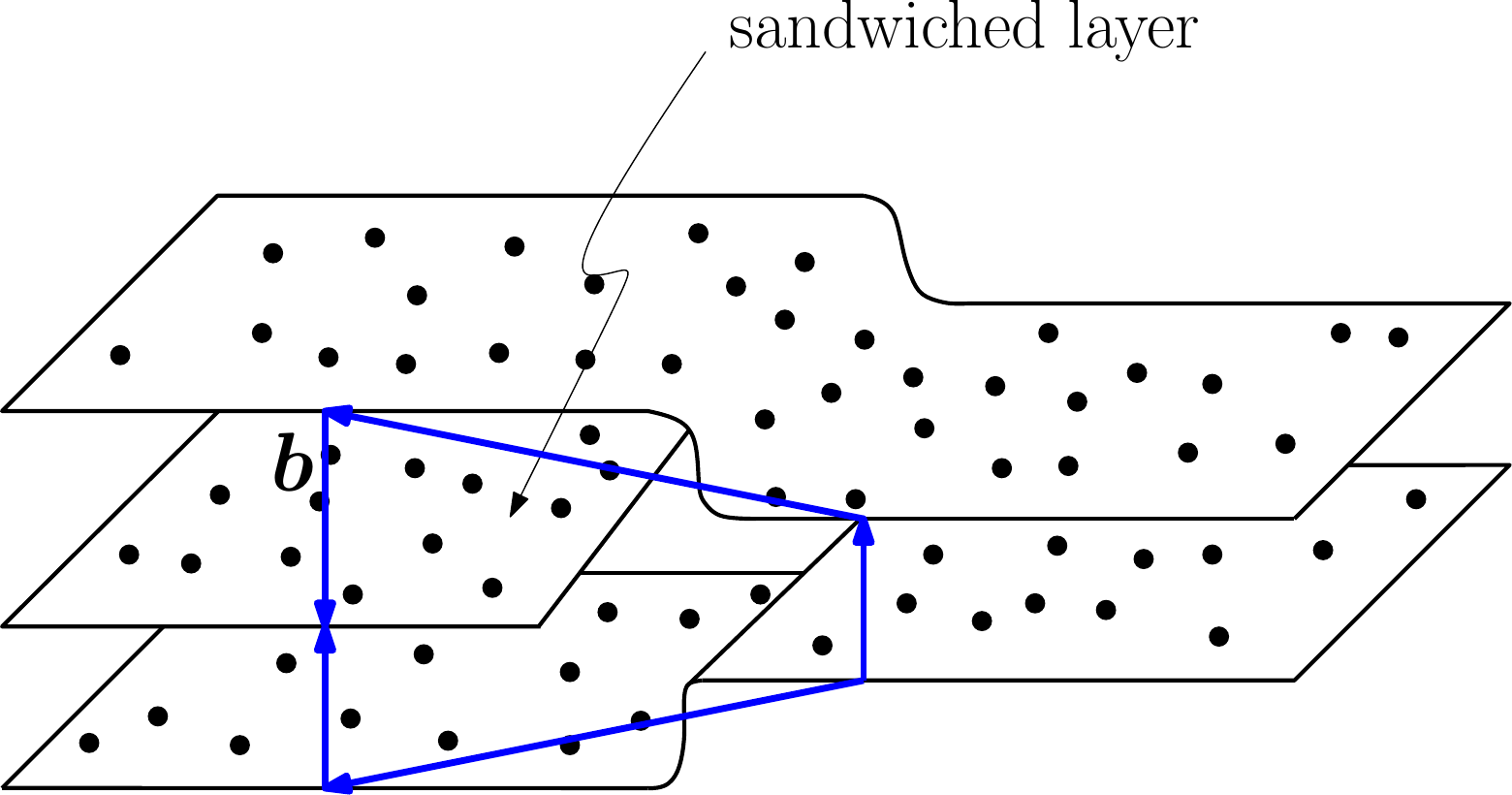}}
\caption{(a) Isolated surface edge dislocation and (b) surface screw dislocation in a 2-dimensional cubic lattice. (c) Isolated edge dislocation in a thin multi-layered structure. The marks on the surface represent lattice points which may carry identical atoms as well as directors.}
\label{dislo}
 \end{figure}

The \textit{translational anomalies} are represented by dislocations. The nature of dislocations in 2-dimensional matter is analogous to that in 3-dimensional materials. Isolated edge and screw surface dislocations are shown in Figures \ref{dislo}(a) and \ref{dislo}(b), respectively, within a 2-dimensional cubic lattice along with the Burgers parallelograms. The Burgers vector, defined as the closure failure of the Burgers parallelogram, is tangential to the surface of the lattice in the former case and transverse in the latter. In these examples, the dislocations appear essentially due to the breaking of the intrinsic translational symmetries of the 2-dimensional matter. On the other hand, in thin multi-layered structures or thin slices of oriented media, dislocations may be present, irrespective of the crystallinity of the material, as a result of either an order-mismatch of individual layers within the stack or as a homogenized or effective limit of all the distributed dislocations within the 3-dimensional slice. The Burgers parallelogram is, naturally, transverse to the representative mid-surface of the stack, in contrast to the examples shown in Figures \ref{dislo}(a,b). The precise type of these dislocations, edge or screw, can be determined from the direction of the Burgers vector. An edge dislocation in a layered medium is shown in Figure \ref{dislo}(c), arising due to the presence of a sandwiched semi-infinite layer between two infinite layers of material \cite[Ch.~VI]{lanlif}.

\begin{figure}[t!]
 \centering
 \subfloat[]{\includegraphics[scale=0.5]{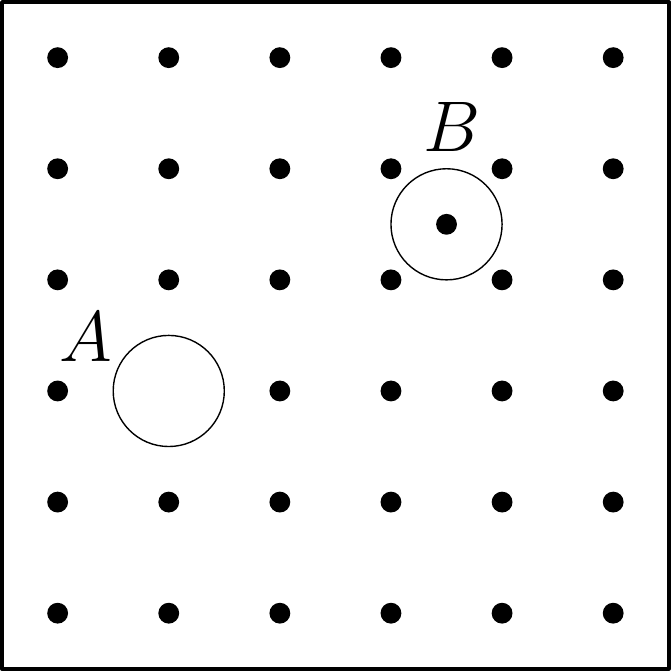}}
 \hspace{5mm}
 \subfloat[]{\includegraphics[scale=0.55]{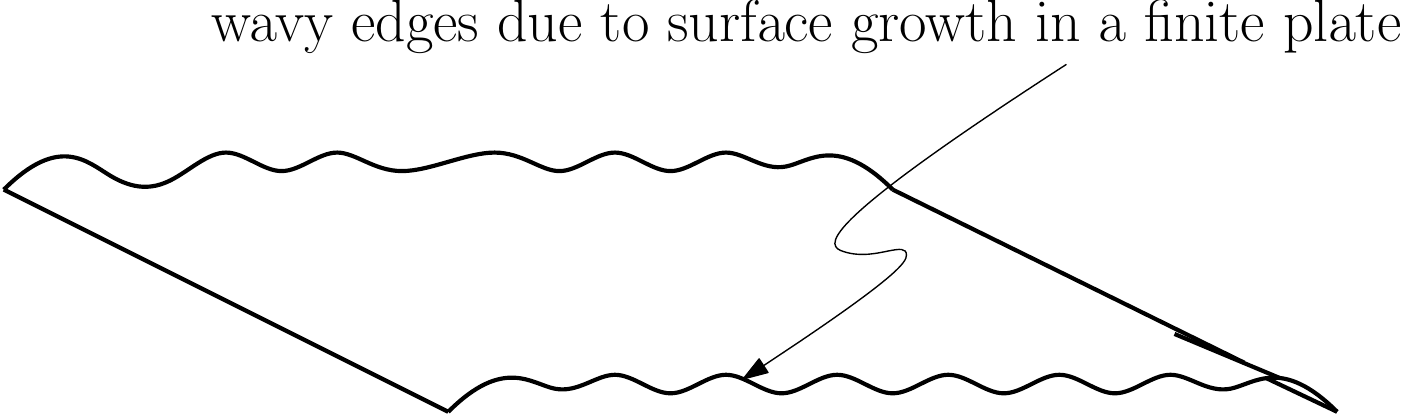}}
  \hspace{5mm}
 \subfloat[]{\includegraphics[scale=0.55]{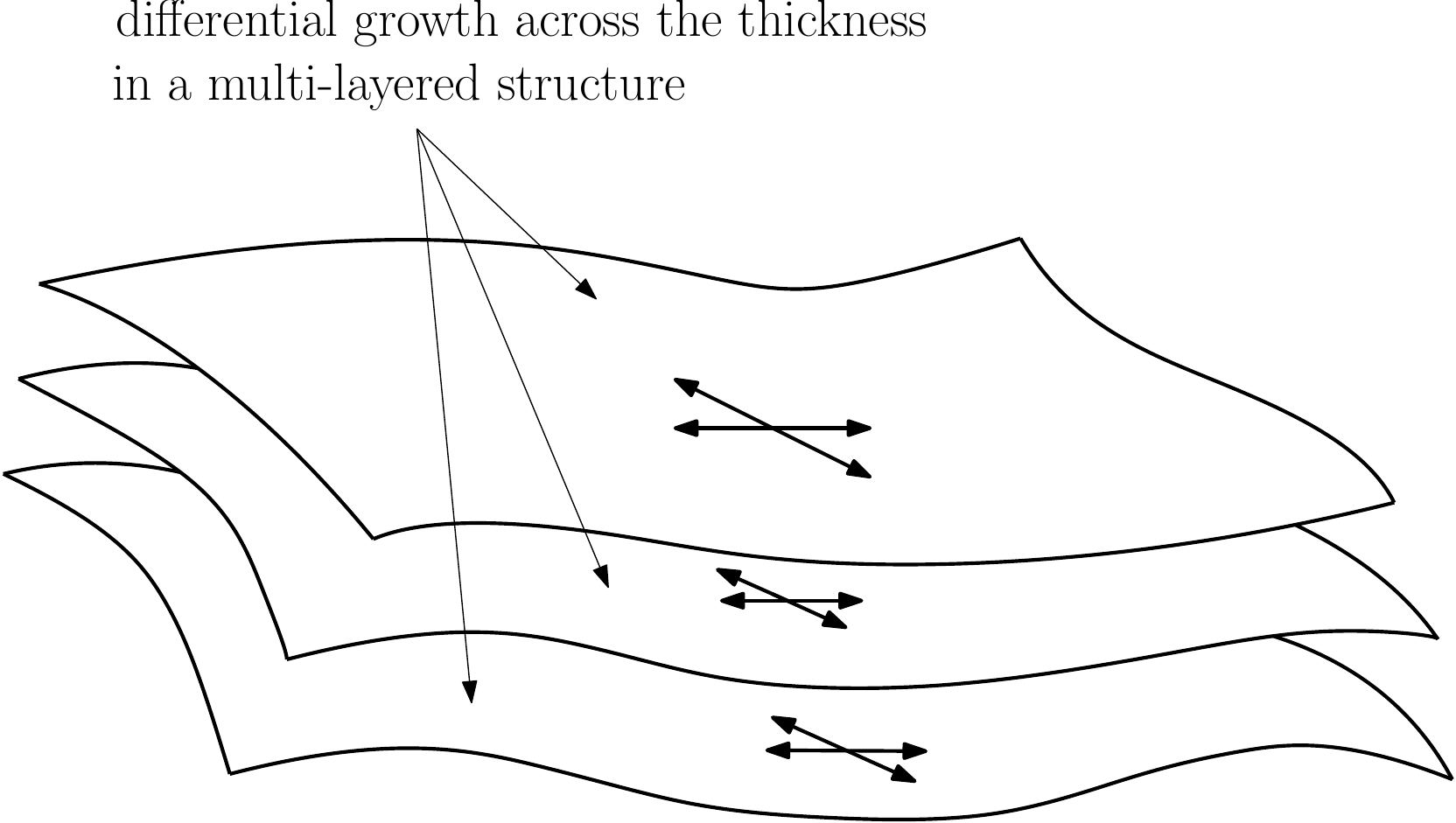}}
\caption{(a) A vacancy at position $A$ and a self-interstitial at position $B$ is a 2-dimensional cubic lattice. (b) Incompatible surface growth of a plate. (c) Differential growth of a thin multi-layered structure.}
\label{pd}
 \end{figure}
 
The \textit{metrical anomalies} bring about ambiguity in the (local) notion of ``length'' and ``angle'' over the surface. Metric anomalies are generated due to intrinsic point imperfections such as vacancies and self-interstitials, see Figure \ref{pd}(a), as well as a result of in-surface thermal deformation and biological growth, see Figure \ref{pd}(b). Note that foreign interstitials fall within the realm of materially non-uniform bodies (e.g., functionally graded materials) where the material constitution changes from point to point; their consideration is outside the scope of the present work. If the distance between the constituent entities in a lattice structure is measured by counting lattice steps, the presence of point defects, such as a vacancy or a self-interstitial, clearly introduces ambiguity in this step counting \cite{kroner81a}. Apart from these pure in-surface metric anomalies, differential growth (or thermal deformation) across the thickness direction within a thin multi-layered structure may result in transverse metric anomalies within an appropriately homogenized 2-dimensional theory, see Figure \ref{pd}(c).

The simple examples described above are sufficient to motivate the non-Euclidean nature of the defects. Recall that, in order to quantify disclinations, we required circumnavigation of a vector along a loop and  rotational mismatch between the initial and the final orientation of the vector. These notions correspond, respectively, to parallelly transporting a vector with respect to an affine connection and to the Riemann-Christoffel curvature associated with the affine connection. The Frank vector $\boldsymbol{\omega}$ uniquely characterizes the Riemann-Christoffel curvature tensor \cite{anth70}. The affine connection has to be necessarily non-Euclidean, since the director fields leading to disclinations are clearly not parallel in the Euclidean sense. Moreover, as the directors may point outside the surface, a differential geometric description of disclinations in structured surfaces would necessarily require embedding the surface into a 3-dimensional space with a specific non-Euclidean connection. In the case of dislocations, the closure failure of the Burgers parallelogram is analogous to the notion of torsion of an affine connection over a manifold which characterizes closure failures of infinitesimal parallelograms \cite{kondo52,bilby55}. Finally, the metric anomalies are characterized by the non-metricity tensor, which quantifies the non-uniformity of the metric tensor with respect to an appropriate affine connection \cite{anth71, kroner59}.  Motivated with these geometric analogies, we are now in a position to pursue a systematic study of geometry of defects in a structured surface. 

\section{Geometry of surface defects}\label{geom}

The mathematical prototype for structured surfaces is a connected, compact 2-dimensional manifold $\omega$, possibly with boundary, which is embeddable (as a topological submanifold) in $\mathbb{R}^3$. Examples of such manifolds, in the orientable category, are sphere, sphere with a finite number of handles added, twisted bands with  $2n\pi$ twists for integers $n$ etc., and in non-orientable category, twisted bands with  $(2n+1)\pi$ twists for integers $n$, e.g., a M{\"o}bius band for which $n$ is zero. We can add boundaries to these manifolds by removing finite number of open discs. The condition of embeddability in $\mathbb{R}^3$ precludes Klein bottle like surfaces and real projective planes. Our prototype manifold $\omega$ is topologically characterized by its orientability, twistedness, Euler characteristic, the number of open discs removed, i.e., the boundaries, and other topological invariants. We will call $\omega$ the {\it body manifold}. 
A fundamental theorem in differential topology (Tubular Neighbourhood Theorem \cite[Theorem~11.4]{bredon}) guarantees the existence of a {\it tubular neighbourhood} $\mathcal{M}:=\{\boldsymbol{y}\in\mathbb{R}^3\,|\,\mbox{dist}(\omega,\boldsymbol{y})<\epsilon,\epsilon>0\}$ of $\omega$ in $\mathbb{R}^3$, for sufficiently small $\epsilon$. Here, $\mbox{dist}(\omega,\boldsymbol{y})$ denotes the minimum Euclidean distance of $\omega$ from $\boldsymbol{y}$. As a bounded open set in $\mathbb{R}^3$, $\mathcal{M}$ naturally admits a manifold structure, with $\omega$ as an embedded submanifold. Existence of $\mathcal{M}$ induces a vector bundle (the normal bundle) structure over $\omega$ \cite{bredon},  which entails a vector field $\boldsymbol{d}:\omega\to\mathbb{R}^3$ defined over $\omega$. Our choice of $\omega$, naturally endowed with a {\it director field} $\boldsymbol{d}$,  is therefore appropriate for modelling structured surfaces. The differential structure, and all the fields to be defined over $\omega$ and $\mathcal{M}$, including $\boldsymbol{d}$, is assumed to be as smooth as the context demands. 

Our strategy for characterising material defects on a structured surface is to first equip $\mathcal{M}$ with a geometrical structure by associating with it a metric and an affine connection. This is then used to induce an appropriate non-Riemannian geometrical structure over $\omega$, where various fundamental geometric objects, such as non-metricity, torsion, and curvature, are interpreted as defect density measures. The induced metric and connection on $\omega$ is sufficient to encode all the information about the material structure of the structured surface. The Binachi-Padova relations are used to obtain several restrictions on defect density fields. With these relations, it is emphasized that the various defect densities are in fact dependent on each other.  The metric associated with $\mathcal{M}$ is also used to induce a Riemannian structure over $\omega$. The relationship of the curvature tensor, associated with the affine connection, with Riemann-Christoffel curvature tensor, obtained from the metric, is derived. These relations will provide the starting point for deducing strain incompatibility equations in the following section. They also lead to the well known local conditions under which $\omega$ is {\it isometrically} embeddable into $\mathbb{R}^3$, a notion that is related to compatibility of the strain fields.

In rest of the paper, lowercase Greek indices $\alpha$, $\beta$, $\gamma$ etc. take values from the set $\{1,2\}$ and lowercase Roman indices $i$, $j$, $k$ etc., from the set $\{1,2,3\}$. Einstein's summation convention hold over repeated indices unless specified otherwise. Round and square brackets enclosing indices indicate symmetrization and anti-symmetrization, respectively, with respect to them. The superscript $(-1)$ is used to denote the inverse of an invertible matrix, whereas the superscript $T$ is used to denote the transpose.

\subsection{Geometry on $\omega$ induced from the non-Riemannian structure on $\mathcal{M}$: the material space}\label{embed:nr}

Let the 3-dimensional embedding manifold $\mathcal{M}$ be equipped with an affine connection $\mathfrak{L}$ and a metric $\boldsymbol{g}$. Consider a chart $(V,\theta^i)$ of $\mathcal{M}$ with $U:=V\cap\omega\ne\emptyset$ such that the coordinates $\theta^\alpha$ defined over $V \subset \mathcal{M}$ lie along $U$ with $\zeta:=\theta^3\equiv 0$ at $U$. Such a coordinate system $\theta^i$ is called {\it adapted} to $U\subset\omega$. The restriction of the natural basis vector fields $\boldsymbol{G}_i$ over $V$ to $U$ will be denoted by $\boldsymbol{A}_i$, i.e., $\boldsymbol{A}_i(\theta^\alpha):=\boldsymbol{G}_i(\theta^\alpha,\zeta=0)$, hence $\boldsymbol{A}_3$ is transverse to $U$. The coefficients of $\mathfrak{L}$ and the covariant components of $\boldsymbol{g}$ are denoted by $L^i_{jk}$ and $g_{ij}$, respectively, with respect to $\boldsymbol{G}_i$. The covariant derivative of a sufficiently smooth vector field $\boldsymbol{u}=u^i (\theta^i) \boldsymbol{G}_i:V\to T_X V$, $X\in V$, with respect to $\mathfrak{L}$, is denoted by
\begin{equation}
 u^i_{;j}:=u^i_{,j}+L^i_{jk} u^k.
\end{equation}
The notation $\nabla$ is used for the surface covariant derivative of a tangent vector field $\boldsymbol{v}=v^\alpha(\theta^\alpha)\boldsymbol{A}_\alpha:U\to T_Y U$, $Y\in U$, with respect to the projection of $\mathfrak{L}$ on $U$, i.e., a connection with coefficients $L^\mu_{\alpha\nu}\big|_{\zeta=0}$,
\begin{equation}
 \nabla_\alpha v^\mu:=v^\mu_{,\alpha}+L^\mu_{\alpha \nu}\big|_{\zeta=0} v^\nu.
\end{equation}
Here, the subscript $(\cdot)_{,i}$ denotes ordinary partial derivative with respect to $\theta^i$. A vector field $\boldsymbol{u}$ along a curve over $V$ is called {\it parallel} with respect to $\mathfrak{L}$ if, and only if, its covariant derivative along the curve vanishes identically.

The body manifold $\omega$, equipped with connection $\mathfrak{L}$ and metric $\boldsymbol{g}$ from the embedding space $\mathcal{M}$, forms the {\it material space} $(\omega; \mathfrak{L}, \boldsymbol{g})$ of the structured surface. We will call $\mathfrak{L}$ the {\it material connection} and $\boldsymbol{g}$ the {\it material metric}. The ``material'' nature of these mathematical objects is due to the fact that the geometric quantities derived from $\mathfrak{L}$ and $\boldsymbol{g}$, when restricted to $\zeta=0$, represent various material inhomogeneities or defects within the material structure of the structured surface. As we will see immediately below, the non-metricity tensor is a measure of distributed metric anomalies, the torsion tensor is a measure of distributed translational anomalies (dislocations), and the Riemann-Christoffel curvature tensor is a measure of distributed rotational anomalies (disclinations). Most importantly, we assume $\mathfrak{L}$ to be such that the non-metricity, torsion, and curvature tensors associated with $\mathfrak{L}$ are uniform in the $\zeta$ coordinate and equal to their respective values at $\zeta=0$, i.e., at $U\subset\omega$. This assumption alludes to the applicability of our model to thin multi-layered structures, or thin slices of defective media, represented as homogenized 2-dimensional surfaces. It should also be noted that we are only looking at local defects and not the ones which could arise out of various topological anomalies for multiply connected and non-orientable surfaces.

\subsubsection{Non-metricity of the material connection: metric anomalies}
\begin{figure}[t!]
 \centering
\subfloat[]{\includegraphics[scale=0.4]{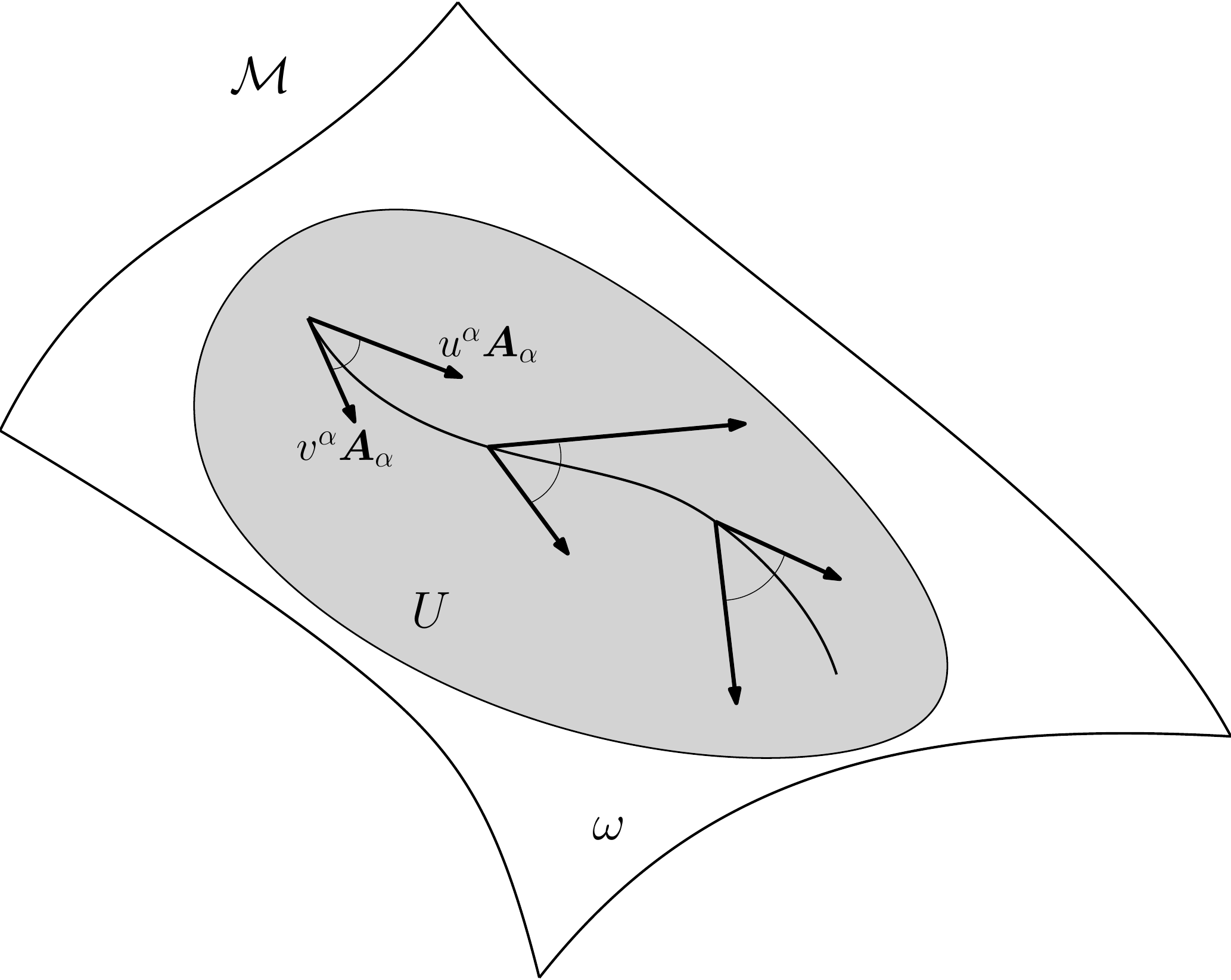}}
\hspace{5mm}
\subfloat[]{\includegraphics[scale=0.4]{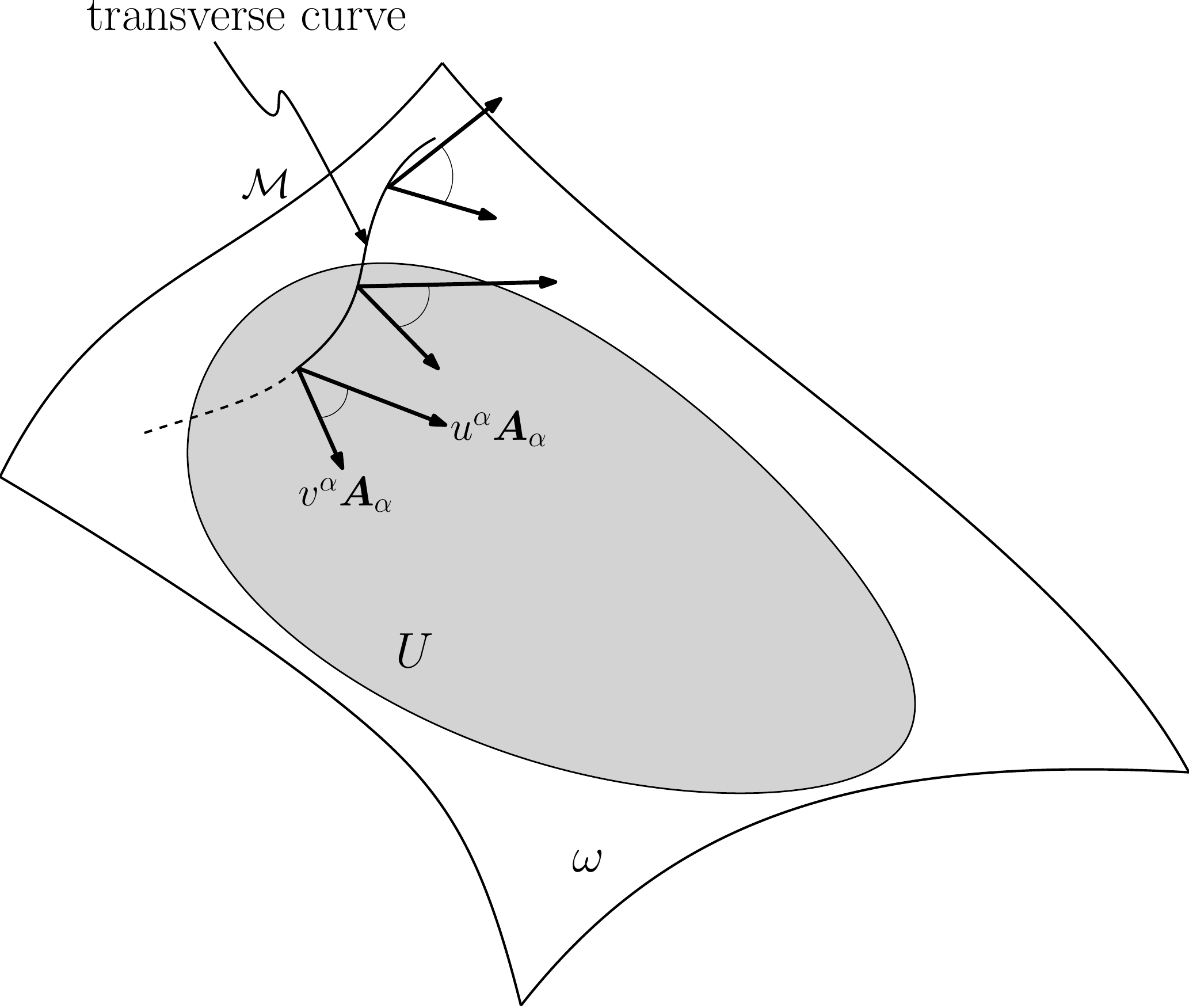}}
\vspace{0mm}
\subfloat[]{\includegraphics[scale=0.4]{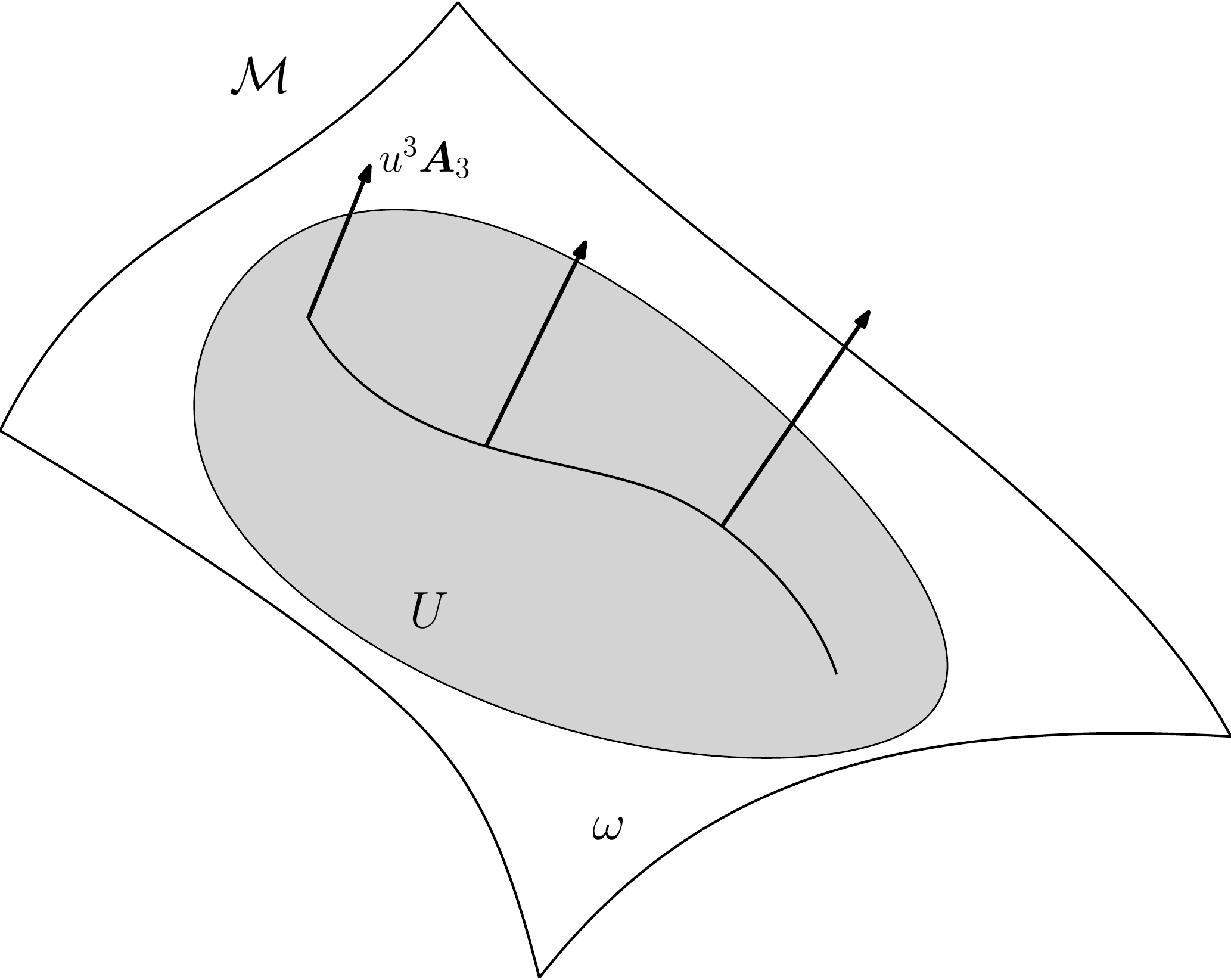}}
 \caption{(a) Change in angle between two tangent vectors to $U$ due to non-zero $Q_{\mu\alpha\beta}$. (b) Change in angle between two vectors along a transverse curve due to non-zero $Q_{3ij}$. (c) Change in length of a transverse vector along a surface curve due to non-zero $Q_{\mu i3}$.}
 \label{non-metricity}
\end{figure}

The third order non-metricity tensor of the material space, measuring non-uniformity of the metric $\boldsymbol{g}$ with respect to the connection $\mathfrak{L}$, has covariant components $\tilde Q_{kij}$ defined as 
\begin{equation}
\tilde Q_{kij} := -g_{ij;k}=-g_{ij,k}+L^p_{k i}g_{pj}+L^p_{k j}g_{ip}.
\end{equation}
The negative sign in the definition is purely conventional. We assume that $\tilde Q_{kij}(\theta^\alpha,\zeta)= \tilde Q_{kij}(\theta^\alpha,0)=: Q_{kij}(\theta^\alpha)$. The pure in-surface components $Q_{\alpha\mu\nu}$ provide measure for the distributed surface metric anomalies, whereas components $Q_{kij}$, with either of $k$, $i$ or $j$ taking the value 3, indicate the presence of out-of-surface metric anomalies, e.g., thickness-wise growth. A non-zero $Q_{\alpha\mu\nu}$ lead to variation in angle between tangent vectors during parallel transport with respect to the projected connection $L^\alpha_{\beta\gamma}\big|_{\zeta=0}$, see Figure \ref{non-metricity}(a). Indeed, the inner product $\boldsymbol{g}\big|_{\zeta=0}(\boldsymbol{u},\boldsymbol{w})=a_{\alpha\beta}u^\alpha v^\beta$ of two tangent vectors $\boldsymbol{u}=u^\alpha\boldsymbol{A}_\alpha$ and $\boldsymbol{v}=v^\alpha\boldsymbol{A}_\alpha$, where $a_{\alpha\beta} (\theta^\alpha):= g_{\alpha\beta}(\theta^\alpha,\zeta=0)$, changes under parallel transport with respect to $L^\alpha_{\beta\gamma}\big|_{\zeta=0}$ from the initial point $C^\alpha(0)$ to any generic point $C^\alpha(s)$, along some parametrized curve $\mathcal{C}=C^\mu(s)\boldsymbol{A}_\mu(\theta^\alpha(s))$ lying over $U$, by the amount
\begin{eqnarray}
 a_{\alpha\beta}u^\alpha v^\beta(s) - a_{\alpha\beta}u^\alpha v^\beta(0) &=&   \int^s_{0} (a_{\alpha\beta}u^\alpha v^\beta)_{,\mu} (\tau) \, \dot{C}^\mu(\tau)\,d\tau \nonumber\\
 &=&  - \int^s_{0} Q_{\mu\alpha\beta}(\theta^\alpha(\tau)) u^\alpha(\tau) v^\beta (\tau)\, \dot{C}^\mu(\tau)\,d\tau. \label{inner}
\end{eqnarray}
Here, we have used, $u^\alpha_{;\mu}\big|_{\zeta=0} \,\dot C^\mu \equiv 0$ and $v^\beta_{;\mu}\big|_{\zeta=0}\,\dot C^\mu \equiv 0$ throughout $\mathcal{C}$, as they are parallelly transported fields along $\mathcal{C}$, where $\dot C^\mu(s)$ denotes the ordinary derivative of $C^\mu(s)$ with respect to its argument. In structured surfaces, as we have earlier discussed in Section \ref{natdef}, this variation in inner product, characterized above in terms of a non-trivial $Q_{\alpha\mu\nu}$, may arise from a distribution of point imperfections in the arrangement of molecules or atoms over the surface, e.g., vacancies and self-interstitials in 2-dimensional crystals, inserting (or removing) a lipid molecule into (or out of) a crystalline arrangement of identical molecules over a monolayer, thermal deformation of the surface, biological growth of cell membranes, leaves etc. The remaining components $Q_{3ij}=-g_{ij;3}\big|_{\zeta=0}$ and $Q_{\mu i 3}=-g_{i3;\mu}\big|_{\zeta=0}$ measure the non-uniformity of the material metric in the $\zeta$-direction, i.e., along the thickness of the structured surface, and the change in length of transverse vectors along the surface, respectively, see Figures \ref{non-metricity}(b) and \ref{non-metricity}(c). These provide faithful representations for differential growth along the thickness in thin multi-layered structures discussed in Section \ref{natdef} and illustrated in Figure \ref{pd}(c).

\subsubsection{Torsion of the material connection: dislocations}
\begin{figure}[t!]
 \centering
 \subfloat[]{\includegraphics[scale=0.4]{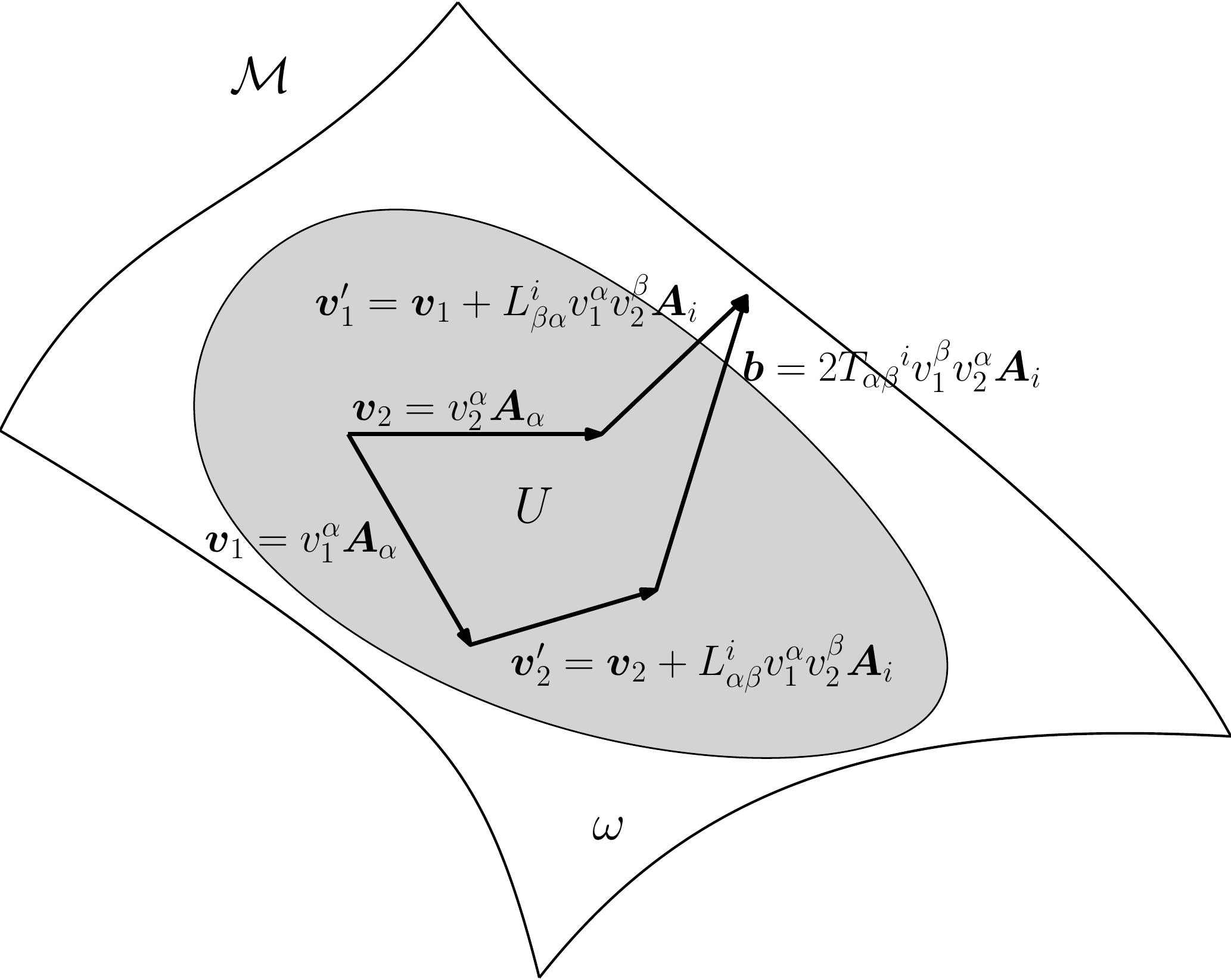}}
  \subfloat[]{\includegraphics[scale=0.4]{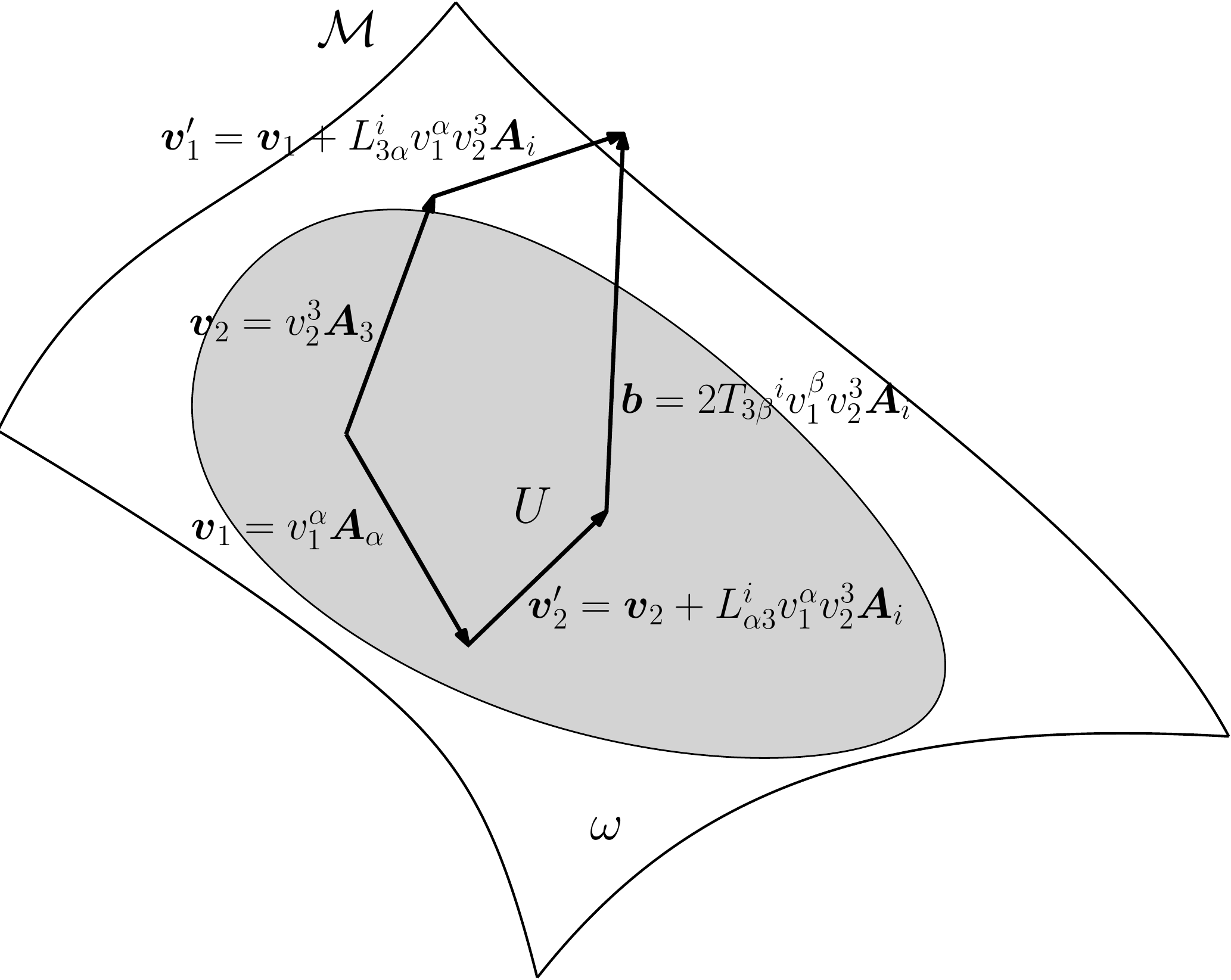}}
 \caption{(a) Closure failure of an infinitesimal in-surface parallelogram due to the $T_{\alpha\beta}{}^i$ components of the torsion tensor. (b) Closure failure of an infinitesimal transverse parallelogram due to the $T_{\alpha 3}{}^i$ components of the torsion tensor.}
 \label{torsion}
\end{figure}

Consider two tangent vectors $\boldsymbol{v}_1=v_1^i \boldsymbol{A}_i, \boldsymbol{v}_2=v_2^i\boldsymbol{A}_i$ at some point $Y$ on $U$. Translating $\boldsymbol{v}_1$ parallelly along $\boldsymbol{v}_2$ and $\boldsymbol{v}_2$ along $\boldsymbol{v}_1$ with respect to $\mathfrak{L}$, we obtain the vectors
\begin{equation}
  \boldsymbol{v}'_1 = \boldsymbol{v}_1+L^i_{jk}\big|_{\zeta=0}v_1^k v_2^j\boldsymbol{A}_i~~\text{and}~~\boldsymbol{v}'_2 = \boldsymbol{v}_2+L^i_{kj}\big|_{\zeta=0}v_1^k v_2^j\boldsymbol{A}_i,
\end{equation}
respectively. The closure failure of the parallelogram is given by (see Figure \ref{torsion})
\begin{equation}
 \boldsymbol{b} = \boldsymbol{v}_2+\boldsymbol{v}'_1-\boldsymbol{v}_1-\boldsymbol{v}'_2
 = 2 T_{jk}{}^i(\theta^\alpha)\,v_1^k v_2^j\, \boldsymbol{A}_i,
\end{equation}
where the functions 
\begin{equation}
T_{jk}{}^i(\theta^\alpha):=L^i_{[jk]}\big|_{\zeta=0}
\end{equation}
constitute the components of the third-order torsion tensor (anti-symmetric in the lower indices) over $U$. Let $\tilde{T}_{jk}{}^i:=L^i_{[jk]}$. We assume that $\tilde{T}_{jk}{}^i(\theta^\alpha,\zeta) = \tilde{T}_{jk}{}^i(\theta^\alpha,0)$, which in turn is same as  $T_{jk}{}^i(\theta^\alpha)$. Associated with the torsion tensor, we have the second-order axial tensor 
\begin{equation}
 \alpha^{ij}(\theta^\alpha):=\frac{1}{2}\varepsilon^{ikl}(\theta^\alpha)T_{kl}{}^j(\theta^\alpha).
\end{equation}
Here, $\varepsilon^{ijk}(\theta^\alpha):=g^{-\frac{1}{2}}e^{ijk}$, where $e^{ijk}=e_{ijk}$ is the 3-dimensional permutation symbol and $g:=\mbox{det}[g_{ij}\big|_{\zeta=0}]$. For later use, we define  $\varepsilon_{ijk}(\theta^\alpha):=g^{\frac{1}{2}}e_{ijk}$. The components $\alpha^{ij}$ provide measures for a variety of dislocation distributions over the structured surface. Taking $v^3_1=v^3_2=0$ (i.e., $\boldsymbol{v}_1$ and $\boldsymbol{v}_2$ tangential to $U$, see Figure \ref{torsion}(a)), and comparing with Figures \ref{dislo}(a,b), it is immediate that
\begin{equation}
 J^\alpha:=\alpha^{3\alpha}=\frac{1}{2}\varepsilon^{\mu\nu 3} T_{\mu\nu}{}^\alpha
\end{equation}
represent a distribution of in-surface edge dislocations and 
\begin{equation}
 J^3:=\alpha^{33}=\frac{1}{2}\varepsilon^{\mu\nu 3} T_{\mu\nu}{}^3
\end{equation}
a distribution of in-surface screw dislocations (cf.~\cite{povs85,povs91}). Next, taking $v_1^3=v_2^\alpha=0$ (i.e., $\boldsymbol{v}_1$ tangential and $\boldsymbol{v}_2$ transverse to $U$, see Figure \ref{torsion}(b)), and comparing with Figure \ref{dislo}(c), it is evident that the components $\alpha^{\mu k}:=\frac{1}{2} \varepsilon^{3\alpha\mu} T_{3\alpha}{}^k$ represent the out-of-surface dislocations in thin multi-layered oriented media such as those discussed in Section \ref{natdef}.

\subsubsection{Curvature of the material connection: disclinations}

The components of the fourth order Riemann-Christoffel curvature tensor of the material connection $\mathfrak{L}$ are given by
\begin{equation}
\tilde\Omega_{klj}{}^i:= L^i_{lj,k}- L^i_{kj,l}+L^h_{lj} L^i_{kh}-L^h_{kj} L^i_{lh}. \label{curv}
\end{equation}
The functions $\tilde\Omega_{kl j}{}^i$ measure, in the linear approximation, the change that a vector, $\boldsymbol{v}\in T_X V$, $X\in V$, suffers under parallel transport with respect to $\mathfrak{L}$ along an infinitesimal loop $\mathcal{C}$ based at $X$ and lying within $V$:
\begin{equation}
\delta v^i \approx -\frac{1}{2} \tilde\Omega_{kl j}{}^i(X)v^j \oint_{\mathcal{C}} \theta^k d\theta^l,
\label{curv-loop}
\end{equation}
where $v^i$ are the components of the initial vector with respect to the basis $\boldsymbol{G}_i(X)$; the integral represents the infinitesimal area bounded by the loop $\mathcal{C}$. The above formula in fact holds true for any general loop (not necessarily infinitesimal) $\mathcal{C}$ in $V$. We define the purely covariant components $\tilde\Omega_{kl ji}$ by lowering the fourth index with the material metric $g_{ij}$ as $\tilde\Omega_{klji}:=g_{ip} \tilde\Omega_{klj}{}^p.$ Clearly, $\tilde\Omega_{kl j}{}^i=-\tilde\Omega_{lk j}{}^i$ and $\tilde\Omega_{kl ij}=-\tilde\Omega_{lk ij}$. Moreover, as we did for non-metricity and torsion tensors, we assume $ \tilde\Omega_{kl ij} (\theta^\alpha,\zeta)  = \tilde\Omega_{kl ij} (\theta^\alpha,0) =: \Omega_{kl ij}(\theta^\alpha)$. 

\begin{figure}[t!]
 \centering
  \subfloat[]{\includegraphics[scale=0.4]{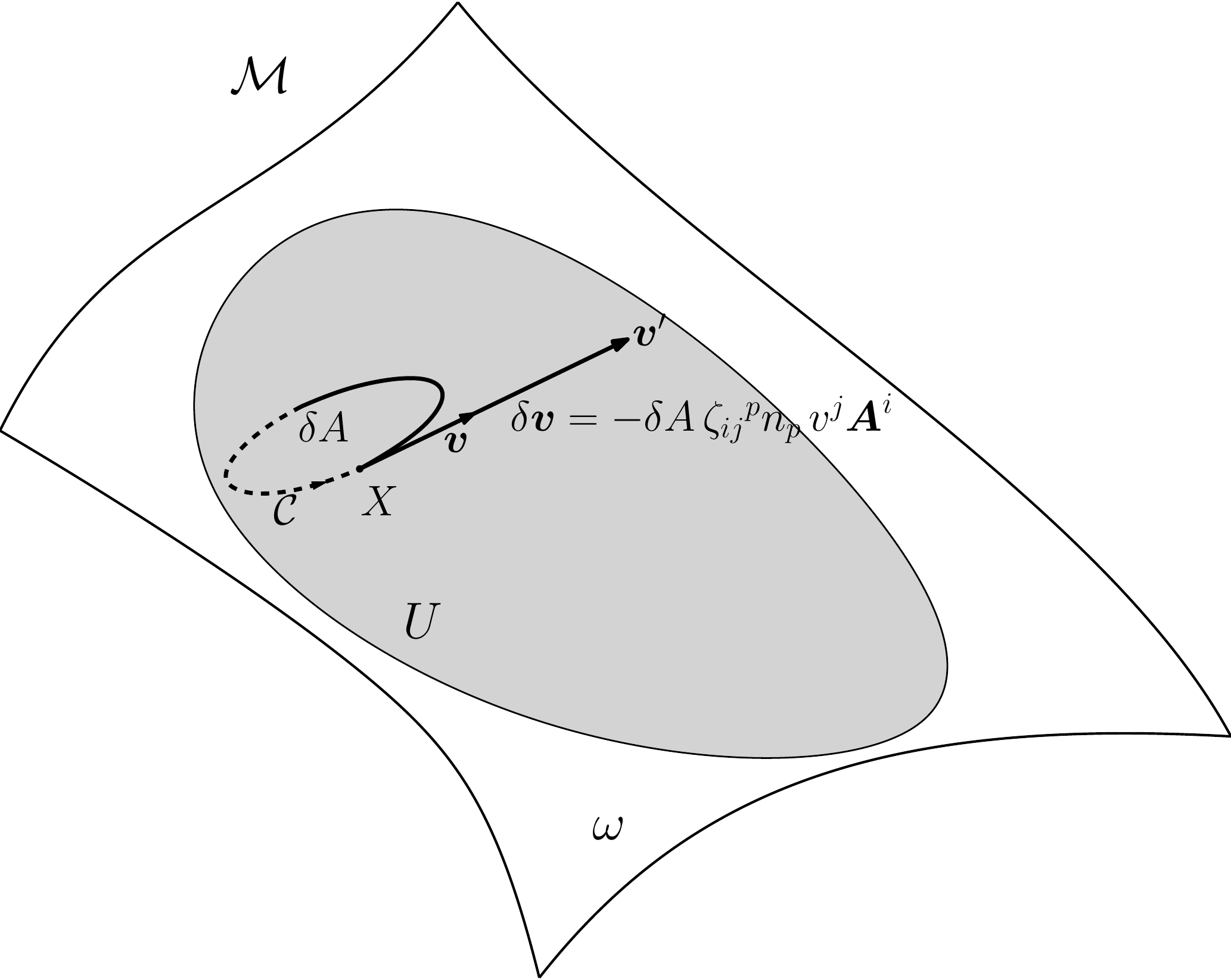}}
 \subfloat[]{\includegraphics[scale=0.4]{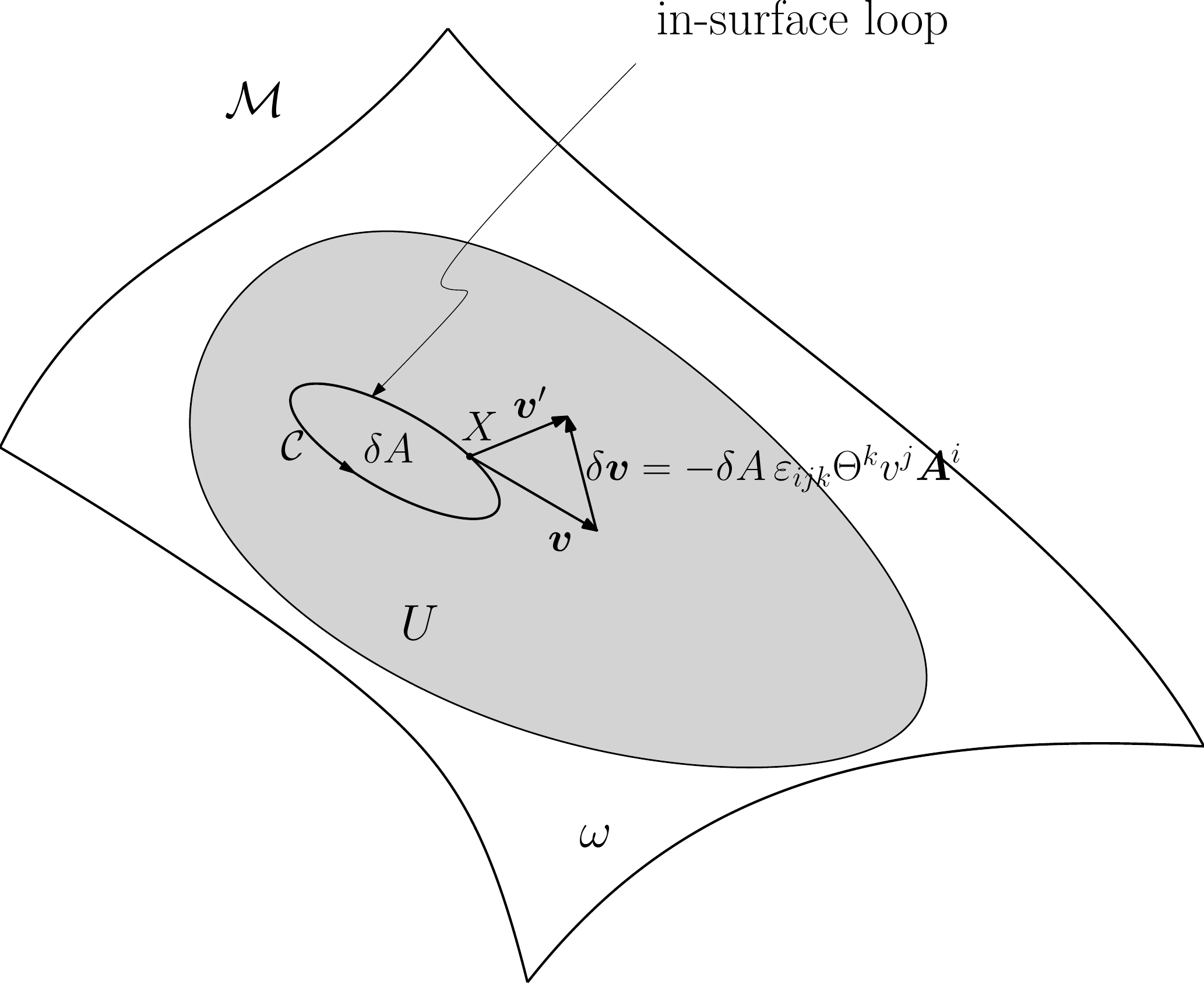}}
 \vspace{0mm}
  \subfloat[]{\includegraphics[scale=0.4]{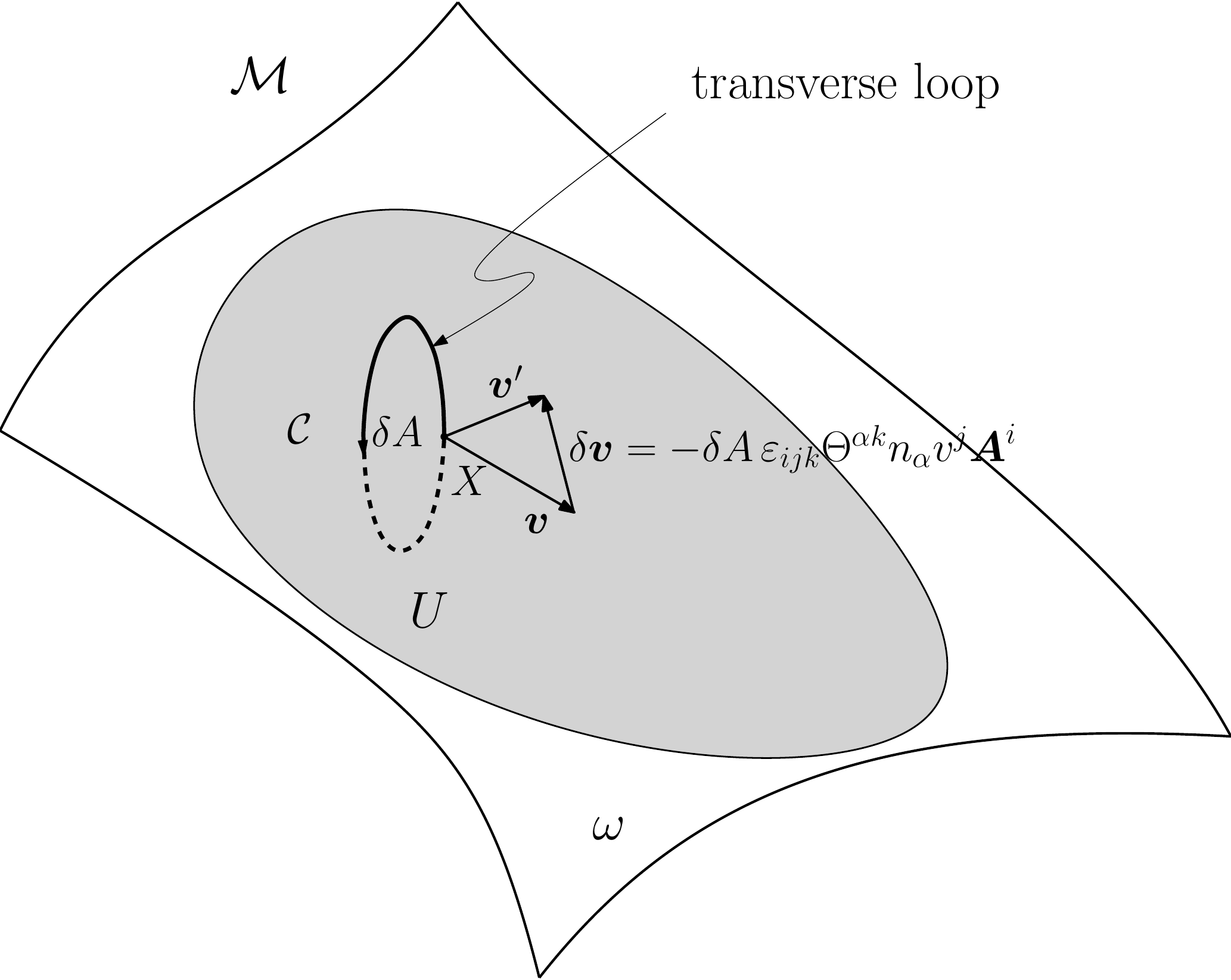}}
 \caption{(a) The symmetric part $\Omega_{ij(kl)}$, characterized by the third-order tensor $\boldsymbol{\zeta}$, measures the stretching in $\boldsymbol{v}\in T_X V$ brought about by curvature of the material space. Here, $\boldsymbol{v}$ is a principal direction of the second order tensor $\boldsymbol{\zeta n}$, where $\boldsymbol{n}$ is the unit normal to the infinitesimal area element $\delta A$ bounded by the loop $\mathcal{C}$. 
 (b) For $\mathcal{C}$ completely lying within $U$, the skew part $\Omega_{\alpha\beta[ij]}$, characterized by the vector field $\Theta^i\boldsymbol{A}_i$, measures the purely rotational part of the change in $\boldsymbol{v}=v^i\boldsymbol{A}_i$ brought about by the curvature tensor. (c) When $\mathcal{C}$ is transverse to $U$, the skew part $\Omega_{\alpha 3[ij]}$, characterized by the second-order tensor field $\Theta^{\alpha q}\boldsymbol{A}_\alpha\otimes\boldsymbol{A}_q$, measures the purely rotational part of the change in $\boldsymbol{v}=v^i\boldsymbol{A}_i$ brought about by the curvature tensor. }
 \label{curvature}
\end{figure}

It is useful to decompose the components $\Omega_{kl ij}(\theta^\alpha)$ into skew and symmetric parts \cite{povs91}
 \begin{equation}
  \Omega_{klij}=\varepsilon_{pkl}\varepsilon_{qij}\Theta^{pq}+\varepsilon_{pkl}\zeta_{ij}{}^p,
  \label{curv:decomposition}
 \end{equation}
where
\begin{equation}
\Theta^{pq}:=\frac{1}{4}\varepsilon^{pij}\varepsilon^{qkl}\Omega_{ijkl}~~\text{and}~~\zeta_{ij}{}^p:=\frac{1}{2}\varepsilon^{pkl}\Omega_{kl(ij)}
\end{equation} 
are components of the second-order tensor field $\boldsymbol{\Theta}=\Theta^{pq}\boldsymbol{A}_p\otimes\boldsymbol{A}_q$ and the third-order tensor field $\boldsymbol{\zeta}=\zeta_{ij}{}^k\boldsymbol{A}^i\otimes\boldsymbol{A}^j\otimes\boldsymbol{A}_k$. They represent, respectively,
 the skew part  and the symmetric part of $\Omega_{kl ij}$ with respect to the last two indices. A geometric interpretation of these two fundamental tensors is as follows (see Figure \ref{curvature}). Let the infinitesimal loop $\mathcal{C}$ in \eqref{curv-loop} be based at $X\in U\subset\omega$. Then the change $\delta\boldsymbol{v}$ that a vector $\boldsymbol{v}\in T_X V$ undergoes when parallelly transported along $\mathcal{C}$, in the linear approximation, can be characterized by a second-order tensor $\boldsymbol{\beta}=\beta_{ij}\boldsymbol{A}^i\otimes\boldsymbol{A}^j$, i.e.,  $\delta\boldsymbol{v}=\boldsymbol{\beta v}$, where 
\begin{equation}
 \beta_{ij}:=-\frac{\delta A}{2}\Omega_{klij}\varepsilon^{rkl}\,n_r
       = -\delta A\,\varepsilon_{qij}\, \Theta^{pq}\, n_p-\delta A\,\zeta_{ij}{}^p\,n_p.\label{change}
\end{equation}
Here, $\delta A$ is a measure of the infinitesimal area bounded by $\mathcal{C}$ and $\boldsymbol{n}=n_r\boldsymbol{A}^r$ its unit normal. The first term $W_{ij}:=-\delta A\,\varepsilon_{qij}\, \Theta^{pq}\, n_p$ in the above expression is skew with axial vector $w^q=\Theta^{pq}\,n_p\,\delta A$. It represents the rotation that $\boldsymbol{v}$ has experienced under parallel transport about the axis $\boldsymbol{A}_p$, for each fixed $p$, probed by the three Euler angles $\Theta^{pq}$. Thus, $\boldsymbol{\Theta}$ is the measure of the rotation of $\boldsymbol{v}$ about the axis $\boldsymbol{n}$. The second term $S_{ij}:=-\delta A\,\zeta_{ij}{}^p\,n_p$, on the other hand, is symmetric; it represents a stretching, with the three principal values of the tensor $\boldsymbol{\zeta n}=\zeta_{ij}{}^p n_p \boldsymbol{A}^{i}\otimes\boldsymbol{A}^j$ as measures of the stretch along their respective (linearly independent) principal directions. The tensor $\boldsymbol{\zeta}$ can be shown to be related to the metrical properties of $\mathcal{M}$ as it gives rise to a smeared out anomaly within the material structure which causes elongation or shortening of material vectors under parallel transport along loops (see \cite{rcgupta16} for details), as shown in Figure \ref{curvature}(a). We will assume $\boldsymbol{\zeta}\equiv \boldsymbol{0}$ in rest of the paper since, at present, we do not know of any defects in 2-dimensional materials which they would otherwise represent. Some consequences of this assumption will be discussed in the next section. The curvature tensor $\Omega_{kl ij}$ is then fully characterized in terms of the non-trivial independent components $\Omega_{[kl] [ij]}$, i.e., the second-order tensor $\boldsymbol{\Theta}$.

We distinguish between two families of local rotational anomalies characterized by $\boldsymbol{\Theta}$. Consider, first, the infinitesimal loop $\mathcal{C}$ completely lying within $U$, see Figure \ref{curvature}(b). Then, the $i$ and $j$ indices in $\Omega_{ijkl}$ can assume only values 1 and 2, and the resulting angular mismatch after parallel transport of arbitrary vectors is characterized by three fields 
\begin{equation}
 \Theta^q(\theta^\alpha):=\Theta^{3q}(\theta^\alpha)=\frac{1}{4}\varepsilon^{3\alpha\beta}\varepsilon^{qkl}\Omega_{\alpha\beta kl}.
 \label{wedge-ds}
\end{equation}
These provide a measure for the distributed rotational anomalies within the material structure of the base manifold $\omega$. Drawing analogy with Figure \ref{discli}, it is clear that the out-of-surface component $\Theta^3$ provides a measure for the density of distributed wedge disclinations over the structured surface, see Figures \ref{discli}(a,c), irrespective of its crystallinity, whereas the in-surface components $\Theta^\mu$ characterize either the distributed intrinsic orientational anomalies, in case of intrinsically crystalline surfaces, or distributed twist disclinations, in case of directed surfaces (as shown in Figure \ref{discli}(b)). Next, we consider $\mathcal{C}$, based at $X\in U$, to lie transversely to $U$, see Figure \ref{curvature}(c). Then one of the indices $i$ and $j$ in $\Omega_{ijkl}$ will take the value 3, and the resulting angular mismatch after parallel transport of arbitrary vectors is characterized by the remaining six independent components of $\boldsymbol{\Theta}$:
\begin{equation}
\Theta^{\alpha q}=\frac{1}{4}\varepsilon^{\alpha\mu 3}\varepsilon^{qkl}\Omega_{\mu 3 kl}.
\label{twist-ds}
\end{equation}
Recalling our discussion in Section \ref{natdef} on disclinations in thin multi-layered structures of oriented media, see also Figure \ref{discli-layered}, we conclude that these components provide a measure for a variety of homogenized/effective rotational anomalies of the distributed disclinations across the thickness of the multi-layered structured surface. Out of these six functions, $\Theta^{11}$ and $\Theta^{22}$ are of wedge type, and $\Theta^{12}$, $\Theta^{21}$, and $\Theta^{\alpha 3}$ are of twist type. As we will see shortly, these functions are in fact dependent on each other in very thin monolayer structures where the dislocation densities $\alpha^{\mu k}$ vanish altogether.

\begin{table}[t!]
\centering
\begin{tabular}{|c|c|}
\hline
{\it Geometric objects} & {\it Defect densities} \\
\hline
\hline
$Q_{\mu\alpha\beta}$ & In-surface metric anomalies; Figures \ref{pd}(a,b) and \ref{non-metricity}(a) \\
$Q_{3 ij}$ and $Q_{\alpha i3}$ & Out-of-surface metric anomalies; Figures \ref{pd}(c) and \ref{non-metricity}(b,c) \\
$J^\mu:=\alpha^{3\mu}$ & In-surface edge dislocations; Figures \ref{dislo}(a) and \ref{torsion}(a)\\
$J^3:=\alpha^{33}$   & In-surface screw dislocations; Figures \ref{dislo}(b) and \ref{torsion}(a)\\
$\alpha^{\mu k}$     & Out-of-surface dislocations; Figures \ref{dislo}(c) and \ref{torsion}(b)\\
$\Theta^3:=\Theta^{33}$ & In-surface wedge disclinations; Figures \ref{discli}(a,c) and \ref{curvature}(b)\\
$\Theta^\mu:=\Theta^{3\mu}$ & \makecell{In-surface twist disclinations or intrinsic orientational anomalies;  \\ Figures \ref{discli}(b) and \ref{curvature}(b)} \\
$\Theta^{\mu k}$ & Disclinations associated with transverse loops; Figures \ref{discli-layered} and \ref{curvature}(c) \\
$\zeta_{ij}{}^k$ & Metrical disclinations; Figure \ref{curvature}(a)\\
\hline
\end{tabular}
\caption{Non-Riemannian geometric objects on $\omega$ and the defects they characterize in structured surfaces.}
\label{tbl1}
\end{table}

We have summarized the set of all defect densities in Table \ref{tbl1}.

\subsubsection{Bianchi-Padova relations}\label{sec-bprel}
The tensors of non-metricity, torsion, and curvature of a non-Riemannian space cannot be arbitrary due to geometric restrictions. Besides the restrictions $\tilde Q_{k[ij]}=0$, $\tilde T_{(ij)}{}^k=0$, and $\tilde\Omega_{(ij)kl}=0$, which follow from their definitions, they satisfy the following system of differential relations, known as the Bianchi-Padova relations \cite[p. 144]{schouten}:

\begin{subequations}
 \begin{align}
  2\tilde T_{[jk}{}^l{}_{;i]}&=\Omega_{[ijk]}{}^l+4\tilde T_{[ij}{}^p\,\tilde T_{k]p}{}^l,\label{diff:1}\\
  \tilde \Omega_{[jk|l|}{}^p{}_{;i]}&=2 \tilde T_{[ij}{}^q\,\tilde \Omega_{k]ql}{}^p,\,\textrm{and}\label{diff:2}\\
  \tilde Q_{[j|kl|;i]}&= \tilde T_{ij}{}^p\,\tilde Q_{pkl}-\tilde \Omega_{ij(kl)}.\label{diff:3}
 \end{align}
\label{bp-identities}%
\end{subequations}
In the above expressions, anti-symmetrization with respect to three indices is defined as
\begin{equation}
 A_{[nml]\cdots}{}^{\cdots}:=\frac{1}{6}(A_{nml\cdots}{}^{\cdots}+A_{lnm\cdots}{}^{\cdots}+A_{mln\cdots}{}^{\cdots}-A_{lmn\cdots}{}^{\cdots}-A_{nlm\cdots}{}^{\cdots}-A_{mnl\cdots}{}^{\cdots}).
\end{equation}
The enclosed indices within two vertical bars in the subscript are to be exempted from anti-symmetrization. Clearly, $A_{[\alpha\beta\mu]\cdots}{}^{\cdots} = 0$ and $A_{[nnl]\cdots}{}^{\cdots} = 0$ (no summation on $n$).
 Additionally, there is a fourth Bianchi-Padova relation \cite[p. 145]{schouten}, purely algebraic in nature, based on the following identity satisfied by the components of any fourth-order tensor $\tilde\Omega_{ijkl}$ with $\tilde\Omega_{(ij)kl} = 0$:
 \begin{equation}
   \tilde\Omega_{ijkl}-\tilde\Omega_{klij}=-\frac{3}{2}\big(\tilde\Omega_{[jik]l}+\tilde\Omega_{[jlk]i}+\tilde\Omega_{[lik]j}+\tilde\Omega_{[ijl]k}\big) +\tilde\Omega_{kj(li)}+\tilde\Omega_{ik(lj)}+\tilde\Omega_{jl(ik)}+\tilde\Omega_{li(jk)}+\tilde\Omega_{lk(ji)}+\tilde\Omega_{ij(lk)}.\label{bprel-4th}
 \end{equation}
After substituting relations \eqref{diff:1} and \eqref{diff:3} into \eqref{bprel-4th}, it boils down to an expression for $\tilde\Omega_{ijkl}-\tilde\Omega_{klij}$ in terms of $\tilde T_{ij}{}^k$, $\tilde Q_{kij}$, $T_{[jk}{}^l{}_{;i]}$, and $ \tilde Q_{[j|kl|;i]}$. For a torsion-free, metric-compatible connection (i.e., a Levi-Civita connection), this implies the familiar symmetry $\tilde\Omega_{ijkl}=\tilde\Omega_{klij}$. However, as shown below, this particular symmetry is achieved in very thin structured surfaces under much less restrictive conditions. 
The first three Bianchi-Padova relations, restricted to a surface, have been considered previously by Povstenko \cite{povs91}, but without studying any of the implications, some of which are noted below.

\textit{Consequences of the first Bianchi-Padova relation}: Equation \eqref{diff:1} is non-trivial only when at least one of the indices $i$, $j$ and $k$ assume the value 3, since otherwise $A_{[\alpha\beta\mu]} = 0$. Recalling our assumption that that $\tilde T_{ij}{}^k$ is uniform with respect to the $\zeta$ coordinate, \eqref{diff:1} reduces to
\begin{equation}
 4 \nabla _{[\beta}T_{|3|\alpha]}{}^l{}=-(\Omega_{\alpha\beta 3}{}^l + \Omega_{3\alpha\beta}{}^l - \Omega_{3\beta\alpha}{}^l) - 4(T_{\alpha\beta}{}^\mu T_{3\mu}{}^l + T_{3\alpha}{}^p T_{\beta p}{}^l - T_{3\beta}{}^p T_{\alpha p}{}^l).
 \label{rel-3}
\end{equation}
Furthermore, if we assume that the structured surface is sufficiently thin with no dislocations associated with the transverse Burgers parallelograms, i.e., $\alpha^{\mu k} = 0$ (the in-surface dislocations $J^i$ can still be present), then \eqref{rel-3} simplifies into a system of algebraic equations:
\begin{equation}
 \Omega_{\alpha\beta 3 l}= \Omega_{3\beta\alpha l} - \Omega_{3\alpha\beta l}. \label{rel-1}
\end{equation}
For $l=3$, we obtain $\Omega_{3\beta\alpha 3} =  \Omega_{3\alpha\beta 3}$, since $\Omega_{\alpha\beta 33}=0$ (from $\boldsymbol{\zeta} = \boldsymbol{0}$). This is equivalent to $\Omega_{\beta 3\alpha 3} =\Omega_{\alpha 3\beta 3}$, or
\begin{equation}
\Theta^{\alpha\beta} = \Theta^{\beta\alpha}. \label{bp-rel1}
\end{equation}
For $l=\mu$, \eqref{rel-1} can be rewritten as $\Omega_{\alpha\beta 3 \mu}= \Omega_{3\beta \alpha\mu} - \Omega_{3\alpha\beta\mu}$, or equivalently
\begin{equation}
 \Theta^{3\mu}=\Theta^{\mu 3}\label{bp-rel2}.
\end{equation}
Combining the above two relations we can therefore infer that, for vanishing $\alpha^{\mu k}$, the disclination density tensor $\boldsymbol{\Theta}$ is symmetric. Moreover, due to \eqref{wedge-ds}, $\Theta^\mu =\Theta^{\mu 3}$, i.e., the pure in-surface disclination densities $\Theta^\mu$ (which may either characterize densities of twist disclinations in directed surfaces or intrinsic orientational anomalies in hemitropic surfaces) should be identical to the wedge disclination densities $\Theta^{\mu 3}$ associated with transverse loops, e.g., in multi-layered surfaces as discussed in Section \ref{natdef}; in particular, they should vanish in sufficiently thin structured surfaces, e.g., in 2-dimensional crystals, where both $\Theta^{\mu 3}$ and $\alpha^{\mu k}$ will be absent. We note that, in contrast, for 3-dimensional solids, the symmetry of the disclination density tensor is implied only under vanishing of the full torsion and the non-metricity tensor. It is worthwhile to reemphasize that the assumption $\alpha^{\mu k} = 0$  is realistic only in sufficiently thin structures (biological membranes, graphene sheets, etc.), which, otherwise, can support only surface edge and screw dislocations (characterized by $J^i$).  

On the other hand, if we consider multi-layered or moderately thin structures of oriented media, where the assumption of vanishing $\alpha^{\mu k}$ is no longer physical, and assume that they do not contain any disclinations and metric anomalies, and also that $J^i$ and $\alpha^{\mu k}$ are small (of the same order), then \eqref{rel-3} yields
\begin{equation}
 \nabla_\mu \alpha^{\mu k} =0.
\end{equation}
This is a conservation law for the $\alpha^{\mu k}$-type dislocations enforcing that they must always form loops or leave the surface. In either case, whether the $\alpha^{\mu k}$-dislocations are absent or not, there is no restriction on the distribution of in-surface dislocations $J^i$. This again is in contrast to 3-dimensional solids, where the first Bianchi-Padova relation provides a conservation law for all dislocation densities \cite{povs91,rcgupta16}.

\textit{Consequences of the second Bianchi-Padova relation}: Equation \eqref{diff:2}, in the absence of both $\alpha^{\mu q}$-type dislocations and metric anomalies ($Q_{ijk} = 0$), in addition to $\boldsymbol{\zeta} = \boldsymbol{0}$, reduces to a simple conservation law
\begin{equation}
 \nabla_\mu\Theta^{\mu k} = 2\varepsilon_{3\mu\nu}J^\mu \Theta^{\nu k},
\end{equation}
to be satisfied by disclinations characterized by $\Theta^{\mu k}$, as well as $\Theta^{k\mu}$ owing to the symmetries $\Theta^{\mu k}=\Theta^{k\mu}$ (Equations \eqref{bp-rel1} and \eqref{bp-rel2}), and surface edge dislocations. Assuming that $J^\alpha$ and $\Theta^{\mu k}=\Theta^{k\mu}$ are small, and of the same order, we obtain

\begin{equation}
 \nabla_\mu \Theta^{\mu k} = \nabla_\mu \Theta^{k\mu} =0.
\end{equation}
These are linear conservation laws for the respective disclinations, requiring their lines to either form loops or leave the surface. Note that there is no restriction on $\Theta^{3}$ (wedge disclinations), in contrast to what one would expect for 3-dimensional solids.

\textit{Consequences of the third Bianchi-Padova relation}: We use \eqref{diff:3} to obtain a simple representation for non-metric tensor. With $\boldsymbol{\zeta} = \boldsymbol{0}$,  \eqref{diff:3} can be rewritten as
\begin{equation}
 \big(\tilde Q_{jkl,i}+L^p_{jk}\tilde Q_{ipl}+L^p_{jl}\tilde Q_{ipk}\big)_{[ji]}=0.
 \label{bprel2}
\end{equation}
It can be shown by direct substitution that a non-trivial solution of \eqref{bprel2} is given by
\begin{equation}
 \tilde Q_{kij}=-2 \tilde q_{ij;k},
 \label{irrot}
\end{equation}
where $\tilde q_{ij}=\tilde q_{ji}$ are arbitrary symmetric functions over $V$. It is a consequence of the fundamental existence theorem of linear differential systems that in absence of disclinations (i.e., $\Omega_{ijkl}=0$) over a simply connected $U$ (hence $V$), if the matrix field $\bar g_{ij}:=g_{ij}-2\tilde q_{ij}$ is positive-definite for symmetric functions $\tilde q_{ij}=\tilde q_{ji}$, then $ \tilde Q_{kij}=-2 q_{ij;k}$ is the only solution to \eqref{bprel2} over $V$. This result is proved in \cite{rcgupta16}. As the density of metric anomalies is assumed to be uniform with respect to the $\zeta$ coordinate, we will interpret this representation of the metric anomalies in absence of disclinations over simply connected patches over $\omega$ as
\begin{equation}
 Q_{kij}(\theta^\alpha)=-2 \tilde q_{ij;k}\big|_{\zeta=0}.
 \label{irrot-1}
\end{equation}
The symmetric matrix field $\tilde q_{ij}$ is known as quasi-plastic strain \cite{anth71}. In absence of disclinations, the positive-definite symmetric matrix field $\bar g_{ij}$ can be used to define an {\it auxiliary material space} $(\omega, \mathfrak{L},\bar{\boldsymbol{g}})$, equipped with the original material connection $\mathfrak{L}$ but a metric $\bar{\boldsymbol{g}}$. The non-metricity of the auxiliary material space vanishes identically by definition.
 The second-order tensor field $\boldsymbol{q}:=q_{\mu\nu}\boldsymbol{A}^\mu\otimes\boldsymbol{A}^\nu$, where $q_{\mu\nu}(\theta^\alpha):=\tilde q_{\mu\nu}(\theta^\alpha,\zeta=0)$, characterizing pure in-surface metric anomalies in the absence of disclinations, has the unique decomposition
\begin{equation}
 q_{\mu\nu}=\lambda a_{\mu\nu}+\mathfrak{q}_{\mu\nu},
\end{equation}
where $\lambda:=\frac{1}{2}q^\mu{}_{\mu}=\frac{1}{2}a^{\mu\alpha}q_{\alpha\mu}$ is the trace of $\boldsymbol{q}$ and $\mathfrak{q}_{\mu\nu}$ is the deviatoric part of $q_{\mu\nu}$ (i.e., $\mathfrak{q}^\mu{}_{\mu}=0$). The first term represents isotropic metric anomalies and the second represents anisotropic metric anomalies \cite{rcgupta16}. When $\boldsymbol{q}$ is purely isotropic, i.e., $q_{\mu\nu}=\lambda a_{\mu\nu}$, it is straightforward to obtain $Q_{\alpha\mu\nu}=-\mu_{,\alpha}a_{\mu\nu}$, where $\mu:=\ln(1+2\lambda)$. The surface metric of the auxiliary material space for isotropic metric anomalies is, hence, conformal to the surface metric of the original material space, $\bar a_{\mu\nu}=(1+2\lambda)a_{\mu\nu}$. This formulation is readily applicable to model various real-life surface metric anomalies such as 2-dimensional anisotropic biological growth, thermal expansion, distributed point defects, etc.

\textit{Consequences of the fourth Bianchi-Padova relation}: The fourth Bianchi-Padova relation imposes interdependence on the disclination density measures $\Theta^{pq}$, out of which the interdependence between the two distinct families of disclinations characterized by $\Omega_{\alpha\beta ij}$ and $\Omega_{\alpha 3 ij}$, derived in the following, are in particular interesting.
Assuming $\boldsymbol{\zeta} = \boldsymbol{0}$, the in-surface components of \eqref{bprel-4th} require
$\Omega_{\alpha\beta\mu\nu}-\Omega_{\mu\nu\alpha\beta} =  0$,
since $A_{[\alpha\beta\mu]\cdots} = 0$, which is the trivial relation $\Theta^{33}=\Theta^{33}$. Next, if we also assume that the metric anomalies are absent, i.e., $Q_{kij} = 0$, then \eqref{bprel-4th}, together with \eqref{diff:1}, yields
\begin{equation}
 \Omega_{\alpha j\mu 3}-\Omega_{\mu 3\alpha j}=-3\big(T_{[3\mu|\alpha|;j]}-2T_{[j 3}{}^i T_{\mu]i\alpha}  + T_{[\alpha\mu|j|;3]}-2T_{[3\alpha}{}^i T_{\mu]ij}   + T_{[j 3|\mu|;\alpha]}-2T_{[\alpha j}{}^i T_{3]i\mu}\big).
 \label{bprel-4th-1}
\end{equation}
Here, $T_{ijp}    := T_{ij}{}^{k} g_{kp}$. After substituting $T_{ijk;3} =  0$, as per our assumption on $\mathfrak{L}$, and $T_{3\alpha}{}^i =  0$, or equivalently $\alpha^{\mu k} =  0$, the right hand side of the above relation vanishes identically, thereby enforcing the symmetries
\begin{equation}
 \Omega_{\alpha\beta\mu 3}-\Omega_{\mu 3\alpha\beta}=0~~\text{and}~~\Omega_{\alpha 3\mu 3}-\Omega_{\mu 3\alpha 3}=0.
\end{equation}
In terms of disclination densities, these are, respectively, $\Theta^{3\mu}=\Theta^{\mu 3}$ and $\Theta^{\nu\mu}=\Theta^{\mu \nu}$. Interestingly, we reached the same conclusion from the first Bianchi-Padova relation. We will of course obtain a non-trivial consequence of the fourth Bianchi-Padova identity whenever $\alpha^{\mu k}\neq 0$.

The results of this section are summarized in Table \ref{tbl2}. 

\begin{table}[t!]
\centering
\begin{tabular}{|c|c|}
\hline
\makecell{\it Symmetries and conservation laws \\ \it from Bianchi-Padova relations with $\zeta_{ij}{}^k\equiv 0$} & {\it Implications on defect densities}  \\
\hline
\hline
{$\alpha^{\mu k} = 0\Rightarrow \Theta^{ij}=\Theta^{ji}$}  & \makecell{The two distinct families of disclinations\\ $\Theta^{3i}$ and $\Theta^{\alpha i}$ are dependent on each other}\\
\hline
\makecell{\{$\Theta^{ij} = 0, ~Q_{kij} = 0$, and $\alpha^{\mu k}$, $J^i$  small\} $\Rightarrow$ \\ $\nabla_{\mu}\alpha^{\mu k}=0$} & \makecell{$\alpha^{\mu k}$-dislocations either form \\ loops
   or leave the surface} \\
\hline
\makecell{\{$\alpha^{\mu k} = 0,~ Q_{kij} = 0$, $J^\alpha$ and $\Theta^{\mu k}=\Theta^{k\mu}$ small\} \\ $\Rightarrow$ $\nabla_\mu \Theta^{\mu k}=\nabla_\mu\Theta^{k\mu}=0$} &  \makecell{Disclinations associated with the transverse  \\ loops,
 either form loops or leave the surface}\\
\hline
\makecell{On simply connected domains on $\omega$,  \\ with $\Theta^{ij} = 0$, $Q_{kij}=-2\tilde q_{ij;k}\big|_{\zeta=0}$} & \makecell{Non-metricity $Q_{kij}$ can be represented in \\  terms of a symmetric second-order tensor}   \\
\hline
\end{tabular}
\caption{Symmetries, conservation laws, and representations of defect density fields imposed by the Bianchi-Padova relations.}
\label{tbl2}
\end{table}

\subsection{The induced Riemannian structure}\label{irs}
The coefficients $L^i_{jk}$ of any general affine connection of a manifold $\mathcal{M}$, with non-trivial torsion $\tilde T_{ij}{}^p$ and non-metricity $\tilde Q_{kij}$, can be decomposed as \cite[p.~141]{schouten}
\begin{equation}
 L^i_{jk}=\Gamma^i_{jk}+\tilde W_{jk}{}^i,
 \label{con:1}
\end{equation}
where the functions $\Gamma^i_{jk}$ are coefficients of the Levi-Civita connection (torsion-free, metric-compatible) induced by the metric $g_{ij}$:
\begin{equation}
 \Gamma^i_{jk}:=\frac{1}{2}g^{ip}(g_{pk,j}+g_{pj,k}-g_{jk,p}),
 \label{chris-deformed1}
\end{equation}
with $[g^{ij}]:=[g_{ij}]^{-1}$, and
\begin{subequations}
\begin{align}
\tilde W_{ij}{}^k &:=  \tilde C_{ij}{}^k+ \tilde M_{ij}{}^k, \label{con:11a}\\
\tilde C_{ij}{}^k &:= g^{kp}\big(-\tilde T_{ip j}+\tilde T_{p ji}-\tilde T_{jip}), \label{con:11b}\\
\tilde M_{ij}{}^k &:= \frac{1}{2} g^{kp} \big(\tilde Q_{i p j}-\tilde Q_{pji}+\tilde Q_{ji p}\big). \label{con:11c}
\end{align}
\label{con:11}%
\end{subequations}
The functions $\tilde C_{ij}{}^k$ form the components of the contortion tensor, whereas the tensor $\tilde M_{ij}{}^k$ is an equivalent measure of non-metricity. The covariant components
\begin{equation}
\tilde R_{kljp}:= g_{pi}\big(\Gamma^i_{lj,k}- \Gamma^i_{kj,l}+\Gamma^h_{lj} \Gamma^i_{kh}-\Gamma^h_{kj} \Gamma^i_{lh}\big) 
\label{curv:riem}
\end{equation}
of the Riemann-Christoffel curvature tensor of the Levi-Civita connection $\Gamma$ and components $\tilde \Omega_{kljp}$ of the material curvature are related as \cite[p.~141]{schouten}
\begin{equation}
\tilde R_{ijpl}=\tilde \Omega_{ijpl} -2\tilde \partial_{[i}\tilde W_{j]pl} -2\tilde W_{[i|m l|}\, \tilde W_{j]p}{}^m,
\label{identity1}
\end{equation}
where $\tilde W_{ijp}:=\tilde W_{ij}{}^k g_{kp}$ and $\tilde \partial$ denotes covariant differentiation with respect to the Levi-Civita connection $\Gamma$. From the general symmetry relations, $\tilde R_{ijkl}=\tilde R_{klij}=-\tilde R_{jikl}$, of the Riemannian curvature induced by a metric, it is evident that it has only six independent components characterized by $\tilde R_{\alpha\beta\mu\nu}$, $\tilde R_{\alpha\beta\mu 3}$ and $\tilde R_{\alpha 3\beta 3}$. The only non-trivial relations out of \eqref{identity1}, when restricted to $\zeta=0$, are
\begin{subequations}
 \begin{align}
  R_{\alpha\beta\mu\nu}&=\Omega_{\alpha\beta\mu\nu} -2\partial_{[\alpha}W_{\beta]\mu\nu} -2W_{[\alpha|i \nu|}\, W_{\beta]\mu}{}^i,\label{incompatibility-relations-a}\\
R_{\alpha\beta\mu 3}&=\Omega_{\alpha\beta\mu 3} -2\partial_{[\alpha}W_{\beta]\mu 3} -2W_{[\alpha|i 3|}\, W_{\beta]\mu}{}^i,~\text{and}~~\label{incompatibility-relations-b}\\
R_{\alpha 3\mu 3}&=\Omega_{\alpha 3\mu 3} -\partial_{\alpha}W_{3\mu 3} -2W_{[\alpha|i 3|}\, W_{3]\mu}{}^i\label{incompatibility-relations-c}.
 \end{align}
\label{incompatibility-relations}%
\end{subequations}
Here, $R_{ijkl}(\theta^\alpha):=\tilde R_{ijkl}(\theta^\alpha,0)$, $W_{ijk}(\theta^\alpha):=\tilde W_{ijk}(\theta^\alpha,0)$, $W_{ij}{}^k(\theta^\alpha):= \tilde W_{ij}{}^k(\theta^\alpha,0)$, and $\partial$ denotes covariant differentiation with respect to the projected Levi-Civita connection on $U$, consisting of components $\Gamma_{\alpha\beta}^\mu\big|_{\zeta=0}$. The relations \eqref{incompatibility-relations} are central to the theory of mechanics of defects as they are directly related to the strain incompatibility equations which we discuss next. Indeed, once we have identified the material metric $\boldsymbol{g}$ in terms of strain fields on the structured surface, \eqref{incompatibility-relations} constitute a system of PDEs for the strain fields, with defect densities as source terms. It should be noted that the components  $\Omega_{\mu 3\alpha\beta}$ do not appear in any of the equations \eqref{incompatibility-relations}. This is because, according to the fourth Bianchi-Padova relation \eqref{bprel-4th-1}, they can be written in terms of $ \Omega_{\alpha\beta\mu 3}$ and other defect measures, and hence are not independent quantities.


\section{Strain incompatibility relations for structured surfaces}\label{strain}

In this section, we begin by introducing the notion of strain for a structured surface. The complete set of strains represent essentially the kinematical nature of shell theory that is being employed to describe structured surfaces. The strain fields also provide us with fundamental variables for construction of constitutive responses of the continuum. Once the strain fields are fixed, we look for the necessary and sufficient (compatibility) conditions for the existence of a local isometric embedding of the surface. Finally, we discuss how various defect densities become sources of strain incompatibility precluding the existence of the local isometric embedding. This will then set the stage for posing complete boundary-value-problems for internal stress distribution and natural shapes of defective structured surfaces, as will be discussed subsequently. 

\subsection{Strain measures and strain compatibility} \label{comp}
Let us assume that there exist a local isometric embedding $\boldsymbol{R}:\omega\to\mathbb{R}^3$ of $\omega$ into $\mathbb{R}^3$. Let $\boldsymbol{A}_{\alpha}:=\boldsymbol{R}_{,\alpha}$ and $\boldsymbol{N}:={\boldsymbol{A}_1\times\boldsymbol{A}_2}/{|\boldsymbol{A}_1\times\boldsymbol{A}_2|}$. The first and second fundamental forms associated with this embedding (over a local patch) are therefore $A_{\alpha\beta} = \boldsymbol{A}_{\alpha} \boldsymbol\cdot \boldsymbol{A}_{\beta}$ and $B_{\alpha\beta} = - \boldsymbol{N}_{,\beta}\boldsymbol{\cdot}\boldsymbol{A}_{\alpha}$, respectively. We consider the following sufficiently smooth fields, defined over $\boldsymbol{R}(U)$, as descriptors of strain on the structured surface: (i) a symmetric tensor $E_{\alpha\beta}$, representing the in-surface strain field for measuring the local changes in length and angle; (ii) a tensor $\Lambda_{\alpha\beta}$ for transverse bending strains; (iii) two vectors $\Delta_\alpha$ and $\Lambda_\alpha$ for measuring transverse shear and normal bending strains, respectively; and (iv) a scalar $\Delta$ for normal expansion. We now pose the central question for conditions of local strain compatibility.

Given sufficiently smooth strain fields (i)-(iv) over a fixed local isometric embedding $\boldsymbol{R}(U)$ of a 2-dimensional manifold $\omega$, with first and second fundamental forms $A_{\alpha\beta}(\theta^\alpha)$ and $B_{\alpha\beta}(\theta^\alpha)$, respectively, what are the conditions to be satisfied for there to exist a sufficiently smooth local isometric embedding $\boldsymbol{r}:U\subset\omega\to\mathbb{R}^3$, with first and second fundamental forms $a_{\alpha\beta}$ and $b_{\alpha\beta}$ suitably constructed out of the given fields, along with a sufficiently smooth director field $\boldsymbol{d}:\boldsymbol{r}(U)\to\mathbb{R}^3$, so that the equations 
\begin{subequations}
\begin{align}
E_{\alpha\beta} &= \frac{1}{2}(\boldsymbol{a}_{\alpha}\boldsymbol{\cdot}\boldsymbol{a}_{\beta} - \boldsymbol{A}_{\alpha}\boldsymbol{\cdot}\boldsymbol{A}_{\beta}) =\frac{1}{2}(a_{\alpha\beta}-A_{\alpha\beta}), \label{def1}\\
\Delta_{\alpha} &= \boldsymbol{d}\boldsymbol{\cdot}\boldsymbol{a}_{\alpha}- \boldsymbol{N}\boldsymbol{\cdot}\boldsymbol{A}_{\alpha}  = d_\alpha,\label{def2}\\
\Delta &= \boldsymbol{d}\boldsymbol{\cdot}\boldsymbol{a}_3- \boldsymbol{N}\boldsymbol{\cdot}\boldsymbol{N}= d_3-1,\label{def3}\\
\Lambda_{\alpha\beta} &= \boldsymbol{d}_{,\beta}\boldsymbol{\cdot}\boldsymbol{a}_{\alpha}- \boldsymbol{N}_{,\beta}\boldsymbol{\cdot}\boldsymbol{A}_{\alpha} = \partial_\beta d_{\alpha}-d_3\,b_{\alpha\beta}+B_{\alpha\beta},~\text{and}\label{def4}\\
\Lambda_{\beta} &=  \boldsymbol{d}_{,\beta}\boldsymbol{\cdot}\boldsymbol{a}_{3}- \boldsymbol{N}_{,\beta}\boldsymbol{\cdot}\boldsymbol{N} =  d_{3,\beta}+d_\mu\,b^\mu_\beta,\label{def5}
 \end{align}
 \label{def-strain}%
 \end{subequations}
are satisfied on $U$? Here $a_{\alpha}:=\boldsymbol{r}_{,\alpha}$, $\boldsymbol{a}_3:={\boldsymbol{a}_1\times\boldsymbol{a}_2}/{|\boldsymbol{a}_1\times\boldsymbol{a}_2|}$, while $\partial$ denotes covariant derivative with respect to the surface Christoffel symbols induced by the metric $a_{\alpha\beta}$ on the deformed base configuration. Clearly, the strain fields measure deformation of the structured surface from its reference configuration $(\boldsymbol{R}(U),\boldsymbol{N}(U))$ to the deformed configuration $(\boldsymbol{r}(U),\boldsymbol{d}(U))$.
The necessary and sufficient conditions, to be satisfied by the given strain fields, so that a local deformed configuration of the structured surface does exist such that equations \eqref{def-strain} are satisfied, are called local strain compatibility conditions. These are nothing but the integrability conditions for $\boldsymbol{r}$ and $\boldsymbol{d}$, as inferred from the system of PDEs in \eqref{def-strain}. Such compatibility conditions in the context of thin shells have been derived earlier by Epstein \cite{eps1} and more recently by the present authors \cite{rcgupta15}. The discussion below follows the latter.

The local strain compatibility conditions, over a simply-connected open set $W\subset U$, are given by
\begin{subequations}
\begin{align}
 a_{\alpha\beta} &:= A_{\alpha\beta}+2E_{\alpha\beta}~\text{is positive-definite}, \label{compatibility1}\\
 \Delta &\ne -1,\label{compatibility2}\\
 \Lambda_\beta&-\Delta_{,\beta}-\Delta_\mu b^\mu_\beta =0,\label{compatibility3}\\
 \Lambda_{[\alpha\beta]}&-\partial_{[\beta}\Delta_{\alpha]}=0,\label{compatibility4}\\
 \partial_1 b_{21}&-\partial_2 b_{11} =0,\label{compatibility5}\\
 \partial_1 b_{22}&-\partial_2 b_{12} =0,~\text{and} \label{compatibility6}\\
 K_{1212}- (b_{12}^2  &- b_{11}b_{22}) = 0,\label{compatibility7}
 \end{align}
 \label{compatibility}
\end{subequations}
where 
\begin{equation}
 b_{\alpha\beta}:=\frac{\partial_{(\beta}\Delta_{\alpha)} -\Lambda_{(\alpha\beta)}+B_{\alpha\beta}}{\Delta+1}
 \label{second-ff}
\end{equation}
is symmetric and $K_{1212}$ is the only independent component of the Riemann-Christoffel curvature of the projected Levi-Civita connection on $U$. We assume $\Delta\ne -1$ for \eqref{second-ff} to be a valid  definition. This would physically mean that directors are nowhere tangential to the base surface (see Remark \ref{dirtang} for the situation otherwise). Equations \eqref{compatibility5}, \eqref{compatibility6}, and \eqref{compatibility7} are the well-known Codazzi-Mainardi and Gauss equations for $a_{\alpha\beta}$ and $b_{\alpha\beta}$. Whenever these conditions are satisfied by the strain fields, there exists a sufficiently smooth local isometric embedding $\boldsymbol{r}:W\to\mathbb{R}^3$, with first and second fundamental form given by $a_{\alpha\beta}$ and $b_{\alpha\beta}$, respectively, and a director field $\boldsymbol{d}:\boldsymbol{r}(U)\to\mathbb{R}^3$ given by $d_\alpha=\Delta_\alpha, ~d_3=\Delta+1$, such that the PDEs \eqref{def-strain} are identically satisfied everywhere on $W$.

We now prove this result. Using the given strain fields $E_{\alpha\beta}$, $\Lambda_{\alpha\beta}$, $\Lambda_{\alpha}$, $\Delta_\alpha$ and $\Delta$, we construct a material metric $\boldsymbol{g}$ with components
\begin{equation}
g_{\alpha\beta}:= a_{\alpha\beta}+\zeta\,P_{\alpha\beta} +\zeta^2 \,Q_{\alpha\beta},\,\,\,g_{\alpha 3}=g_{3\alpha}:= \Delta_\alpha+\zeta\,U_\alpha,\,\,\,g_{33}:= V,
\label{metric:defn}
\end{equation}
where $a_{\alpha\beta}$ is as defined in \eqref{compatibility1} and
\begin{subequations}
\begin{align}
 P_{\alpha\beta} &:= 2(\Lambda_{(\alpha\beta)}-B_{\alpha\beta}),\\
Q_{\alpha\beta} &:= a^{\sigma\gamma}\,(\Lambda_{\sigma\alpha}-B_{\sigma\alpha})(\Lambda_{\gamma\beta}-B_{\gamma\beta})+\Lambda_\alpha\,\Lambda_\beta,\\
 U_\alpha &:= a^{\sigma\gamma}\,\Delta_\sigma (\Lambda_{\gamma\alpha}-B_{\gamma\alpha})+\Lambda_\alpha (\Delta+1),~\text{and}\\
 V &:= a^{\alpha\beta}\,\Delta_\alpha \Delta_\beta+(\Delta+1)^2.
 \end{align}
 \label{metric-definition}%
\end{subequations}
In the above, $[a^{\alpha\beta}]:=[a_{\alpha\beta}]^{-1}$ exists if we assume $E_{\alpha\beta}$ to be such that $a_{\alpha\beta}$ is positive-definite. Note that, since $\omega$ is bounded and $g_{ij}$ is continuous in $\theta^\alpha$ and $\zeta$, $g_{ij}$ will be positive-definite on $V:=U\times(-\epsilon,\epsilon)\subset\mathcal{M}$ for sufficiently small $\epsilon$. Our result is valid for this sufficiently small $\epsilon$ and we {\it a priori} construct $\mathcal{M}$ such that $\epsilon$ conforms to this small value throughout. For a technical discussion on the issue of smallness of $\epsilon$ and positive definiteness of $g_{ij}$, refer to the proof of Theorem 2.8-1 in \cite{ciar1}. The `sufficient thinness' of the structured surface is encoded in the definition \eqref{metric:defn} which describes how the 2-dimensional strain fields can be used to construct a 3-dimensional metric on the tubular neighbourhood $\mathcal{M}$ of $\omega$. The parameter $\epsilon$ can be thought of as a physical length scale inherent to the description of the structured surface, e.g., thickness of a shell structure or the length of the individual molecules (not necessarily transverse to the surface) in lipid membranes.  The 3-dimensional metric $\boldsymbol{g}$ is of second-order in the transverse coordinate $\zeta$ and this dependence brings out the non-locality in the kinematics of the structured surface, taking into account the transverse shear and normal distortion of the attached directors. The form of the metric in  \eqref{metric:defn} is a generalization of the metric with components
 \begin{equation}
  g_{\alpha\beta}(\theta^\alpha,\zeta) = a_{\alpha\beta} - 2\zeta b_{\alpha\beta}+ \zeta^2 a^{\mu\nu}b_{\mu\alpha}b_{\nu\beta}, \,\,\,g_{\alpha 3}=g_{3\alpha}= 0,\,\,\,g_{33}= 1 \label{metric:kl}
 \end{equation}
defined in the proof of Theorem 2.8-1 in \cite{ciar1}, which was otherwise restricted to Kirchhoff-Love theory (i.e., $\Lambda_\alpha=\Delta_\alpha=\Delta=0$), where $b_{\alpha\beta}:=-\Lambda_{(\alpha\beta)}+B_{\alpha\beta}$, and hence ignored any normal distortion or transverse shearing effect of the directors.

The coefficients of the Levi-Civita connection of the metric \eqref{metric:defn}, defined by $\Gamma^q_{ij}:=g^{pq}\Gamma_{ijp}$ where $[g^{pq}]:=[g_{pq}]^{-1}$ and $\Gamma_{ijp}:=\frac{1}{2}(g_{ip,j}+g_{jp,i}-g_{ij,p})$, can be calculated as
\begin{subequations}
\begin{align}
\Gamma_{333} &=0,~\Gamma_{33\rho} =U_\rho-\frac{1}{2}V_{,\rho},~\Gamma_{3\rho 3} =\Gamma_{\rho 33}=\frac{1}{2}V_{,\rho},\label{a}\\
\Gamma_{3\rho\sigma} &=\Gamma_{\rho 3\sigma}=\Delta_{[\sigma,\rho]}+\frac{1}{2}P_{\rho\sigma}+\zeta\,\bigg(U_{[\sigma,\rho]}+Q_{\rho\sigma}\bigg),\\
\Gamma_{\rho\sigma 3} &=\Delta_{(\sigma,\rho)}-\frac{1}{2}P_{\rho\sigma}+\zeta\,\bigg(U_{(\sigma,\rho)}-Q_{\rho\sigma}\bigg),~\text{and}\\
\Gamma_{\rho\sigma \delta} &=s_{\rho\sigma\delta}+\frac{\zeta}{2}\bigg(P_{\rho\delta,\sigma}+P_{\sigma\delta,\rho}-P_{\sigma\rho,\delta}\bigg)               +\frac{\zeta^2}{2}\bigg(Q_{\rho\delta,\sigma}+Q_{\sigma\delta,\rho}-Q_{\sigma\rho,\delta}\bigg).
\end{align}
\label{levivicita}%
\end{subequations}
The local strain compatibility conditions are the conditions for the embedding space $\mathcal{M}$ to be Euclidean, i.e., the Riemann-Christoffel curvature $\tilde R_{ijkl}$ of the metric \eqref{metric:defn} to become identically zero. However, as shown in \cite{rcgupta15}, in order to ensure compatibility of the 2-dimensional strain fields, it is enough to impose that $\tilde R_{ijkl}\big|_{\zeta=0}=R_{ijkl}(\theta^\alpha)=0$. The curvature $R_{ijkl}$ has six independent components such that $R_{ijkl}=0$ if and only if $R_{1212}=0$, $R_{12\sigma 3}=0$, and $R_{\rho 3 \sigma 3}=0$. After some manipulations, it can be shown that
\begin{equation}
 R_{1212} = K_{1212}-(b^2_{12}-b_{11}b_{22}),\label{rel0}
\end{equation}
where the functions $K_{\beta\alpha\mu\nu} :=a_{\rho\nu}(\Gamma^\rho_{\alpha\mu,\beta}-\Gamma^\rho_{\beta \mu,\alpha}+\Gamma^\delta_{\alpha\mu} \Gamma^\rho_{\beta\delta} - \Gamma^\delta_{\beta\mu} \Gamma^\rho_{\alpha\delta})$
constitute the covariant components of the Riemann-Christoffel curvature of the projected Levi-Civita connection $\Gamma^\mu_{\alpha\nu}\big|_{\zeta=0}$ on $U$. These, by definition, have the symmetries $K_{\alpha\beta\mu\nu}=-K_{\alpha\beta\nu\mu}=K_{\mu\nu\alpha\beta}$ and, hence, have only one independent component $K:=\frac{1}{4}\varepsilon^{\alpha\beta}\varepsilon^{\mu\nu}K_{\alpha\beta\mu\nu}$,
the Gaussian curvature induced by the surface metric $a_{\alpha\beta}$, where  $\varepsilon^{\alpha\beta}:=a^{-\frac{1}{2}}e^{\alpha\beta}$ and $e^{\alpha\beta}=e_{\alpha\beta}$ denotes the 2-dimensional permutation symbol; also, $a:=\mbox{det} [a_{\alpha\beta}]$. It is easily seen that $K_{1212}=4a K$.  Consequently, $R_{1212}=0$, in conjunction with \eqref{rel0}, can be used to infer \eqref{compatibility7}, which is the single independent Gauss' equation satisfied by $a_{\alpha\beta}$ and $b_{\alpha\beta}$. Further, we can evaluate
\begin{equation}
R_{121 3}=a^{2\beta}\Delta_\beta\bigg(K_{2 1 21}-\big(b^2_{1 2}-b_{1 1} b_{2 2}\big)\bigg)-(\Delta+1)\big(\partial_2 b_{1 1}-\partial_1 b_{1 2}\big)~\text{and}
\label{rel1}
\end{equation}
\begin{equation}
R_{122 3}=-a^{1\beta}\Delta_\beta\bigg(K_{1 2 12}-\big(b_{12}^2-b_{1 1} b_{2 2}\big)\bigg)-(\Delta+1)\big(\partial_2 b_{2 1}-\partial_1 b_{2 2}\big).
\label{rel2}
\end{equation}
Substituting \eqref{compatibility7} in \eqref{rel1} and \eqref{rel2}, the condition $R_{12\sigma 3}=0$ yields \eqref{compatibility5} and \eqref{compatibility6},
which are the two independent Codazzi-Mainardi equations satisfied by $a_{\alpha\beta}$ and $b_{\alpha\beta}$. The Gauss and Codazzi-Mainardi equations satisfied over a simply-connected domain $W\subset\omega$ ensure the existence of a local isometric embedding $\boldsymbol{r}:W\to\mathbb{R}^3$, with first fundamental form $a_{\alpha\beta}$ and second fundamental form $b_{\alpha\beta}$, modulo isometries of $\mathbb{R}^3$.

Finally, we calculate
\begin{eqnarray}
R_{\rho 3 \sigma 3}
 &=&(\Delta+1)\, \partial_{(\sigma} I_{\rho)}-\Lambda_{(\rho}I_{\sigma)}-a^{\alpha\beta}\Delta_\alpha\bigg\{b_{\beta(\rho}I_{\sigma)}+\frac{1}{2}e_{\beta(\rho}I_{\sigma)}J- b_{\rho\sigma}\,I_\beta\bigg\} \nonumber\\
 && + \bigg( a^{\alpha\beta}(\Delta + 1)^2 + a^{\alpha\mu} a^{\beta\nu} \Delta_\mu \Delta_\nu \bigg) (J)^2 e_{\alpha\rho} e_{\beta\sigma},\label{incomp-3}
 \end{eqnarray}
where
\begin{equation}
I_\beta:=\Lambda_\beta-\Delta_{,\beta}-\Delta_\alpha a^{\alpha\gamma} b_{\gamma\beta}~~\text{and}~~J:=\frac{\varepsilon^{\alpha\beta}\big(\partial_{[\beta}\Delta_{\alpha]}-\Lambda_{[\alpha\beta]}\big)}{2(\Delta+1)}.
\label{comp4}
\end{equation}
The condition $R_{\rho 3 \sigma 3}=0$ is therefore a set of three coupled first order homogeneous non-linear partial differential algebraic equations for three unknowns $I_\alpha$ and $J$. It has been shown previously \cite{rcgupta15} that the only physically meaningful solution is the trivial set $I_\alpha=0$ and $J=0$; the non-zero solutions become unstable under generic perturbations of zero initial condition. These equalities are equivalent to \eqref{compatibility3} and \eqref{compatibility4}, respectively. They ensure the existence of a well-defined director field $\boldsymbol{d}:\boldsymbol{r}(W)\to\mathbb{R}^3$, defined by $d_\alpha=\Delta_\alpha$ and $d_3=\Delta+1$ (see \eqref{def2} and \eqref{def3}), which satisfies \eqref{def4} and \eqref{def5} identically over any simply-connected open set $W\subset\omega$. This finishes our proof.

\begin{rem}  (Existence of global isometric embedding)
{\rm
If $\omega$ is a simply-connected manifold such that it can be covered by a single chart (manifolds homeomorphic to an open disc), the simply-connected open set $W$ in the preceding discussion can be extended to the whole manifold $\omega$, i.e., equations \eqref{compatibility} are sufficient to be satisfied point-wise over $\omega$ by various strain measures such that there exists a global deformed state $(\boldsymbol{r}(\omega),\boldsymbol{d}(\omega))$ satisfying \eqref{def-strain}.
On the other hand, if $\omega$ is simply-connected but cannot be covered by a single chart (e.g., manifolds homeomorphic to a sphere), then equations \eqref{compatibility} are sufficient to be satisfied by the given strain fields over every open set $W_i$ in some open covering $\{W_i\}_{i\in I}$ of $\omega$, where $I$ is some indexing set (i.e., $\bigcup_{i\in I} W_i = \omega$), such that there exists a global deformed state $(\boldsymbol{r}(\omega),\boldsymbol{d}(\omega))$ satisfying \eqref{def-strain}.
}
\end{rem}

\begin{rem}  (Strain fields in structured surfaces with tangential director field)
{\rm
When the director fields are everywhere tangential to their respective base surfaces, we choose the reference director field $\boldsymbol{D}$ to be some known tangent vector field over $\boldsymbol{R}(\omega)$ (rather than the normal field $\boldsymbol{N}$). The relations \eqref{def-strain} are now replaced by 
\begin{subequations}
\begin{align}
E_{\alpha\beta} &= \frac{1}{2}(\boldsymbol{a}_{\alpha}\boldsymbol{\cdot}\boldsymbol{a}_{\beta} - \boldsymbol{A}_{\alpha}\boldsymbol{\cdot}\boldsymbol{A}_{\beta}) =\frac{1}{2}(a_{\alpha\beta}-A_{\alpha\beta}), \label{def2-1}\\
\Delta_{\alpha} &= \boldsymbol{d}\boldsymbol{\cdot}\boldsymbol{a}_{\alpha}- \boldsymbol{D}\boldsymbol{\cdot}\boldsymbol{A}_{\alpha}  = d_\alpha-D_\alpha,\label{def2-2}\\
\Lambda_{\alpha\beta} &= \boldsymbol{d}_{,\beta}\boldsymbol{\cdot}\boldsymbol{a}_{\alpha}- \boldsymbol{D}_{,\beta}\boldsymbol{\cdot}\boldsymbol{A}_{\alpha} = \partial_\beta d_{\alpha}-\bar\nabla_\beta D_\alpha, ~\text{and}~~\label{def2-3}\\
\Lambda_\beta &= \boldsymbol{d}_{,\beta}\boldsymbol{\cdot}\boldsymbol{n} - \boldsymbol{D}_{,\beta}\boldsymbol{\cdot}\boldsymbol{N}=d_\mu b^\mu_\beta - D_\mu B^\mu_\beta, \label{def2-4}
 \end{align}
 \label{def-strain-2}%
 \end{subequations}
where $\bar\nabla$ denotes the covariant derivative with respect to the induced Levi-Civita connection by the metric $A_{\alpha\beta}$ on the reference embedding $\boldsymbol{R}(\omega)$. 
The integrability conditions, obtained from the above PDEs, for unknown $\boldsymbol{r}$ and $\boldsymbol{d}$, given $A_{\alpha\beta}$, $B_{\alpha\beta}$, $D_\alpha$, and the strain fields, provide the local strain compatibility conditions. To derive local compatibility relations, we note that the metric of the deformed surface is completely determined by \eqref{def2-1}, $a_{\alpha\beta}:=A_{\alpha\beta} + 2E_{\alpha\beta}$,  with $E_{\alpha\beta}$ such that $a_{\alpha\beta}$ is positive-definite; this is same as before. However, we no longer have a straight forward formula for the functions $b_{\alpha\beta}$, contrary to the case with non-tangential directors. As a candidate for the second fundamental form of the deformed surface, we choose any $b_{\alpha\beta}$ that solves the algebraic equation
\begin{equation}
 (\Delta_\mu + D_\mu) b^\mu_\beta = \Lambda_\beta + D_\mu B^\mu_\beta,
\end{equation}
which is arrived after eliminating $d_\alpha$ between \eqref{def2-2} and \eqref{def2-4}. The Codazzi-Mainardi and Gauss' equations involving $a_{\alpha\beta}$ and $b_{\alpha\beta}$ provide the first set of strain compatibility conditions, ensuring the existence of a local embedding $\boldsymbol{r}:U\subset\omega\to\mathbb{R}^3$ with metric and curvature given by $a_{\alpha\beta}$ and $b_{\alpha\beta}$, respectively, modulo isometries of $\mathbb{R}^3$. The other strain compatibility condition is given by 
\begin{equation}
 \Lambda_{\alpha\beta} = \partial_\beta(\Delta_\alpha+D_{\alpha}) - \bar\nabla_\beta D_\alpha,
\end{equation}
obtained by eliminating $d_\alpha$ between \eqref{def2-2} and \eqref{def2-3}. This ensures the existence of a tangential director field $\boldsymbol{d}:\boldsymbol{r}(U)\to\mathbb{R}^3$ such that equations \eqref{def-strain-2} are satisfied. 
} \label{dirtang}
\end{rem}

\subsection{Strain incompatibility arising from defects}

It is well-known that distributed defects within the material structure are inherent sources of strain incompatibility and, hence, internal stress \cite{kroner81a}. A continuous distribution of material anomalies gives rise to non-trivial strain fields over a structured surface which are, in general, incompatible. This means that the fields $a_{\alpha\beta}$ and $b_{\alpha\beta}$, constructed out of the strain fields that solve the {\it strain incompatibility relations}, do not correspond to the first and second fundamental form of any realizable isometric embedding of $\omega$ into $\mathbb{R}^3$, not even locally. Hence, all the strain compatibility conditions must be violated in presence of defects. The local strain incompatibility relations are given by the equations \eqref{incompatibility-relations}, where the surface strain incompatibility measures $R_{ijkl}$ apearing on the left-hand-side are defined by the expressions \eqref{rel0}-\eqref{incomp-3} in terms of the strain fields, and on the right-hand-side appear the source terms of various defect densities. These relations are the following:
\begin{equation}
 K_{1212}-[b^2_{12}-b_{11}b_{22}]= g\Theta^3 -2\partial_{[1}W_{2]12} -2W_{[1|i 2|}\, W_{2]1}{}^i, \label{strain-incompatibility-relations1}
\end{equation}
\begin{equation}
 a^{2\beta}\Delta_\beta\bigg(K_{1212}-\big[b^2_{1 2}-b_{1 1} b_{2 2}\big]\bigg)-(\Delta+1)\big[\partial_2 b_{1 1}-\partial_1 b_{1 2}\big]= -g\Theta^2-2\partial_{[1}W_{2]1 3} -2W_{[1|i 3|}\, W_{2]1}{}^i, \label{strain-incompatibility-relations2}
\end{equation}
\begin{equation}
 -a^{1\beta}\Delta_\beta\bigg(K_{1 2 12}-\big[b_{12}^2-b_{1 1} b_{2 2}\big]\bigg)-(\Delta+1)\big[\partial_2 b_{2 1}-\partial_1 b_{2 2}\big]= g\Theta^1-2\partial_{[1}W_{2]2 3} -2W_{[1|i 3|}\, W_{2]2}{}^i,~\text{and} \label{strain-incompatibility-relations3}
\end{equation}

\begin{eqnarray}
(\Delta+1)\, \partial_{(\sigma}I_{\rho)}-\Lambda_{(\rho}I_{\sigma)}-a^{\alpha\beta}\Delta_\alpha\bigg\{b_{\beta(\rho}I_{\sigma)}+\frac{1}{2}e_{\beta(\rho}I_{\sigma)}J- b_{\rho\sigma}\,I_\beta\bigg\}& &\nonumber \\
+\bigg( a^{\alpha\beta}(\Delta+1)^2+a^{\alpha\mu}a^{\beta\nu}\Delta_\mu \Delta_\nu \bigg)\,(J)^2 \varepsilon_{\alpha\rho}\varepsilon_{\beta\sigma} 
&=&\varepsilon_{\rho 3 \nu }\varepsilon_{\sigma 3 \mu}\Theta^{\nu\mu}-\partial_{\rho}W_{3\sigma 3} -2W_{[\rho|i 3|}\, W_{3]\sigma}{}^i.\nonumber\\
\label{strain-incompatibility-relations4}
\end{eqnarray}

The above are the local strain incompatibility relations for a continuously defective structured surface in their full generality. We recall, from Section \ref{irs}, that the functions $W_{ij}{}^k$ are defined in terms of dislocation densities and metric anomalies as $W_{ij}{}^k=C_{ij}{}^k+M_{ij}{}^k$, where the components $C_{ij}{}^k$ of contortion tensor are algebraic functions of the dislocation densities $J^i$ and $\alpha^{\mu k}$, and the components $M_{ij}{}^k$ are algebraic functions of the densities of metric anomalies $Q_{kij}$, see \eqref{con:11}. In the absence of dislocations and metric anomalies, i.e., when $W_{ij}{}^k\equiv 0$, clearly, the density of wedge disclinations $\Theta^3$ act as the single source to the incompatibility of the Gauss equation \eqref{strain-incompatibility-relations1}, while the densities of twist disclinations/intrinsic orientational anomalies $\Theta^\mu= \Theta^{\mu 3}$ are the single source terms to the incompatible Codazzi-Mainardi equations \eqref{strain-incompatibility-relations2} and \eqref{strain-incompatibility-relations3}; the symmetric disclination density fields $\Theta^{\mu\nu}$ are sources to non-trivial $I_\alpha$ and $J$. In many applications, to follow in the next section, we will restrict attention to sufficiently thin structured surfaces, e.g., purely disclinated nematic membranes, monolayer bio-membranes, 2-dimensional crystals etc. In such cases the $\Theta^{\mu\nu}$-disclinations and $\alpha^{\mu k}$-dislocations are naturally absent. We will discuss further simplifications of \eqref{strain-incompatibility-relations1}-\eqref{strain-incompatibility-relations4} in Section \ref{klincomp} under some realistic assumptions of smallness/vanishing of certain strain fields and defect densities. We will also obtain certain forms of these relations that have already appeared in the literature. Finally, note that, the disclination densities $\Theta^{\mu 3}$ seem to be absent from the above relations. This is so because they are not independent but expressible in terms of $\Theta^\mu$ and other defect densities as a consequence of the fourth Bianchi-Padova relation.


\section{Internal stress and natural shapes}\label{bvp-stress-shape}
A central problem in mechanics of solids is, for a given distribution of material defects, to determine the internal stress field and the deformed shape of the defective body with respect to a fixed reference configuration. The notion of defects is to be understood in the sense of material anomalies, as discussed in Section \ref{natdef}, which lead to an inhomogeneous material response in an otherwise materially uniform body. In particular, if we assume stress to be purely elastic in origin, then, in general, there is no one-to-one mapping from the current configuration of the defective body, which is realized as a connected set in the physical space, to its defect-free natural (stress-free) state. This means that the natural state of the defective material body cannot be realized as a connected set in the physical space. It also entails an incompatible elastic strain field, which appears as the energetic dual of stress, with sources of incompatibility derived from various defect densities. The absence of an elastic deformation map also implies that there is no one-to-one (plastic deformation) map which connects the reference configuration to the natural state. A  plastic strain field, whose incompatibility is again related to defect densities, can be derived from the difference of metric tensors associated with the natural and the reference configurations. The formulation of a well-posed boundary-value-problem for stress and deformed shape, for given defect densities, requires a prescription on how the strain fields -- total, elastic, and plastic -- are all related to each other. 

The problem of relating the three configurations (reference, natural, and current) is usually addressed by assuming a multiplicative decomposition of the total deformation gradient into elastic and plastic distortion tensors. The total deformation gradient tensor is the gradient of the total deformation map (a bijective map between the reference and the current configuration) and yields a compatible total strain tensor. The elastic and plastic distortion tensors map tangent spaces from the natural configuration to the current configuration and from the reference configuration to the natural configuration, respectively. However, in the presence of disclination density, the elastic and plastic distortion tensors are not well-defined \cite{rcgupta15}. The ambiguity arises due to the rotational part of the tensors becoming multi-valued. Nevertheless, the multiplicative decomposition can be used for isotropic materials where both elastic and plastic rotations do not play any role in the final boundary-value-problem \cite{steigmannetal14}. The need for a multiplicative decomposition can also be circumnavigated if we assume an additive decomposition of the total strain into elastic and plastic counterparts. In such a situation, we do not require the notion of elastic and plastic distortion tensors at all. For 3-dimensional elastic solids, the additive decomposition of strain is essentially based on the smallness of both deformation and plastic strain (to the same order). The resulting theory is necessarily applicable to small deformation problems \cite{dewit81}. On the other hand, an additive decomposition of strains, with the notion of strain as defined in the beginning of Section \ref{comp}, in the context of 2-dimensional structured surfaces is less restrictive. It in fact allows for moderately large rotations in the deformation while keeping small strains \cite{NaghdiVongsarnpigoon1983}. This is important for structured surfaces since, unlike 3-dimensional bodies, they are very much likely to accommodate internal stresses by escaping into the third dimension via moderately large rotations.  The nature of the assumed additive decomposition, which allows for a separation of order of the in-surface stretching and the bending mode of deformation for structured surfaces, will be discussed in detail in the following. 

The decomposition of strain field gives way to formulating the boundary-value-problem. We consider a reference configuration for the structured surface where directors are aligned with the normal; the case where directors are tangential can be treated following Remark \ref{dirtang}. The plastic strains satisfy the incompatibility equations \eqref{strain-incompatibility-relations1}-\eqref{strain-incompatibility-relations4}. The elastic strains will satisfy a different form of incompatibility equations with the reference configuration replaced by current configuration in the derivation of these equations. For this difference, they are much more involved since the current configuration is itself unknown (with directors not necessarily coinciding with the normal); we do not use incompatibility relations for elastic strain in our framework. The (plastic) strain incompatibility relations are combined with the additive decomposition, the constitutive laws (for relating elastic strains with stresses and moments), and the equilibrium equations, to yield the full boundary-value-problem for the determination of stress field and natural shape of the structured surface for a given distribution of defects. We will proceed to do so in the following under the Kirchhoff-Love deformation constraint on the director field, requiring them to coincide with the local normal field in the current configuration. This is done only in order to present a simplified theory, while postponing further generalizations to future works.

\subsection{Kinematics of Kirchhoff-Love shells with small strain accompanied by moderate rotation}

\begin{figure}[t!]
 \centering
 \includegraphics[scale=0.85]{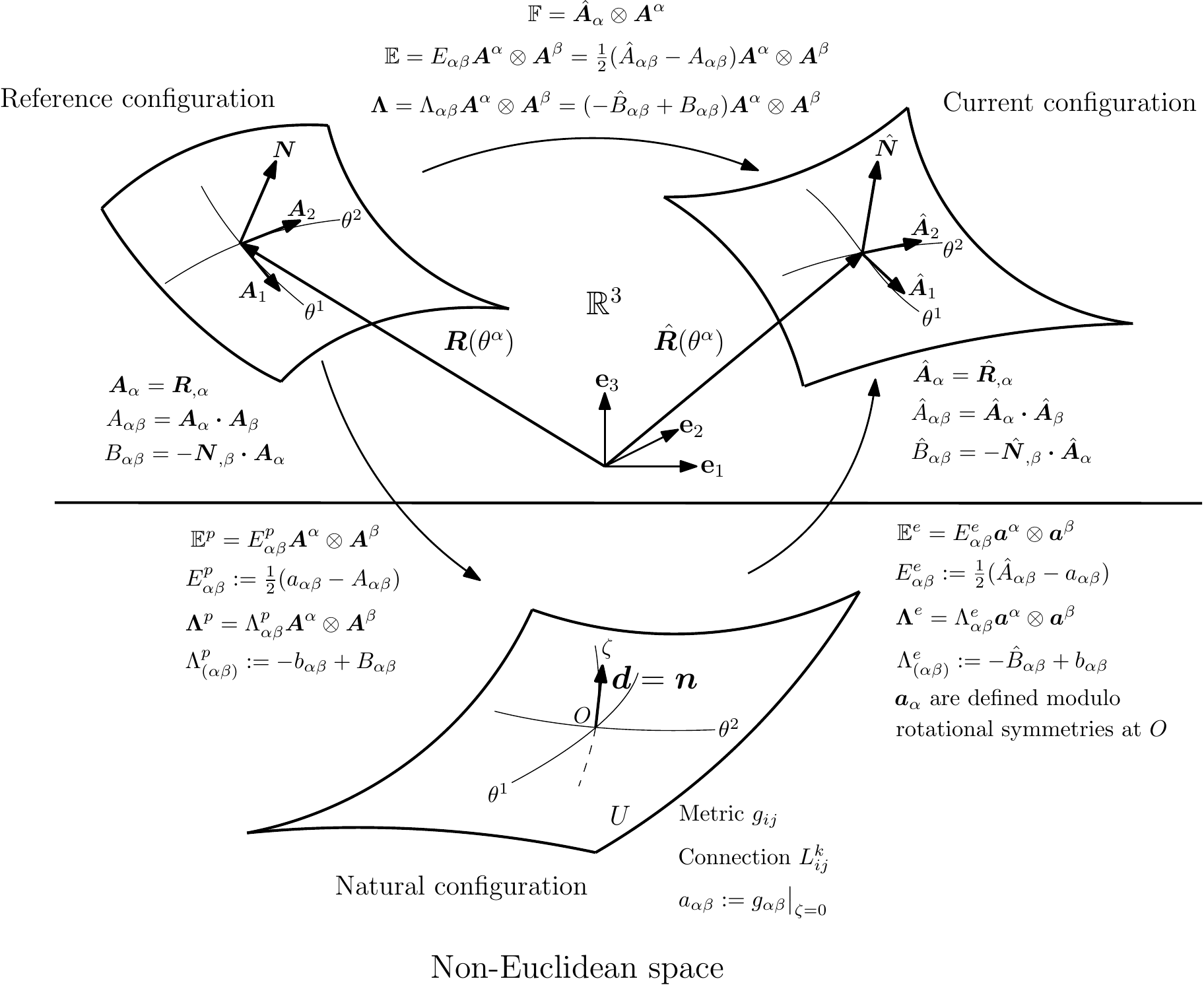}
 \caption{Kinematics of the elastic-plastic decomposition of the total deformation measures in Kirchhoff-Love shells. The only non-trivial disclinations are represented by $\Theta^3$, and hence we have a well defined normal over the surface in the natural configuration.}
 \label{decomposition}
\end{figure}

Following Section \ref{comp}, we consider the fixed reference configuration of the Kirchhoff-Love structured surface to be given by a local isometric embedding $\boldsymbol{R}:U\to\mathbb{R}^3$, where $U$ is a simply-connected open set of $\omega$; also, as before, we take $(\theta^\alpha,\zeta)$ as the adapted coordinates on $U$. The tangent spaces of $\boldsymbol{R}(U)$ are spanned by the natural base vectors $\boldsymbol{A}_\alpha=\boldsymbol{R}_{,\alpha}$. The first and second fundamental forms associated with the reference surface are given by $A_{\alpha\beta}=\boldsymbol{A}_\alpha\boldsymbol{\cdot}\boldsymbol{A}_\beta$ and $B_{\alpha\beta}=-\boldsymbol{N}_{,\alpha}\boldsymbol{\cdot}\boldsymbol{A}_\beta$, respectively, where $\boldsymbol{N}:={\boldsymbol{A}_1\times\boldsymbol{A}_2}/{|\boldsymbol{A}_1\times\boldsymbol{A}_2|}$ is the local unit normal. We will assume the adapted coordinates $(\theta^\alpha,\zeta)$ to be convected by deformation of the surface. The natural base vectors $\op{\boldsymbol{A}}_{\alpha}$ on the tangent spaces of the current configuration $\op{\boldsymbol{R}}:U\to\mathbb{R}^3$, a different isometric embedding of $U$, are given by $\op{\boldsymbol{A}}_\alpha=\op{\boldsymbol{R}}_{,\alpha}$. The first and second fundamental forms associated with the current configuration are $\hat{A}_{\alpha\beta}=\op{\boldsymbol{A}}_\alpha\boldsymbol{\cdot}\op{\boldsymbol{A}}_\beta$ and $\hat{B}_{\alpha\beta}=-\op{\boldsymbol{N}}_{,\alpha}\boldsymbol{\cdot}\op{\boldsymbol{A}}_\beta$, respectively, where $\hat{\boldsymbol{N}}:={\op{\boldsymbol{A}}_1\times\op{\boldsymbol{A}}_2}/{|\op{\boldsymbol{A}}_1\times\op{\boldsymbol{A}}_2|}$. The reference and the current configurations are shown in Figure \ref{decomposition}. The pairs $(A_{\alpha\beta},B_{\alpha\beta})$ and $(\hat{A}_{\alpha\beta},\hat{B}_{\alpha\beta})$ individually satisfy the Gauss and Codazzi-Mainardi equations owing to the existence of the isometric embeddings $\boldsymbol{R}$ and $\op{\boldsymbol{R}}$. The total surface deformation gradient tensor maps the tangent spaces of $\boldsymbol{R}(U)$ to those of $\op{\boldsymbol{R}}(U)$ and is given by $\mathbb{F}=\op{\boldsymbol{A}}_\alpha\otimes\boldsymbol{A}^\alpha$. The total surface strain and the total bending strain tensors, defined as
$\mathbb{E}=E_{\alpha\beta}\boldsymbol{A}^\alpha\otimes\boldsymbol{A}^\beta=\frac{1}{2}(\hat{A}_{\alpha\beta}-A_{\alpha\beta})\boldsymbol{A}^\alpha\otimes\boldsymbol{A}^\beta$, and $\boldsymbol{\Lambda}=\Lambda_{\alpha\beta}\boldsymbol{A}^\alpha\otimes\boldsymbol{A}^\beta=(-\hat{B}_{\alpha\beta}+B_{\alpha\beta})\boldsymbol{A}^\alpha\otimes\boldsymbol{A}^\beta$, respectively, measure the relative first and second fundamental forms of the current configuration with respect to the reference configuration of the structured surface. Other strain measures, introduced in the beginning of Section \ref{comp}, are identically zero under the Kirchhoff-Love constraint (which imposes the director field to coincide with the unit normal field).

The elastic surface strain tensor $\mathbb{E}^e$ and the elastic bending strain tensor $\boldsymbol{\Lambda}^e$ are defined as energetic dual of surface stress tensor and bending moment tensor, respectively, see Section \ref{matreskir}. On the other hand, the plastic surface strain and and the plastic bending strain tensors can be defined as $\mathbb{E}^p=E^p_{\alpha\beta}\boldsymbol{A}^\alpha\otimes\boldsymbol{A}^\beta$  and $\boldsymbol{\Lambda}^p=\Lambda^p_{\alpha\beta}\boldsymbol{A}^\alpha\otimes\boldsymbol{A}^\beta$, respectively, where $E^p_{\alpha\beta}:=\frac{1}{2}({a}_{\alpha\beta}-A_{\alpha\beta})$ and $\Lambda^p_{(\alpha\beta)}:=-b_{\alpha\beta}+B_{\alpha\beta}$. Also, $a_{\alpha\beta}$ and $b_{\alpha\beta}$ are, respectively, the first and (non-Riemannian) second fundamental form of $U$ in the material space $(\omega,\mathfrak{L},\boldsymbol{g})$; both of them are determined from the metric and the connection of the material space. Here, and henceforth, we will use superscripts $e$ and $p$ to denote elastic and plastic variables, respectively; in particular, they should not be read as indices.  
We note that it is only in the absence of disclinations and intrinsic orientational anomalies that there exist well-defined crystallographic vector fields $\boldsymbol{a}_\alpha:=\mathbb{F}^p\boldsymbol{A}_\alpha$ over the tangent spaces of local intermediate configuration $U$, where $\mathbb{F}^p$ is the (single-valued) plastic distortion field, see Remark \ref{onlydisl} for details. The elastic strain tensors can then be written as $\mathbb{E}^e=E^e_{\alpha\beta}\boldsymbol{a}^\alpha\otimes\boldsymbol{a}^\beta$ and $\boldsymbol{\Lambda}^e=\Lambda^e_{\alpha\beta}\boldsymbol{a}^\alpha\otimes\boldsymbol{a}^\beta$, where $E^e_{\alpha\beta}:=\frac{1}{2}(\hat{A}_{\alpha\beta}-a_{\alpha\beta})$ and $\Lambda^e_{(\alpha\beta)}:=-\hat{B}_{\alpha\beta}+b_{\alpha\beta}$. Here, $\boldsymbol{a}^\alpha :=a^{\alpha\beta}\boldsymbol{a}_\beta$ are the dual crystallographic vector fields on the material space, with $[a^{\alpha\beta}]:=[a_{\alpha\beta}]^{-1}$. The crystallographic vector fields $\boldsymbol{a}_\alpha$ are no longer well-defined in the presence of rotational anomalies. 

We now discuss the additive decomposition of total strain tensors in elastic and plastic parts. Introduce a small parameter $\epsilon:={h}/{R}$, where $h$ is the maximum thickness of the structured surface and $R$ is the minimum radius of curvature that $U$ can assume in all possible deformations. Let $\mathbb{E}$, $\mathbb{E}^p$, $\mathbb{E}^e$, and their first and second spatial derivatives be of order $O(\epsilon)$, and $\boldsymbol{\Lambda}$, $\boldsymbol{\Lambda}^p$, $\boldsymbol{\Lambda}^e$, and their first spatial derivatives be of order $O(\epsilon^{\frac{1}{2}})$. Here, following Landau's notation, for $\boldsymbol{f}:\mathbb{R}\to\mathbb{R}^k$, we write $\boldsymbol{f}(s)=O(s^r)$ if and only if there exist positive constants $M$ and $\delta$ such that $||\boldsymbol{f}(s)||_{\mathbb{R}^k}\le M|s|^r$ for all $|s|<\delta$, where $r$ is any real number. Following Naghdi and Vongsarnpigoon \cite{NaghdiVongsarnpigoon1983}, we emphasize that the resulting theory, where the surface and bending strains follow these separated orders, allows for {\it small surface strain accompanied by moderate rotation}. We postulate that the following decompositions for the total surface and bending strains hold:
\begin{subequations}
\begin{align}
 \mathbb{E}&=\mathbb{E}^e+\mathbb{E}^p~\text{of order}~O(\epsilon),\label{decomposition:strain-tensor}\\
 \boldsymbol{\Lambda}&=\boldsymbol{\Lambda}^e+\boldsymbol{\Lambda}^p~\text{of order}~O(\epsilon^{\frac{1}{2}}).\label{decomposition:curvature-tensor}
\end{align}
\end{subequations}
The first decomposition, which is of order $O(\epsilon)$, is the standard additive decomposition for small strains used commonly in small deformation theories. The second decomposition, of order $O(\epsilon^{\frac{1}{2}})$, is non-standard and captures moderately large rotation relative to in-surface stretching. It can be shown that the resulting deformation, which is of order $O(\epsilon^{\frac{1}{2}})$, is more general than infinitesimally small deformation theory of structural shells and, at the same time, stricter than a fully non-linear finite deformation theory \cite{NaghdiVongsarnpigoon1983}. Also, as $\boldsymbol{A}_\alpha\approx\op{\boldsymbol{A}}_\alpha\approx\boldsymbol{a}_\alpha$, upto order $O(\epsilon^{\frac{1}{2}})$ (cf.~\cite{NaghdiVongsarnpigoon1983}), we have
\begin{subequations}
\begin{align}
 E_{\alpha\beta}&={E}^e_{\alpha\beta}+{E}^p_{\alpha\beta}~\text{of order}~O(\epsilon),\label{decomposition:strain-components}\\
 {\Lambda}_{\alpha\beta}&={\Lambda}^e_{\alpha\beta}+{\Lambda}^p_{\alpha\beta}~\text{of order}~O(\epsilon^{\frac{1}{2}}).\label{decomposition:curvature-components}
\end{align}
\end{subequations}
Note that, as $\Lambda_{\alpha\beta}$ is symmetric, necessarily $\Lambda^p_{[\alpha\beta]}=-\Lambda^e_{[\alpha\beta]}$. These approximated decompositions with the mentioned order hold for sufficiently thin structured surfaces where the bending mode dominates over surface stretching for a given loading (internal or external).

\begin{rem}
{\rm According to a well-known result in differential geometry, whenever disclinations and metric anomalies are identically absent, the material connection and material metric can be written as
\begin{equation}
 L^k_{ij}=(F^{p^{-1}})^{kq} F^p_{qi,j}~\text{and}~\boldsymbol{g}=\mathbf{F}^{p^T}\mathbf{F}^p\label{puredislocation}
\end{equation}
in terms of an invertible second order tensor field $\mathbf{F}^p:=F^p_{ij}\boldsymbol{G}^i\otimes\boldsymbol{G}^j$, the plastic distortion field, defined over simply connected subsets $V\subset\mathcal{M}$  \cite{rcgupta16}. The well-defined surface plastic distortion tensor $\mathbb{F}^p:=\mathbf{F}^p\big|_{\zeta=0}$ maps the reference base vectors $\boldsymbol{A}_\alpha$ to the crystallographic base vectors $\boldsymbol{a}_\alpha:=\mathbb{F}^p\boldsymbol{A}_\alpha$ over the material space. The plastic rotation $\mathbf{Q}^p$ in the polar decomposition $\mathbb{F}^p=\mathbb{R}^p\mathbb{U}^p$, where $\mathbb{U}^p$ ($ = \left(\mathbb{E}^p\right)^{1/2}$) is the in-surface plastic stretch, is derivable by solving a first order PDE involving the in-surface plastic strain \cite{shield73}. Let $\boldsymbol{a}^\alpha:=a^{\alpha\beta}\boldsymbol{a}_\beta$ be the dual crystallographic base vectors, $\boldsymbol{n}:={\boldsymbol{a}_1\times\boldsymbol{a}_2}/{|\boldsymbol{a}_1\times\boldsymbol{a}_2|}$ the local unit normal field, $\boldsymbol{D}_\alpha:=(\Lambda^p_{\sigma\alpha}-B_{\sigma\alpha})\boldsymbol{a}^\sigma + \Lambda^p_\alpha \boldsymbol{n}$, $\boldsymbol{d}:=\Delta^p_\alpha \boldsymbol{a}^\alpha + (\Delta^p+1)\boldsymbol{n}$, $\boldsymbol{g}_\alpha(\theta^\alpha,\zeta):=\boldsymbol{a}_\alpha(\theta^\alpha)+\zeta\boldsymbol{D}_\alpha(\theta^\alpha)$, and $\boldsymbol{g}_3(\theta^\alpha,\zeta):=\boldsymbol{d}(\theta^\alpha)$. Clearly, $\mathbf{F}^p=\boldsymbol{g}_i\otimes\boldsymbol{G}^i$, as can be seen by comparing $\boldsymbol{g}$ obtained from \eqref{puredislocation}$_2$ with the expression \eqref{metric:defn} \cite{rcgupta15}. The dislocation densities $J^i$ and $\alpha^{\mu k}$ can then be directly read off from their definitions in terms of the torsion $T_{ij}{}^k(\theta^\alpha):=L^k_{[ij]}\big|_{\zeta=0}=\big((F^{p^{-1}})^{kq} F_{q[i,j]}\big)\big|_{\zeta=0}$. Therefore, in the absence of disclinations and metric anomalies, the dislocation density fields are expressible in terms of plastic distortion $\mathbb{F}^p$ and other strain fields. An analogous description of the above results can also be given in terms of the elastic distortion field.} \label{onlydisl}
\end{rem}

\subsection{Strain incompatibility relations for sufficiently thin Kirchhoff-Love shells with small strain accompanied by moderate rotation} \label{klincomp} 

We assume that disclination densities with components $\Theta^{\mu\nu}$, $\Theta^{\mu 3}$, and $\Theta^\mu$, and dislocation densities with components $\alpha^{\mu k}$, are identically zero. This is reasonable under the Kirchhoff-Love constraint, where direction field coincides with the orientation field, and if we restrict ourselves to sufficiently thin structured surfaces. The allowable defects are then the in-surface wedge disclinations $\Theta^3$, the surface screw and wedge dislocations $J^i$, and the metric anomalies $Q_{kij}$. For a non-trivial wedge disclination density $\Theta^3$, and with other disclinations absent, the rotational ambiguity in the crystallographic basis vector fields $\boldsymbol{a}_\alpha$ always falls within the rotational symmetry group of the base material at the respective points on $U$. In other words, the rotation part of the plastic distortion field (see Remark \ref{onlydisl}) is well-defined modulo the rotations from the symmetry group. As a result, under the considered assumption on the nature of allowable defects, the fields $\boldsymbol{a}_\alpha$ are well-defined with known rotational ambiguity. The normal $\boldsymbol{n}$ at each point in the natural configuration is then well-defined.

The Kirchhoff-Love constraint simplifies the formalism of Section \ref{strain}. We now revisit the strain incompatibility relations derived therein for the plastic strain tensor. The corresponding result for the elastic strain tensor is more involved since the reference configuration, used in case of plastic strains, has to be replaced with current configuration, which is itself unknown. The material metric has a simple block diagonal form, given in \eqref{metric:kl} but now in terms of the plastic strain components, which can be deduced from \eqref{metric:defn} using $\Lambda_\alpha^p=\Delta_\alpha^p=\Delta^p=0$. Also, from \eqref{comp4}, we can infer that $I_\alpha = 0$ and $J = - (1/2) \varepsilon^{\alpha \beta} \Lambda_{[\alpha \beta]}^p$. The local strain incompatibility relations \eqref{strain-incompatibility-relations1}-\eqref{strain-incompatibility-relations4}, under these considerations, and with only $\Theta^3$, $J^i$, and $Q_{kij}$ as non-trivial defect measures, are reduced to
\begin{eqnarray}
  K_{1212}-[b_{11}b_{22}-b^2_{12}]&=& a\Theta^3 -2\partial_{[1}W_{2]12} -2W_{[1|i 2|}\, W_{2]1}{}^i \label{strain-incompatibility-relations1-kl},\\
-\partial_2 b_{1 1}+\partial_1 b_{1 2}&=& -2\partial_{[1}W_{2]1 3} -2W_{[1|i 3|}\, W_{2]1}{}^i, \label{strain-incompatibility-relations2-kl}\\
-\partial_2 b_{2 1}+\partial_1 b_{2 2}&=&  -2\partial_{[1}W_{2]2 3} -2W_{[1|i 3|}\, W_{2]2}{}^i,~\text{and}\label{strain-incompatibility-relations3-kl}\\
a^{\alpha\beta}\varepsilon_{\alpha\rho}\varepsilon_{\beta\sigma} a^{-1} (\Lambda^p_{[12]})^2 &=& -\partial_{\rho} M_{3\sigma}{}^3 - (C_{\rho\alpha}{}^3+M_{\rho\alpha}{}^3)(C_{3\sigma}{}^\alpha+M_{3\sigma}{}^\alpha)\nonumber\\
&&-M_{\rho 3}{}^3 M_{\sigma 3}^3 + M_{3\alpha}{}^3 W_{\rho\sigma}^\alpha + M_{33}{}^3 W_{\rho\sigma}^3, \label{strain-incompatibility-relations4-kl}
\end{eqnarray}
where, recall that, $W_{ij}{}^k$ are components of a tensor defined as a sum of the contortion and non-metricity tensors, see \eqref{con:11a}, as $W_{ij}{}^k=C_{ij}{}^k+M_{ij}{}^k$; the purely covariant components are $W_{ijk} = g_{mk}|_{\zeta=0} W_{ij}{}^m$. For $\alpha^{\mu k} = 0$ and metric given by \eqref{metric:kl}, the components of the contortion tensor, defined in \eqref{con:11b}, take a simple form:  
\begin{subequations}
\begin{align}
& C_{3\beta}{}^3 = C_{\beta 3}{}^3= C_{3 3}{}^i = C_{3\beta 3} = C_{\beta 3 3}= C_{3 3 i} = 0,~C_{3\beta}{}^\alpha = C_{\beta 3}{}^\alpha = a^{\alpha\nu}\varepsilon_{\nu\beta}J^3,\\
& C_{3\beta\alpha} = C_{\beta 3\alpha} = C_{\alpha\beta}{}^3 =C_{\alpha\beta 3} = \varepsilon_{\alpha\beta}J^3,\\
&C_{\alpha\beta\mu} = J^\sigma\big( a_{\sigma\beta} \varepsilon_{\mu\alpha}  + a_{\sigma\alpha} \varepsilon_{\mu\beta} + a_{\sigma\mu} \varepsilon_{\alpha\beta}\big), ~\text{and}~C_{\alpha\beta}{}^\mu = a^{\mu\nu} C_{\alpha\beta\nu}.
\end{align}
\end{subequations}
Here, $\varepsilon_{\alpha\beta}:=a^{\frac{1}{2}}e_{\alpha\beta}$. On the other hand, the tensor associated with non-metricity has components
\begin{subequations}
\begin{align}
 & M_{33}{}^3 = M_{333} = \frac{1}{2}Q_{333},~M_{33\alpha} = \frac{1}{2}(2Q_{3\alpha 3}-Q_{\alpha 33}),~M_{33}{}^\alpha = a^{\alpha\beta} M_{33\beta},\\
 & M_{3\alpha}{}^3 = M_{\alpha 3}{}^3 =M_{3\alpha 3} = M_{\alpha 33} =  \frac{1}{2} Q_{\alpha 33},\\
&  M_{3\alpha}{}^{\beta} = M_{\alpha 3}{}^\beta = \frac{1}{2} a^{\beta\nu} (Q_{3\nu\alpha}-Q_{\nu\alpha 3} + Q_{\alpha 3 \nu}),\\
 & M_{\alpha\beta}{}^3 = M_{\alpha\beta 3} = \frac{1}{2}(Q_{\alpha 3\beta}-Q_{3\beta\alpha} + Q_{\beta\alpha 3})\\
&  M_{\alpha\beta\mu} =  \frac{1}{2}(Q_{\alpha\mu\beta}-Q_{\mu\beta\alpha}+ Q_{\beta\alpha\mu}),~\text{and}~M_{\alpha\beta}{}^\mu = a^{\mu\nu}M_{\alpha\beta\nu}.
 \end{align}
\end{subequations}
These are to be substituted into \eqref{strain-incompatibility-relations1-kl}-\eqref{strain-incompatibility-relations4-kl} to obtain the incompatibility relations in terms of defect densities. As discussed above, the only non-trivial disclination density is $\Theta^3$, and the only non-trivial dislocation densities are $J^i$; there are no restrictions on the non-metricity densities. Note that the in-surface metric anomalies do not appear in \eqref{strain-incompatibility-relations4-kl}. Therefore, in the absence of out-of-surface metric anomalies, the right side of \eqref{strain-incompatibility-relations4-kl} reduces to $a^{\alpha\beta}\varepsilon_{\alpha\rho}\varepsilon_{\beta\sigma} (J^3)^2$ which implies that $|\Lambda^p_{[12]}|=a^{\frac{1}{2}}|J^3|$. The skewness of plastic bending strain is then completely characterized in terms of in-surface screw dislocations.
Finally, whenever both disclinations and metric anomalies are absent,  \eqref{strain-incompatibility-relations1-kl}-\eqref{strain-incompatibility-relations4-kl} reduce to
\begin{eqnarray}
  K_{1212}-[b_{11}b_{22}-b^2_{12}]&=&2\sqrt{a}\, a_{\sigma[1} \partial_{2]}J^\sigma -2C_{[1|\mu 2|}\, C_{2]1}{}^\mu -a(J^3)^2 \label{strain-incompatibility-relations1-kl-1},\\
-\partial_2 b_{1 1}+\partial_1 b_{1 2}&=& \partial_1(\sqrt{a}J^3) + a J^2 J^3, \label{strain-incompatibility-relations2-kl-2}\\
-\partial_2 b_{2 1}+\partial_1 b_{2 2}&=&  \partial_2(\sqrt{a}J^3) + a J^1 J^3~\text{and}\label{strain-incompatibility-relations3-kl-3}\\
a^{\alpha\beta}\varepsilon_{\alpha\rho}\varepsilon_{\beta\sigma} a^{-1} (\Lambda^p_{[12]})^2 &=& a^{\alpha\beta}\varepsilon_{\alpha\rho}\varepsilon_{\beta\sigma} (J^3)^2.\label{strain-incompatibility-relations3-kl-4}
\end{eqnarray}

We will next reduce the incompatibility relations \eqref{strain-incompatibility-relations1-kl}-\eqref{strain-incompatibility-relations4-kl} under further kinematical assumptions. In particular, we restrict ourselves to plate like structures. Accordingly, we take the reference surface to be flat, i.e. $B_{\alpha\beta}=0$, and identify the curvilinear coordinates $(\theta^1,\theta^2)$ with the Cartesian coordinates, i.e. $A_{\alpha\beta}=\delta_{\alpha\beta}$. The covariant derivatives then get replaced by ordinary partial derivatives.
 
{\it Pure bending of elastic plates}:
The deformation of a perfectly flexible flat surface, e.g., a thin sheet of paper, involves large bending strain with vanishingly small in-surface stretching. Therefore, we have $a_{\alpha\beta}=A_{\alpha\beta}=a^{\alpha\beta}=A^{\alpha\beta}=\delta_{\alpha\beta}$, and hence $a=A=1$, $\Gamma_{\alpha\beta}^\mu = 0$, $K = 0$. The local strain incompatibility equations \eqref{strain-incompatibility-relations1-kl}-\eqref{strain-incompatibility-relations4-kl}, under these simplifications, take the form 
\begin{eqnarray}
 (\Lambda^p_{(12)})^2-\Lambda^p_{11}\Lambda^p_{22} &=& \Theta^3 -2\partial_{[1}W_{2]12} -2W_{[1|i 2|}\, W_{2]1}{}^i,\\
\Lambda^p_{1 1,2}- \Lambda^p_{(1 2),1}&=& -2W_{[2|1 3|,1]} -2W_{[1|i 3|}\, W_{2]1}{}^i, \\
 \Lambda^p_{(12),2}- \Lambda^p_{2 2,1}&=&  -2W_{[2|2 3|,1]} -2W_{[1|i 3|}\, W_{2]2}{}^i,~\text{and}\\
e_{\alpha\rho}e_{\alpha\sigma} (\Lambda^p_{[12]})^2 &=& -M_{3\sigma}{}^{3}{}_{,\rho} - (C_{\rho\alpha}{}^3+M_{\rho\alpha}{}^3)(C_{3\sigma}{}^\alpha+M_{3\sigma}{}^\alpha)\nonumber\\
&&-M_{\rho 3}{}^3 M_{\sigma 3}^3 + M_{3\alpha}{}^3 W_{\rho\sigma}^\alpha + M_{33}{}^3 W_{\rho\sigma}^3.
\end{eqnarray}
These provide a complete system of partial differential algebraic equations for the plastic bending strain ${\boldsymbol{\Lambda}}^p$ with various defect densities as source terms. In absence of disclinations and metric anomalies, these further reduce down to (compare with \eqref{strain-incompatibility-relations1-kl-1}-\eqref{strain-incompatibility-relations3-kl-4})
\begin{eqnarray}
 (\Lambda^p_{(12)})^2-\Lambda^p_{11}\Lambda^p_{22} &=& J^1_{,2} -J^2_{,1} -(J^3)^2,\\
\Lambda^p_{1 1,2}- \Lambda^p_{(1 2),1}&=& J^3_{,1} +  J^2 J^3, \\
 \Lambda^p_{(12 ),2}- \Lambda^p_{2 2,1}&=&  J^3_{,2} +  J^1 J^3,~\text{and}\\
|\Lambda^p_{[12]}|=|J^3|.
\end{eqnarray}
These relations have been earlier obtained by Derezin \cite{derezin}. On the other hand, if dislocations and metric anomalies are both absent then, clearly, the plastic bending strain is symmetric and also curl free. Therefore, we can write $\Lambda^p_{\alpha\beta}=w^p_{,\alpha\beta}$ for some scalar field $w^p$ defined over simply connected open sets $U\subset\omega$. The plastic Gaussian curvature of the material space $K^p = (\Lambda^p_{12})^2-\Lambda^p_{11}\Lambda^p_{22}$  is then given by the negative of the wedge disclination density $\Theta^3$; this is the only non-trivial incompatibility equation in this case.

{\it Combined bending and stretching of elastic plates}:
We assume that the wedge disclination density $\Theta^3$, the in-surface dislocation densities $J^i$, and the density of in-surface metric anomalies $Q_{\mu\alpha\beta}$, along with their spatial derivatives upto first order, to be $O(\epsilon)$. We also assume that the density of metric anomalies characterized by $Q_{kij}$, with at least one of the indices $k$, $i$ or $j$ taking the value 3, along with their first spatial derivatives, to be $O(\epsilon^\frac{1}{2})$. This is in accordance with our assumption on the kinematics of the in-surface and out-of-surface deformation. Again, we identify $(\theta^1,\theta^2)$ with the Cartesian coordinates on $\omega$; also $A_{\alpha\beta}=A^{\alpha\beta}=\delta_{\alpha\beta}$, $B_{\alpha\beta}=0$, and $a=1+O(\epsilon)$. A straightforward calculation shows that, upto $O(\epsilon)$,
\begin{equation}
 \Gamma^\tau_{\alpha\beta}=\frac{1}{2}a^{\tau\sigma}(a_{\sigma\beta,\alpha}+a_{\sigma\alpha,\beta}-a_{\alpha\beta,\sigma})\approx A^{\tau\sigma}(E^p_{\sigma\beta,\alpha}+E^p_{\sigma\alpha,\beta}-E^p_{\alpha\beta,\sigma}).
 \label{levicivitaorder}
\end{equation}
Consequently, the local strain incompatibility relations \eqref{strain-incompatibility-relations1-kl}-\eqref{strain-incompatibility-relations4-kl} are reduced to
\begin{subequations}
\begin{align}
  -2E^p_{12,12}+E^p_{11,22}+E^p_{22,11}+(\Lambda^p_{(12)})^2-\Lambda^p_{11}\Lambda^p_{22}&=\Theta^3 -2W_{[2|12|,1]} -2W_{[1|3 2|}\, W_{2]1}{}^3,\\
  \Lambda^p_{1 1,2}-\Lambda^p_{(1 2),1} =-2W_{[2|1 3|,1]} &-2W_{[1|3 3|}\, W_{2]1}{}^3,\\
  \Lambda^p_{(12),2}-\Lambda^p_{2 2,1} =-2W_{[2|2 3|,1]} &-2W_{[1|3 3|}\, W_{2]2}{}^3,~\text{and}\\
  e_{\alpha\rho}e_{\alpha\sigma} (\Lambda^p_{[12]})^2 = - M_{3\sigma}{}^{3}{}_{,\rho} - (C_{\rho\alpha}{}^3+M_{\rho\alpha}{}^3)(C_{3\sigma}{}^\alpha+M_{3\sigma}{}^\alpha)&-M_{\rho 3}{}^3 M_{\sigma 3}^3 + M_{33}{}^3 W_{\rho\sigma}^3,
\end{align}
\label{sslr}%
\end{subequations}
where, in the first and the fourth equation, we have retained terms upto $O(\epsilon)$, and in the second and the third equation, we have retained terms upto $O(\epsilon^{\frac{1}{2}})$ (note that the first term on the right hand side of the last equation is $O(\epsilon^{\frac{1}{2}})$). In the following paragraph, we restrict these equations for a distribution of metric anomalies, while ignoring both disclinations and dislocations.

{\it Pure metric anomalies}:
In the absence of disclinations we can represent metric anomalies in terms of the quasi-plastic strain field $\tilde q_{ij}$ as $Q_{kij}(\theta^\alpha)=-2 \tilde q_{ij;k}\big|_{\zeta=0}$. We assume the following form of $\tilde q_{ij}(\theta^\alpha,\zeta)$:
\begin{equation}
 \tilde q_{\alpha\beta}=q^0_{\alpha\beta}-2\zeta q^\prime_{\alpha\beta}+\zeta^2 \delta^{\mu\nu} q^\prime_{\alpha\mu}q^\prime_{\nu\beta},~\tilde q_{\alpha 3}=\tilde q_{3\alpha}=0,~\text{and}~\tilde q_{33}=1,
\end{equation}
where the symmetric functions $q^0_{\alpha\beta}(\theta^\alpha)$ and $q^\prime_{\alpha\beta}(\theta^\alpha)$ are $O(\epsilon)$ and $O(\epsilon^{\frac{1}{2}})$, respectively. The above is motivated by the form of the material metric for Kirchhoff-Love plates with small surface strain accompanied by moderate rotations, as can be deduced from \eqref{metric:kl}. We obtain $Q_{\mu\alpha\beta}=-2 q^0_{\alpha\beta,\mu}$, $Q_{3\alpha\beta}=4q^\prime_{\alpha\beta}$, and $Q_{k33}=Q_{k\alpha 3}=Q_{k3\alpha}=0$. Accordingly, the functions $q^0_{\alpha\beta}$ measure in-surface metric anomalies, e.g., surface growth, whereas $q^\prime_{\alpha\beta}$ measure the tangential differential surface growth along the thickness direction. Under these assumptions, the strain incompatibility relations \eqref{sslr} are reduced to
\begin{subequations}
\begin{align}
  -2E^p_{12,12}+E^p_{11,22}+E^p_{22,11}+(\Lambda^p_{(12)})^2-\Lambda^p_{11}\Lambda^p_{22}&= (-2q^0_{12,12}+q^0_{11,22}+q^0_{22,11})+4[(q^\prime_{12})^2-q^\prime_{11} q^\prime_{22}],\\
  \Lambda^p_{1 1,2}-\Lambda^p_{(1 2),1} &=2 (q^\prime_{12,1}- q^\prime_{11,2}),\\
  \Lambda^p_{(12),2}-\Lambda^p_{2 2,1} &=2 (q^\prime_{22,1}- q^\prime_{21,2}),~\text{and}\\
  e_{\alpha\rho}e_{\alpha\sigma} (\Lambda^p_{[12]})^2 &=  4q^\prime_{\rho\alpha} q^\prime_{\alpha\sigma},
\end{align}
\label{sslr-pure-metric-anomalies}%
\end{subequations}
where the second and third relations are approximated upto $O(\epsilon^{\frac{1}{2}})$. With the first three relations, we can directly identify the plastic strain components with quasi-plastic strains: $\Lambda^p_{(\alpha\beta)}=-2 q^\prime_{\alpha\beta}$ and $E^p_{\alpha\beta}=q^0_{\alpha\beta}$. The last relation, for $(\rho,\sigma)=(1,1)$, $(2,2)$, and $(1,2)$, leads to $(\Lambda^p_{[12]})^2 =  4((q^\prime_{11})^2 +  (q^\prime_{12})^2)$, $(\Lambda^p_{[12]})^2 =  4((q^\prime_{22})^2 +  (q^\prime_{12})^2)$ and $q^\prime_{12}(q^\prime_{11} +  q^\prime_{22}) = 0$,
respectively. The former two of these imply that $q^\prime_{11}=\pm q^\prime_{22}$. According to the latter one, when $q^\prime_{11}=- q^\prime_{22}\ne 0$, $q^\prime_{12}$ may assume any non-zero value, e.g., in anisotropic tangential differential growth along thickness. For $q^\prime_{11}=q^\prime_{22}\ne 0$, $q^\prime_{12}=0$, a case of isotropic tangential differential growth along thickness, we have $|\Lambda^p_{[12]}|= 2|q^\prime_{11}|= 2|q^\prime_{22}|$. Finally, if $q^\prime_{11}=q^\prime_{22}= 0$, $q^\prime_{12}$ may assume any non-zero value, representing the tangential differential growth of shear type along the thickness, such that $|\Lambda^p_{[12]}|= 2|q^\prime_{12}|$.

As an application we look for conditions on temperature field which would yield compatible thermal strain in thin plates. For isotropic thermal deformation in thin elastic plates, $\tilde{q}_{\mu\nu}(\theta^\alpha,\zeta)={q}_{\mu\nu}(\theta^\alpha)=\alpha T(\theta^\alpha) \delta_{\mu\nu}$, where $\alpha$ is the uniform thermal expansion coefficient and $T$ is the change in temperature. Clearly, from \eqref{sslr-pure-metric-anomalies}, the temperature distribution $T(\theta^\alpha)$ that gives rise to compatible strain fields satisfies the 2-dimensional Laplace equation $T_{,\alpha\alpha}  =0$ \cite[Ch. 14]{barber}.

\subsection{Material response and equilibrium equations for sufficiently thin Kirchhoff-Love shells}\label{matreskir}

We assume that the structured surface is materially uniform, simple, and hyperelastic. Material uniformity requires that there exist locally undistorted states with respect to which the constitutive response function (e.g., stress-strain relation) is independent of the material points on the surface; the stress-free natural configuration provides such an undistorted state. The response of a simple material is local in nature in the sense that the material response at a point depends only on the local state of deformation at that point. These two hypotheses, combined with the principle of material frame indifference, require that the isothermal material response of materially uniform and simple structured surfaces is expressible in terms of the elastic strain field with respect to the natural state of the surface. In particular, a hyperelastic material response is governed by a single scalar energy density function $\psi(E^e_{\alpha\beta},\Lambda^e_{\alpha\beta},\Lambda^e_\alpha,\Delta^e_\alpha,\Delta^e)$ per unit area of the surface in natural configuration. The energy density function will be further required to satisfy appropriate material symmetry restrictions \cite{ericksen70, wangshell, eps2, edeleon, steigmann99, steigmann99a}. Such 2-dimensional strain energy density functions have been established by techniques such as thickness-wise integration of a 3-dimensional material response \cite{steigmann2013b, steigmann2013a}, gamma-convergence \cite{frieseckeetal2003, frieseckeetal2006}, asymptotic expansion \cite{ciar3}, etc. 

The equilibrium equations of a Kirchhoff-Love shell, with strain energy density function $\psi(E^e_{\alpha\beta},\Lambda^e_{\alpha\beta})$, are \cite{steigmann14}
\begin{subequations}
 \begin{align}
  \hat{\partial}_\alpha (\sigma^{\mu\alpha}+M^{\beta\alpha}\Lambda^\mu_\beta)+\hat{\partial}_\beta M^{\beta\alpha}\,\Lambda^\mu_\alpha &= 0~\text{and}\\  
  (\sigma^{\beta\alpha}+ M^{\mu\alpha}\Lambda^\beta_\mu)\Lambda_{\beta\alpha}+\hat{\partial}_{\alpha\beta} M^{\beta\alpha} &= 0,
 \end{align}
 \label{eqb:Kirchhoff-love}%
\end{subequations}
where
\begin{equation}
 \sigma^{\beta\alpha}=\frac{1}{2}\sqrt{\frac{a}{\hat{A}}}\bigg(\frac{\partial\psi}{\partial E^e_{\alpha\beta}}+\frac{\partial\psi}{\partial E^e_{\beta\alpha}}\bigg)~~\text{and}~~
M^{\beta\alpha}=\frac{1}{2}\sqrt{\frac{a}{\hat{A}}}\bigg(\frac{\partial\psi}{\partial \Lambda^e_{\alpha\beta}}+\frac{\partial\psi}{\partial \Lambda^e_{\beta\alpha}}\bigg), \label{constitutiverel}
\end{equation}
respectively, are the tangential surface stress and bending moment measures. Here, $\hat{\partial}$ denotes the covariant derivative on the current configuration and $\hat{A}:=\mbox{det}[\hat{A}_{\alpha\beta}]$.
For a sufficiently thin isotropic Kirchhoff-Love shell, dimensional analysis and representation theorems can be used to express $\psi(E^e_{\alpha\beta},\Lambda^e_{\alpha\beta})$ as \cite{steigmann99}
\begin{equation}
 \psi(E^e_{\alpha\beta},\Lambda^e_{\alpha\beta})=E\bigg(C(i_1,i_2)+h^2\sum^7_{n=3}i_n D_n(i_1,i_2)\bigg), \label{energy-stkls}
\end{equation}
where $E$ is the 2-dimensional Young's modulus of the shell material and $i_1:=E^e_{\alpha\beta}a^{\alpha\beta}$, $i_2:=E^e_{\alpha\beta}E^e_{\mu\nu}a^{\alpha\mu}a^{\beta\nu}$, $i_3:=(\Lambda^e_{\alpha\beta}a^{\alpha\beta})^2$, $i_4:=\Lambda^e_{\alpha\beta}\Lambda^e_{\mu\nu}a^{\alpha\mu}a^{\beta\nu}$, $i_5:=(E^e_{\alpha\beta}\Lambda^e_{\mu\nu}a^{\alpha\mu}a^{\beta\nu})^2$, $i_6:=a^{-1}(\varepsilon^{\alpha\gamma}\Lambda^e_{\alpha\beta}E^e_{\mu\nu}a_{\sigma\gamma}a^{\sigma\mu}a^{\beta\nu})^2$, and $i_7:=E^e_{\alpha\beta}\Lambda^e_{\rho\sigma}\Lambda^e_{\mu\nu}a^{\rho\sigma}a^{\alpha\mu}a^{\beta\nu}$; $C$ and $D_n$ are dimensionless functions. We will now summarize certain special forms of the above relations. Note, that only the symmetric part $\Lambda^e_{(\alpha\beta)}$ of the elastic bending strain contributes to the constitutive response. The skew part, as a consequence of the additive decomposition, is determined by the skew part of the plastic bending strain, $\Lambda^e_{[\alpha\beta]}=-\Lambda^p_{[\alpha\beta]}$.

{\it Pure bending of thin elastic isotropic shells}:
The surface stress components $\sigma^{\alpha\beta}$ in the equilibrium equations \eqref{eqb:Kirchhoff-love} of a Kirchhoff-Love shell undergoing pure bending are to be interpreted as Lagrange multipliers $\bar\sigma^{\alpha\beta}(\theta^\alpha)$ associated with the deformation constraint $E^e_{\alpha\beta}=0$; these are determined \textit{a posteriori} after solving the complete boundary-value-problem \cite{steigmann14}. The bending moment components $M_{\alpha\beta}$, with respect to an adapted Cartesian coordinate system $\theta^i$, are determined from $\eqref{constitutiverel}_2$ and \eqref{energy-stkls},
\begin{equation}
 M^{\alpha\beta}=D\bigg(\nu \Lambda^e_{\mu\mu}A^{\alpha\beta}+(1-\nu) \Lambda^e_{(\mu\nu)}A^{\mu\alpha}A^{\nu\beta}\bigg),
\end{equation}
where $\nu$ is the Poisson's ratio of the shell material and $D$ is the bending rigidity.

{\it Combined bending and stretching of thin elastic isotropic shells}: 
Under the assumption of small elastic surface strain and moderate elastic bending strain, i.e., $E^e_{\alpha\beta}=O(\epsilon)$ and $\Lambda^e_{\alpha\beta}=O(\epsilon^{\frac{1}{2}})$, we have $i_1=O(\epsilon)$, $i_2=O(\epsilon^2)$, $i_3=O(\epsilon)$, $i_4=O(\epsilon)$, $i_5=O(\epsilon^{2.25})$, $i_6=O(\epsilon^{2.25})$, and $i_7=O(\epsilon^2)$. Neglecting the coupling term $i_7$, the 2-dimensional linear stress-strain and bending moment-bending strain relations, upto $O(\epsilon)$ and $O(\epsilon^{\frac{1}{2}})$, respectively, are given by
\begin{equation}
\sigma^{\alpha\beta}=\frac{E}{(1-\nu^2)}\bigg(\nu E^e_{\mu\mu} A^{\alpha\beta}+(1-\nu)E^e_{\mu\nu}A^{\mu\alpha}A^{\nu\beta}\bigg)~~\text{and}~~ M^{\alpha\beta}=D\bigg(\nu \Lambda^e_{\mu\mu}A^{\alpha\beta}+(1-\nu)\Lambda^e_{\mu\nu}A^{\mu\alpha}A^{\nu\beta}\bigg).
\label{Kirchhoff-Love-constitutive}
\end{equation}

{\it Thin isotropic incompressible fluid films and shape equation}: We consider a  thin isotropic incompressible fluid film with strain energy density of  the form \cite{steigmannetal2003}
\begin{equation}
\psi(E^e_{\alpha\beta},\Lambda^e_{\alpha\beta})=W(H^e,K^e;\theta^\alpha)-\mu\left(\sqrt{{\hat A}/{a}}-1\right), 
\end{equation}
where $H^e$ and $K^e$ are the elastic mean and Gaussian curvature of the material space, $H^e:=\frac{1}{2} A^{\mu\nu}\Lambda^e_{\mu\nu}$, $K^e:=\Lambda^e_{11}\Lambda^e_{22}-(\Lambda^e_{(12)})^2$, and $\mu(\theta^\alpha)$ is a constitutively undetermined Lagrange multiplier corresponding to the incompressibility constraint. The explicit dependence of constitutive function $W$ on $\theta^\alpha$ in fact represents its relation to the local reference neighbourhoods through the reference fundamental forms $A_{\alpha\beta}$ and $B_{\alpha\beta}$. The equilibrium equations with zero body force and couple are \cite{steigmannetal2003}
\begin{equation}
\hat{\Delta}(\frac{1}{2}W_{H^e}) + \hat{\partial}_{\alpha\beta}(W_{K^e}){\tilde{B}}^{\alpha\beta} + W_{H^e}(2{H}^2-{K}) + 2{K}{H}W_{K^e} - 2{H}(W-\mu) = 0 \label{shpfm}
\end{equation}
and $\mu_{,\alpha}=W_{,\alpha}$ (for fixed $H^e$ and $K^e$).
Here, $\hat{\Delta}$ denotes the Laplace-Beltrami operator on the current configuration, $\hat{\Delta}(\cdot):=\hat{\partial}_{\alpha\beta}(\cdot){\hat{A}}^{\alpha\beta}$; $W_{H^e}$ and $W_{K^e}$ denote partial derivatives of $W$ with respect to the arguments in the subscripts; ${H}$ and ${K}$ are, respectively, the mean and Gaussian curvature of the total bending strain $\Lambda_{\alpha\beta}$, $H:=\frac{1}{2}\hat{A}^{\alpha\beta}w_{,\alpha\beta}$, $K:=w_{,11}w_{,22}-(w_{,12})^2$; and ${\tilde{B}}^{\alpha\beta}$ is the cofactor of $\hat{B}^{\alpha\beta}$. Equation \eqref{shpfm} is known as the shape equation of the fluid film, which for the special Helfrich energy $W(H^e,K^e)=k (H^e)^2$ yeilds \cite{agrawal2008}
\begin{equation}
 k \hat{\Delta} (H - H^p) + 2k(H - H^p)(2H^2 - K ) - 2k H (H - H^p)^2 + 2\mu H = 0.
\end{equation}
Here, $k$ is a material constant. We have substituted $(H-H^p)$ for $H^e$, where $H^p$ is the plastic mean curvature of the material space, $H^p:=\frac{1}{2} A^{\mu\nu}\Lambda^p_{\mu\nu}$. $H^p$, also understood as the spontaneous curvature in the mechanics of non-uniform fluid films \cite{agrawal2008}. The solution $H=H^p$ and $\mu=0$ of the above equation, implying a global minimum to the total energy, is ruled out in presence of material defects since $H^p$ might not correspond to any realizable surface isometrically embedded in $\mathbb{R}^3$. The parameter $\mu$ is to be determined from the boundary data after the complete boundary-value-problem has been solved. 
The shape equation can be linearized by retaining terms only upto $O(\epsilon^{\frac{1}{2}})$ as $({k}/{2}) w_{,\alpha\alpha\beta\beta} +  \mu w_{,\alpha\alpha} = k H^p_{,\alpha\alpha}$,
where the plastic mean curvature field can be written in terms of various defect densities using the solution $\Lambda^p_{\alpha\beta}$ from the incompatibility relations for pure bending.

\subsection{F{\"o}ppl-von K{\'a}rm{\'a}n equations for continuously defective thin elastic isotropic plates}

We identify $(\theta^1,\theta^2,\zeta)$ with the Cartesian coordinate system on the planar reference configuration and use $\mathbf{e}_i$ to denote the standard basis. As before, $A_{\alpha\beta}=\delta_{\alpha\beta}$ and $B_{\alpha\beta}=0$. The displacement vector field with respect to the reference configuration can be written as $\boldsymbol{u}(\theta^\alpha)=\hat{\boldsymbol{R}}-\boldsymbol{R}=u^\alpha(\theta^\alpha)\mathbf{e}_\alpha+w(\theta^\alpha)\mathbf{e}_3$, where $u^\alpha=u_\alpha$ are the planar displacement components and $w$ is the vertical displacement. Under the assumption of small strain and moderate rotation, it can be shown that the components of the displacement field satisfy $u_{\alpha,\beta}=O(\epsilon)$ and $w_{,\alpha},w_{,\alpha\beta}=O(\epsilon^{\frac{1}{2}})$ \cite{NaghdiVongsarnpigoon1983}. As a result, we can write
\begin{subequations}
 \begin{align}
   E_{\alpha\beta}&= u_{(\alpha,\beta)} + \frac{1}{2} w_{,\alpha}w_{,\beta}~\text{upto}~O(\epsilon)~\text{and}\\
\Lambda_{\alpha\beta}&=w_{,\alpha\beta}~\text{upto}~O(\epsilon^{\frac{1}{2}}).
 \end{align}
\end{subequations}

In the classical F{\"o}ppl-von K{\'a}rm{\'a}n theory for thin elastic plates \cite{mansfield}, the linearized version of the Kirchhoff-Love equilibrium equations \eqref{eqb:Kirchhoff-love}, retained upto $O(\epsilon)$, are posed as the localized in-plane and vertical force balances:
\begin{equation}
  \sigma^{\alpha\beta}_{,\beta} = 0~\text{and}~ 
  \sigma^{\alpha\beta}\Lambda_{\alpha\beta}+M^{\alpha\beta}_{,\alpha\beta} = 0.
 \label{eqb:foppl-vonkarman}
\end{equation}
The first equation is identically satisfied when the stress components $\sigma^{\alpha\beta}=\sigma_{\alpha\beta}$ in the Cartesian frame are expressed in terms of the 2-dimensional Airy stress function $\Phi(\theta^\alpha)$ as $\sigma_{11}=\Phi_{,22}$, $\sigma_{22}=\Phi_{,11}$, and $\sigma_{12}=-\Phi_{,12}$.
The second equation, after plugging in these expressions, the linear elastic bending constitutive relation $M_{\alpha\beta}=D(\nu \Lambda^e_{\mu\mu}\delta_{\alpha\beta}+(1-\nu)\Lambda^e_{(\alpha\beta)})$, and the decomposition \eqref{decomposition:curvature-components}, can be shown to reduce to
\begin{equation}
 D w_{,\alpha\alpha\beta\beta} +[w,\Phi] = D\Omega_p,
 \label{foppl-vonkarman-1}
\end{equation}
where $[A,B] := A_{,11}B_{,22} + A_{,22}B_{,11} - 2 A_{,12}B_{,12}$ and $\Omega_p:=\nu\Lambda^p_{\alpha\alpha,\beta\beta} + (1-\nu)\Lambda^p_{(\alpha\beta),\alpha\beta}$. 
On the other hand, the compatibility relations for the total strain require
 $-2E_{12,12}+E_{11,22}+E_{22,11}+(w_{,12})^2-w_{,11} w_{,22}=0$,
which, after plugging in the additive surface strain decomposition \eqref{decomposition:strain-components}, along with the constitutive relation $\sigma_{\alpha\beta}=\left({E}/{(1-\nu^2)}\right)(\nu E^e_{\mu\mu}\delta_{\alpha\beta}+(1-\nu)E^e_{\alpha\beta})$, yields
\begin{equation}
 \frac{1}{E}\Phi_{,\alpha\alpha\beta\beta} -\frac{1}{2}[w,w]=-\lambda_p,
 \label{foppl-vonkarman-2}
\end{equation}
where $\lambda_p:= -2E^p_{12,12}+E^p_{11,22}+E^p_{22,11}$.
Equations \eqref{foppl-vonkarman-1} and \eqref{foppl-vonkarman-2} are the F{\"o}ppl-von K{\'a}rm{\'a}n equations of a continuously defective elastic plate, cf. \cite{mansfield, lewicka-mahadevan, liang-mahadevan09}. Along with specified traction and vertical displacement over complementary parts of $\partial\omega$, these equations constitute the boundary-value-problem for the deformed shape function $w(\theta^\alpha)$ and airy stress function $\Phi(\theta^\alpha)$, with known $\Omega_p$ and $\lambda_p$. Considering only metric anomalies, in absence of dislocations and disclinations, $\Omega_p=-2 \left(\nu q^\prime_{\alpha\alpha,\beta\beta} + (1-\nu)q^\prime_{\alpha\beta,\alpha\beta} \right)$ and $\lambda_p= -2q^0_{12,12}+q^0_{11,22}+q^0_{22,11}$. For membranes (with zero bending stiffness), where \eqref{foppl-vonkarman-1} is disregarded, the natural shape is determined by the non-homogeneous Monge-Amp\'{e}re equation, cf.~\cite{zubov1},
\begin{equation}
 [w,w]=2\lambda_p.
 \label{foppl-vonkarman-21}
\end{equation}

\section{Conclusion} \label{conclusion}
The central aim of our work is to provide an unambiguous description of local defects, within a non-Euclidean geometric framework, in structured surfaces. Our results are applicable to rapidly growing class of 2-dimensional materials, as well as liquid crystalline surfaces, and also for sufficiently thin sandwiched shell structures. Our notion of defect also includes metric anomalies such as those induced during biological growth and thermal deformation. The differential geometric framework naturally leads us to describe defects as sources of strain incompatibility, which, with suitably described kinematics and material response, is manifested physically as internal stress and deformed shape of the material surface. Therefore, we have a formulation which can be used to describe the macroscopic mechanical response of a wide variety of 2-dimensional structures given a distribution of defects.

The present work has been primarily concerned with {\it local} anomalies in materially uniform simple elastic 2-dimensional bodies. Material defects can also appear as \textit{global} anomalies on structured surfaces. The global defect affects the topology of the surface, rendering them, for instance, multiply connected or non-orientable, as is the case with M{\"o}bius and toroidal surface crystals, etc.~\cite{harris70, harris74, bowickgiomi09}. Consider, as an example, the self assembly of certain copolymers in colloidosomes, where toroidal micelles are energetically more favourable over spherical or cylindrical topologies within a range of certain physical parameters. In order for the phase transformation to occur from the unstable spherical, or cylindrical, to the stable toroidal topology (driven by some internal or external agency),  one or more global defects must be introduced in each spherical/cylindrical droplets of the unstable phase to achieve the new topology \cite{jainbates2003, pochanetal2004}. One of the imminent extensions of our work would be to include these global topological defects and revisit the issues of strain incompatibility, internal stress, and natural shape for globally defective surfaces; and most importantly to characterize the geometric interplay between local and global anomalies in structured surfaces. Another direction in which our work can be fruitfully extended is to describe geometry driven, inherently discontinuous, physical phenomena, e.g., incompatibility induced microstructures, such as wrinkles and phase transformations in active structures \cite{ronceray:thesis}, which necessarily require considerations of generalized function spaces and non-convex material response.

\bibliographystyle{plain}
\bibliography{inhomogeneity}

\end{document}